\documentclass[useAMS,usenatbib]{mn2e} 
\expandafter\let\csname equation*\endcsname\relax 
\expandafter\let\csname endequation*\endcsname\relax 

\usepackage{times}
\usepackage{amsfonts}
\usepackage{graphicx}
\usepackage{amsmath}
\usepackage{amssymb}
\usepackage{color}
\usepackage{subfigure}
\usepackage{natbib}
\usepackage{aas_macros}

\usepackage[pdftex,colorlinks,citecolor=black,linkcolor=black,urlcolor=black]{hyperref}

\def\tit{\textit}

\newcommand{\SG}[1]{}
\newcommand{\TJ}[1]{}

\newcommand{\Q}[1]{}

\DeclareMathOperator{\sgn}{sgn}

\def\beq{\begin{equation}}
\def\eeq{\end{equation}}
\def\bea{\begin{eqnarray}}
\def\eea{\end{eqnarray}}
\def\ben{\begin{enumerate}}
\def\een{\end{enumerate}}
\def\la{\langle}
\def\ra{\rangle}
\def\a{\alpha}
\def\b{\beta}
\def\g{\gamma}
\def\d{\delta}\def\D{\Delta}
\def\e{\epsilon}

\def\l{\lambda}
\def\m{\mu}
\def\n{\nu}
\def\o{\omega}
\def\O{\Omega}

\def\r{\rho}
\def\s{\sigma}

\def\S{\Sigma }

\def\phibar{\bar{\phi}}

\def\half{{\textstyle{\frac{1}{2}}}}
\def\fourth{{\textstyle{\frac{1}{4}}}}
\def\w{\wedge}

\def\bD{{\bf D}}

\def\vphit{\tilde{\vphi}}

\def\bphi{\boldsymbol{\phi}}
\def\btau{\boldsymbol{\tau}}

\def\cL{{\cal L}}
\def\cJ{{\cal J}}

\def\cS{{\cal S}}
\def\cC{{\cal C}}

\def\vphi{\varphi}

\newcommand{\citeeg}[1]{\citep[e.g.,][]{#1}}

\begin{document}

\title{Spacetime approach to force-free magnetospheres}

\author[Samuel E. Gralla and Ted Jacobson]{Samuel E. Gralla$^a$ and Ted Jacobson$^{a,b}$\\
$^a$Maryland Center for Fundamental Physics \& Joint Space-Science Institute\\ 
Department of Physics, University of Maryland, College Park, MD 20742, USA\\
$^b$Institut d'Astrophysique de Paris, 98 bis Boulevard Arago, 75014 Paris, France}

\maketitle

\begin{abstract}
Force-Free Electrodynamics (FFE) describes magnetically dominated relativistic plasma via non-linear equations for the electromagnetic field alone.  Such plasma is thought to play a key role in the physics of pulsars and active black holes.  Despite its simple covariant formulation, FFE has primarily been studied in 3+1 frameworks, where spacetime is split into space and time.  In this article we systematically develop the theory of force-free magnetospheres taking a spacetime perspective.  Using a suite of spacetime tools and techniques (notably exterior calculus) we cover 1) the basics of the theory, 2) exact solutions that demonstrate the extraction and transport of the rotational energy of a compact object (in the case of a black hole, the Blandford-Znajek mechanism), 3) the behavior of current sheets, 4) the general theory of stationary, axisymmetric magnetospheres and 5) general properties of pulsar and black hole magnetospheres.  We thereby synthesize, clarify and generalize known aspects of the physics of force-free magnetospheres, while also introducing several new results.
\end{abstract}

\section{Introduction}

Soon after the discovery of pulsars \citep{hewish-etal1968} it became clear that they must be rapidly rotating, highly magnetized neutron stars \citep{gold1968,pacini1968} whose magnetosphere is filled with plasma \citep{goldreich-julian1969}. The plasma mass density is many orders of magnitude lower than the electromagnetic field energy density, so one may neglect the plasma four-momentum and set the Lorentz four-force density to zero. The resulting autonomous dynamics for the electromagnetic field, known as \textit{force-free electrodynamics} (FFE), forms a foundation for studies of the pulsar magnetosphere.

Quasars were discovered several years before pulsars \citep{schmidt1963}, and while supermassive black holes were soon suspected as the energy source, more than a decade passed before the discovery of a viable mechanism for extracting the energy. The breakthrough was the seminal work of \citealp{blandford-znajek1977} (BZ), who argued that black holes immersed in magnetic fields could have a force-free plasma. BZ showed that the presence of plasma enables a magnetic Penrose process in which even \textit{stationary} fields can efficiently extract energy from a spinning black hole.

Despite this important progress, little further was done on the subject for several years. \citet{macdonald-thorne1982} diagnosed the difficulty as a problem of language. In addition to significantly extending the theory, they recast the work of Blandford and Znajek in a $3+1$ decomposition designed to render the equations and concepts more familiar to astrophysicists. The efficacy of their cure is well-supported by the significant progress on the problem that has been made since then, nearly all of it using the $3+1$ approach.

But even the best medicines can have side effects. From the relativist's point of view, the use of $3+1$ methods obscures intrinsic structures and creates unnecessary complications by introducing artificial ones.  For a subject in which curved spacetime and highly relativistic phenomena play central roles, one might expect that the impressive arsenal of spacetime techniques developed over the last century could be profitably exploited. However, 
very few general relativity theorists have become involved, and little work of this nature has been pursued. It may be that the unfamiliar language and phenomena of plasma physics, together with their casting in $3+1$ language, have made the subject largely inaccessible to relativists. 

The beginnings of the field were in fact rather relativistic in flavor, with \citet{znajek1977}'s use of a null tetrad formalism and Blandford and Znajek's tensor component calculations. Since then however there has been little use of spacetime techniques on black hole and pulsar force-free magnetospheres, notable exceptions being the work of \citet{carter1979} and \citet{uchida1997general,uchida1997symmetry,uchida1997linear1,uchida1997linear2,uchida1998oblique}. Our own involvement began recently when we noticed that some apparently disparate exact solutions shared the property of having four-current along a geodesic, shear-free null congruence \citep{brennan-gralla-jacobson2013}. We made a null current ansatz and immediately found a large class of non-stationary, non-axisymmetric exact solutions in the Kerr spacetime, which can also be used in flat spacetime in modeling pulsar magnetospheres. This rapid progress suggested to us that translation of magnetospheric physics into spacetime language may be more than a matter of words, and that a geometrical perspective on force-free electrodynamics could lead to powerful insights and significant new results.

The present paper has a number of distinct purposes. One is to present the theory of force-free magnetospheres with a spacetime perspective from the ground up. In this way we hope both to introduce relativists to the subject and to introduce plasma astrophysicists to potentially powerful new techniques.  We focus on intrinsic properties, avoiding the introduction of arbitrary structures---such as a time function or a reference frame---that have no intrinsic relation to either the spacetime geometry or the particular electromagnetic field being discussed.  The other purposes of our paper are to present new insights, techniques, and results, as well as the convenient methods of exterior calculus we have made use of.  

The paper is organized into nine sections and five appendices:
\begin{enumerate}
\item[1.] Introduction
\item[2.] Astrophysical setting
\item[3.] Force-free electrodynamics
\item[4.] Poynting flux solutions
\item[5.] Monopole magnetospheres
\item[6.] Current sheets and split monopoles
\item[7.] Stationary, axisymmetric magnetospheres
\item[8.] Pulsar magnetosphere
\item[9.] Black hole magnetosphere
\item[A.] Differential forms
\item[B.] Poynting flux examples
\item[C.] Kerr metric
\item[D.] Euler potentials with symmetry
\item[E.] Conserved Noether current associated with a symmetry
\end{enumerate}
A table of contents with subsections is included at the end of the document.  We now provide a detailed description of the contents of each section.

In section \ref{sec:astro} we sketch the relevant astrophysical settings and discuss the basic reasoning that accounts for the validity of the force-free approximation.  We then present the basic features and mathematical structure of force-free electrodynamics and degenerate electromagnetic fields (section \ref{sec:FFE}). 
This section is primarily a review and synthesis of previous research, focusing on the spacetime approaches of Carter and especially Uchida, who formulated the theory in terms of two scalar Euler potentials. We have found that the use of differential forms (with wedge product and exterior derivative) together with Euler potentials provides an elegant and computationally efficient method to handle the mathematics, and we focus on this approach throughout the paper. Appendix A covers the properties of differential forms needed in the paper. We emphasize the geometrical role of certain time-like 2-surfaces that, for degenerate magnetically dominated fields, extend the notion of field line to a spacetime object. These are called ``flux surfaces" in the literature, but 
we adopt here the more suggestive name ``field sheets".
In particular, we observe that the induced metric on these sheets governs the dynamics for particles and Alfv\'{e}n waves moving in the magnetosphere, and explain how field sheet Killing fields give rise to conserved quantities. We also note that the field equations of FFE amount to the conservation of two ``Euler currents,'' which have not been explicitly discussed before. 

Section \ref{sec:poynting} is devoted to presenting several exact solutions to FFE involving outgoing electromagnetic energy flux (Poynting flux) in flat, Schwarzschild, and Kerr spacetimes. These include a solution in Kerr recently found by \citet{menon-dermer2007}, a time-dependent and non-axisymmetric generalization of that \citep{brennan-gralla-jacobson2013}, as well as a solution sourced from an arbitrary accelerated worldline in flat spacetime \citep{brennan-gralla2014}.  We showcase the remarkably simple expression of these solutions in the language of differential forms, as well as the efficient computational techniques we can use to check that they are force-free. The solutions illustrate how force-free fields can transport energy via Poynting flux in ways that are unfamiliar in (but not completely absent from) ordinary electrodynamics. Appendix B is devoted to examples that further develop insight into the physical nature of this energy transport.

Turning next to the physics of magnetospheres, Section \ref{sec:magnetospheres} builds on the Poynting flux solutions to present several exact solution models with a monopolar central rotating source. (The more realistic case of a split monopole is deferred for clarity to the next section.) We begin with a discussion of the classic Michel solution, which illustrates the basic mechanism of electromagnetic extraction and transport of the rotational energy of a conducting 
magnetized star. We obtain this solution as a superposition of a monopole and an outgoing Poynting flux solution satisfying the perfect conductor boundary condition, and use it to illustrate the nature of field sheet geometry. 
We next show how our time-dependent generalizations can be used to model dynamical pulsar magnetospheres. In particular we debut the ``whirling monopole'', which is the exact monopolar magnetosphere of a conducting star undergoing \textit{arbitrary} time-dependent rigid body motion with a fixed center. Finally we discuss the monopolar approximate solution of Blandford and Znajek for a rotating black hole. We obtain their solution to first order in the spin by promoting the Michel solution to Kerr in a simple way. The result is an exceptionally simple expression for the BZ field in terms of differential forms, from which its force-free nature as well as basic properties (such as its ``rotation frequency'' of one half the horizon frequency) are easily seen.

In Section \ref{sec:sheets} we discuss the role of current sheets in force-free magnetospheres and provide a simple invariant criterion for the shape and time evolution of a current sheet across which the electromagnetic field flips sign. We use this criterion to efficiently reproduce the standard aligned and inclined split-monopole solutions and discuss generalizations, such as a glitching split-monopole pulsar.  We also discuss a more general, reflection split construction in which
the magnetic field has a component normal to the current sheet.

Section \ref{sec:stataxi} is devoted to the general theory of stationary, axisymmetric, force-free magnetospheres in stationary, axisymmetric spacetimes. We make
extensive use of the natural $2+2$ decomposition into ``toroidal" submanifolds spanned
by the angular and time translation Killing vectors and the orthogonal ``poloidal" submanifolds.
\citet{uchida1997symmetry}'s method of determining the general form of Euler potentials for fields with symmetry
is presented using differential forms in Appendix D. We explain how and why the field is characterized by three quantities: the ``magnetic flux function" $\psi$,
the ``angular velocity of field lines" $\Omega_F(\psi)$, and the ``polar current" $I(\psi)$,
derive the general force-free ``stream equation'' relating these quantities, and discuss approaches to solving it.
Expressions for the energy and angular momentum flux are derived, using the corresponding Noether current
3-forms whose derivation is given in Appendix E. We explain how the ``light surfaces'' 
(where the field rotation speed is that of light) are 
causal horizons for particles and Alfven waves, and derive the relationship between the particle and angular momentum flow directions.
We discuss general restrictions on the
topology of poloidal field lines, presenting a new result that smooth closed loops cannot occur and clarifying the circumstances under which field lines cannot cross a light surface twice. Finally, we present the stream equation for the special case where there is no poloidal magnetic field, which has been largely overlooked in previous work.

Section \ref{sec:pulsar-magnetosphere} discusses basic properties of
the pulsar magnetosphere in the case of aligned rotational and magnetic axes, using the stationary, axisymmetric formalism of the previous section. We discuss the co-rotation of the field lines with the star as well as the dichotomy between \textit{closed} field lines that intersect the star twice and \textit{open} field lines that proceed from the star to infinity. We clarify the precise circumstances under which closed field lines must remain within the light cylinder, and discuss other circumstances in which they may extend outside.

Section \ref{BHM} addresses black hole magnetospheres, focusing on stationary, axisymmetric fields in the Kerr geometry. We derive the so-called Znajek horizon regularity condition and 
identify an additional condition required for regularity in the extremal case. We discuss the status of energy extraction as a Penrose process 
and discuss the nature of the two light surfaces. Finally we present the ``no ingrown hair'' theorem of \citet{macdonald-thorne1982}, showing that a black hole cannot have a force-free zone of closed poloidal field lines. We discuss the types of closed field lines that can in fact occur.

We adopt the spacetime  signature $(-,+,+,+)$, choose units with the speed of light $c=1$ and Newton's constant $G=1$, and use latin letters $a,b,c,...$ for abstract tensor indices (there is no use of coordinate indices in the paper). 
For Maxwell's equations we use Heaviside-Lorentz units.

\section{Astrophysical Setting}\label{sec:astro}

Force-free plasmas exist naturally in pulsar magnetospheres, and possibly in several other astrophysical systems.  \citet{goldreich-julian1969} pointed out that the rotation of a magnetized conducting star in vacuum induces an electric field, with the Lorentz scalar $\vec{E} \cdot \vec{B}$ non-zero outside the star.  Undeflected acceleration of charges along the direction of the magnetic field will thus occur.  For typical pulsar parameters the electromagnetic force is large enough to overwhelm gravitational force and strip charged particles off the star. 
Even if strong material forces retain the particles, the large $\vec{E} \cdot \vec{B}$ outside the star will create particles in another way \citep{ruderman-sutherland1975}: any stray charged particles will be accelerated to high energy along curved magnetic field lines, leading to curvature radiation and a cascade of electron-positron pair production. These mechanisms act to fill the pulsar magnetosphere with plasma.

To estimate the density of plasma, note that produced charges act to screen the component of $\vec{E}$ along $\vec{B}$, eventually shutting off production when $\vec{E} \cdot \vec{B}$ becomes small enough.  The number of particles created should thus roughly agree with the minimum amount required to ensure $\vec{E} \cdot \vec{B}=0$.  If the particles co-rotate with the star, the required charge density is the so-called Goldreich-Julian charge density $\rho \propto \Omega B$, where $\Omega$ is the stellar rotation frequency.  The minimum associated particle density occurs for complete charge separation (one sign of charge only at each point), which for typical pulsar parameters corresponds to a plasma rest mass density that is \textit{sixteen to nineteen orders of magnitude} (for protons or electrons respectively) smaller than the electromagnetic field energy.  Even if particle production mechanisms significantly overshoot this density, the criterion for the force-free description is easily satisfied.\footnote{While the bulk of the magnetosphere should be force-free, small violating regions of two types can exist.  First, regions where particles are produced may expel those particles with high velocity, so that plasma density high enough to achieve $\vec{E}\cdot\vec{B}=0$ is never attained in those regions.  Such regions are called \textit{gaps}, and may provide a source of the high-energy particles observed in the pulsar wind.  Second, as we discuss in some detail later, force-free fields tend to produce thin sheets of current where the field is not force-free.}  Detailed calculations support these simple arguments, finding an overshoot of a few orders of magnitude \citeeg{beskin-book2010}.

Force-free models of the pulsar magnetosphere provide a foundation on which studies of pulsar emission processes may be based.  Models of pulsed emission generally involve particles or plasma instabilities streaming outwards along the magnetic field lines of the magnetosphere \citeeg{beskin-book2010}.  Pulsed emission is observed in radio, optical, X-ray and gamma-ray, with some pulsars  active only in a subset of these bands, and with a variety of pulse profiles.  
The challenge of modeling these complex features remains an active field of research.  

The force-free model has also been applied to black holes, beginning with the work of \citealp{blandford-znajek1977} (BZ). Following the observation of \citet{wald1974} that immersing a spinning black hole in a magnetic field gives rise to electric fields with non-zero $\vec{E}\cdot\vec{B}$, BZ argued that a pair-production mechanism could also operate to produce a force-free magnetosphere near a spinning black hole with a magnetized accretion disk. If the whole system is simulated using magnetohydrodynamics (e.g., \citealp{mckinney-tchekhovskoy-blandford2012} and references therein), it is generally found that the plasma density is very low away from the disk (and especially in any jet region), so that the dynamics there is effectively force-free.
Finally, the last few years has seen work on force-free magnetospheres of binary black hole and neutron star systems \citeeg{palenzuela-lehner-liebling2010,alic-etal2012,palenzuela-etal2013,paschalidis-etienne-shapiro2013}, motivated in part by the possibility of observing electromagnetic counterparts to gravitational-wave observations of binary inspiral.  These simulations have shown energy extraction and jet-like features, even in the case of \textit{non-spinning} (but moving) black holes.

\section{Force-free electrodynamics}
\label{sec:FFE}

In this section we introduce the essential properties of force-free electrodynamics in an
arbitrary curved spacetime background and its description
in the language of differential forms. 

An electromagnetic field $F_{ab}$ normally exchanges energy and momentum 
when interacting with charged matter. 
The energy-momentum tensor for the field is given by 
\beq\label{TEM}
T_{ab}^{\rm EM}= F_{ac} F_b{}^c-\frac14 F_{cd} F^{cd} g_{ab},
\eeq
and Maxwell's equations imply that the exchange is expressed by the equation $\nabla^b T_{ab}^{\rm EM}= -F_{ab} j^b$, where 
$j^b$ is the electric 4-current density. 
$F_{ab} j^b$ is the 4-force density, describing the rate of transfer of energy and momentum between the field and the charges. 
Force-free electrodynamics (FFE) describes the electromagnetic field interacting with a plasma 
in a regime in which the transfer of energy and momentum from the field to
the plasma can be neglected, not because the current is unimportant, but because the field 
energy-momentum overwhelms that of the plasma. 
FFE is thus governed by Maxwell's equations together with the
force-free condition 
\beq\label{FF}
F_{ab} j^b=0.
\eeq
In this regime, remarkably, the field can be evolved 
autonomously, without keeping track of any plasma degrees of freedom, as we now explain.

Maxwell's equations take the form
\begin{align}
\nabla_{[a}F_{bc]}&=0,\label{dF=0}\\ 
 \nabla_b F^{ab} &= j^a,\label{sources}
\end{align}
where the square brackets denote antisymmetrization of the indices, $\nabla_b$ is the spacetime
covariant derivative, and we use Heaviside-Lorentz units.  The first equation is equivalent to the statement that $F_{ab}$ is (at least locally) derivable from a potential, $F_{ab}=2\nabla_{[a} A_{b]}$. The second equation relates the field to the electric 4-current density.  In the force-free setting, this second equation is simply used to identify the 4-current, and so the equation imposes no condition on the field.  We may thus eliminate $j^a$ from the description, and FFE becomes the pair of equations
\beq\label{FFF}
\nabla_{[a}F_{bc]}=0, \quad F_{ab}\nabla_c F^{bc}=0.
\eeq
Note that vacuum Maxwell fields trivially satisfy these equations.  In this paper a ``force-free solution'' will always mean a non-vacuum solution of Eqs.~\eqref{FFF}, i.e.\ one with $\nabla_bF^{ab}\neq 0$.  This is the case of relevance to plasma magnetospheres, and it has a rich structure distinct from that of the vacuum case.

\subsection{Determinism}
The FFE equations determine the evolution of the field given initial data, provided the field is 
magnetically dominated i.e. $F_{ab}F^{ab} = 2(B^2 - E^2) >0$.  
To see how this could be, one can make a 3+1 decomposition in flat spacetime.
The force-free condition (\ref{FF}) then takes the form
\beq\label{FFEB} 
\vec E\cdot \vec \jmath = 0,\quad \quad \rho\vec E+\vec \jmath \times \vec B = 0,
\eeq
stating that the work done on the charges and the momentum transfer to the charges both vanish. 
These equations imply the (Lorentz invariant) condition
\beq\label{EdotB}
\vec E\cdot\vec B=0,
\eeq
unless both the charge and 3-current densities vanish.
Provided $|\vec B|\ne0$ (which holds in all frames if the field is magnetically dominated),  (\ref{FFEB}) determines $\vec \jmath_\perp = |B|^{-2}\rho\vec E\times\vec B$,
the component of the 3-current perpendicular to the magnetic field. Moreover, 
Gauss' law $\vec\nabla\cdot \vec E = \rho$ determines the charge density in terms of spatial derivatives
at one time. To determine the component of $\vec\jmath$ parallel to the magnetic field, 
consider Maxwell's time evolution equations, 
\begin{align}
\partial_t\vec B &= -\vec\nabla\times\vec E\label{Faraday}\\
\partial_t\vec E &= \vec\nabla\times\vec B - \vec\jmath.\label{Ampere}
\end{align}
The time derivative of the orthogonality condition (\ref{EdotB}) implies that 
$\vec E \cdot\mbox{(\ref{Faraday})}+\vec B \cdot \mbox{(\ref{Ampere})}$ vanishes,
which determines $\vec \jmath \cdot \vec B$.
Thus the force-free condition implies 
\beq\label{jEB}
\vec \jmath =\frac{1}{B^2}\left[(\vec\nabla\cdot\vec E)\vec E\times\vec B + (\vec B\cdot\vec\nabla\times\vec B -
\vec E\cdot\vec\nabla\times\vec E)\vec B\right]. 
\eeq
With this substitution, equations (\ref{Faraday}, \ref{Ampere}) 
determine the time derivatives of the fields in terms of the field values at one time, and
the initial value constraints $\vec\nabla\cdot \vec B=0$ and $\vec E\cdot\vec B=0$ are preserved by the 
time evolution. The equations are therefore potentially deterministic. It turns out that they are indeed deterministic (i.e., hyperbolic),
provided the (Lorentz invariant) scalar $B^2-E^2$ is positive \citep{komissarov2002,palenzuela-etal2011,pfeiffer-macfadyen2013}.
 That restriction is not surprising, since when this scalar is negative, there exists at each 
point a Lorentz frame in which $\vec B=0$. In such a frame one cannot solve for $\vec \jmath$ at that
point in terms of the fields and their spatial derivatives only. This shows that the character of the
equations is different in the electrically dominated case.

There is no a priori reason to expect that the condition $B^2>E^2$ is preserved under time evolution.  In fact, it is seen numerically that the condition is \textit{not} preserved.  When the condition is violated some other physics must determine the evolution, which is modeled via various prescriptions in numerical codes.  It is generally found that violation occurs only in regions that are stable under the associated prescriptions, and that these regions tend to be compressed and of high current density: they are the \textit{current sheets} discussed below in Sec.~\ref{sec:sheets}.

\subsection{Degenerate electromagnetic fields}
\label{sec:degenerate}

In this subsection we discuss electromagnetic fields satisfying $F_{[ab}F_{cd]}=0$ (equivalently  $\vec{E}\cdot\vec{B}=0$ in flat spacetime), which are called \textit{degenerate}.  All force-free fields are degenerate, but degeneracy 
 can occur more generally, as explained below.

\subsubsection{Field tensor}\label{sec:degen}
The force-free condition (\ref{FF}) implies that $F_{[ab}F_{cd]}j^d=0$. 
Since every totally antisymmetric four-index tensor (in four dimensions) is 
proportional to the volume element $\e_{abcd}$, this implies the degeneracy condition,
\beq\label{FF=0}
F_{[ab}F_{cd]}=0,
\eeq
which is equivalent to (\ref{EdotB}) in flat spacetime.
This in turn implies that $F_{ab}$ itself can be written as the anti-symmetrized 
product of two rank-1 covectors,\footnote{This factorization property holds at each point, but it can happen that there is no pair of smooth tensor fields $\a_a$ and $\b_b$ such that (\ref{simple}) holds everywhere (for an example see the end of Section 3.5 of \cite{penrose-rindler-book1}).}
\beq\label{simple}
F_{ab}=2\alpha_{[a}\beta_{b]}.
\eeq
To see this, consider the contraction $F_{[ab}F_{cd]}v^b w^d$ with two vector fields $v^a$ and $w^a$ such that $F_{ab}v^a w^b\neq0$.  Expanding out the antisymmetrization produces an expression for $F_{ab}$ of the form \eqref{simple}, where the factors $\alpha_a$ and $\beta_a$ are proportional to $F_{ab}v^b$ and $F_{ab}w^b$.

An electromagnetic field can be degenerate without being force-free. Degeneracy occurs any time there is some vector field 
$v^b$ such that $F_{ab}v^b=0$. For instance, in the presence of a ``perfect" conductor (like a metal or a suitable plasma), the electric field in the local rest frame of the conductor vanishes,
\beq\label{perfect}
F_{ab}U^b=0,
\eeq
where $U^a$ is the unit timelike 4-velocity of the conductor's rest frame.
Thus fields in perfect conductors are degenerate.
For an ionic plasma described by ideal magnetohydrodynamics (MHD)
$U^a$ might be the 4-velocity of the ion ``fluid", 
but degeneracy does not require a unique rest frame to be singled out.  
As long as there is enough free charge to screen
the component of the electric field in the direction of the magnetic field, 
$\vec E\cdot\vec B$ will vanish and hence
the field will be degenerate.

Conversely, a degenerate field $F_{ab}=2\a_{[a}\b_{b]}$ 
always admits at each point a two-dimensional
space of vectors that annihilate it (in the sense that $F_{ab} v^a=0$).
This space, called the \textit{kernel} of $F_{ab}$,  consists of the 
intersection of the three-dimensional kernels of $\a_a$ and $\b_a$.  The covectors $\alpha_a$ and $\beta_a$ themselves span a (co)plane, and any two linearly independent, suitably scaled covectors in the coplane may be chosen.  Taking $\a_a$ and $\b_a$ to be orthogonal, the square of the field tensor is then
\beq
F^2 = F_{ab}F^{ab} = 2(B^2 - E^2) = 2\a^2\b^2.
\eeq
The sign of this Lorentz scalar determines whether the
field is magnetically dominated, electrically dominated, or null. Since there do not exist two orthogonal timelike vectors, this is positive if and only if both $\a$ and $\b$ are spacelike.

For a magnetically dominated field the $\a$-$\b$ plane is thus spacelike
and the kernel, which is orthogonal to $\a$ and $\b$, is timelike.
There is a 1-parameter family of 4-velocities $U^a$ lying in this timelike kernel, each of which
defines a Lorentz frame in which the electric field vanishes (\ref{perfect}). The
orthogonal projection of a preferred frame $t^a$ into the kernel of $F$ selects one of these,
$U_t^a$, whose velocity relative to $t^a$ is known as the \textit{drift velocity}. This relative velocity
is the minimum for all $U^a$ in the kernel of $F$, and is given by $\vec{E}\times\vec{B}/B^2$
in the frame $t^a$.

For a field with $E^2=B^2$, either or $\a$ or $\b$ must be null, so the $\a$-$\b$ plane is null
and so is the kernel (with the same null direction).
For an electrically dominated field the $\a$-$\b$ plane is timelike,
so the kernel is spacelike, and there is always a Lorentz
frame in which the magnetic field vanishes (since the kernel of $*F$ is timelike).

\subsubsection{Stress Tensor}

For a non-null degenerate field, one can
decompose the spacetime metric into
a metric $h_{ab}$ on the kernel
of $F_{ab}$ that vanishes on vectors
orthogonal to the kernel
(so $h_{ab}\alpha^a = h_{ab}\beta^b=0$)
and a metric $h^\perp_{ab}$ that vanishes
on vectors in the kernel, $g_{ab}=h_{ab} + h^\perp_{ab}$.
Using these the stress tensor \eqref{TEM} can be expressed as
\beq\label{Tdeg}
T_{ab} = \frac14 F^2(h^\perp_{ab} - h_{ab}).
\eeq
This can be quickly verified by noting that the right hand side
is the only symmetric tensor built from the available
ingredients that is traceless and satisfies $T_{ab}h^{ab} =
-\frac12 F^2$, which holds because $F_{bc}h^{ab}=0$.  

In the magnetic case and in a $3+1$ decomposition, Eq.~\eqref{Tdeg} may be interpreted in terms of the standard concepts magnetic pressure and magnetic tension.  Choose any frame in which there is no electric field, i.e., any unit timelike $U^a$ in the kernel of $F$.  Let $s^a$ be the unit orthogonal spacelike vector in the kernel.  The magnetic field in this frame is directed along $s^a$, and we denote its magnitude by $B$.  If $\gamma_{ab}$ is the spatial metric orthogonal to $U^a$, then the stress tensor \eqref{Tdeg} may be written $T_{ab} = \tfrac{1}{2} B^2 (U_a U_b + \gamma_{ab} - 2 s_a s_b)$.  From each term, respectively, we identify the energy density of $\tfrac{1}{2}B^2$, an isotropic magnetic pressure of $\tfrac{1}{2}B^2$, and a magnetic tension  of $B^2$ along the magnetic field lines.

\subsubsection{Field sheets}
\label{fieldsheets}

When a degenerate field $F_{ab}$ satisfies the Maxwell equation $\nabla_{[c}F_{ab]}=0$ (\ref{dF=0}), 
the kernels of $F_{ab}$ are \textit{integrable}, i.e.\ tangent to two-dimensional submanifolds.  (A proof of this will be given in the next subsection.)  In the magnetic case ($F^2>0$) these submanifolds are timelike, and their intersection with a spacelike hypersurface gives the magnetic field lines defined by the observers orthogonal to the hypersurface.\footnote{Curiously, for any degenerate field, the magnetic field defined by \textit{arbitrary} observer, $B^d = \half \epsilon^{abcd} F_{ab} U_c$, lies in the kernel of $F$, since $F_{[ab}F_{d]e}=F_{[ab}F_{de]}=0$.}
Each submanifold can thus be thought of as the \textit{spacetime evolution} of a field line,
which we will call a \textit{field sheet}.\footnote{The submanifolds were called ``flux surfaces" by Carter and Uchida.  We prefer the name field sheet because of the connotation of time evolution suggested by the similarity with the term ``worldsheet'', which is appropriate in the magnetic case.  We note also that the term ``flux surface" is commonly used in another sense, to describe a spacelike 2-surface to which the magnetic field is everywhere tangent.}  
While the field lines depend on the arbitrary choice of spacelike hypersurface or observers, the field sheets are 
an intrinsic aspect of the degenerate structure of the field.
The force-free condition \eqref{FF} amounts to the statement that the current four-vector 
$j^a$ is tangent to the field sheets. This generalizes to dynamical fields in curved spacetime
the statement that, in a force-free plasma with zero electric field in flat spacetime,
the current is tangent to the magnetic field lines.

The field sheets can be used to understand and describe particle and wave motion in the underlying plasma in a manner that does not require choosing an arbitrary frame. In that application the \textit{field sheet metric}, induced by the spacetime metric, plays a central role.  We now discuss two examples of this viewpoint: the propagation of charged particles and Alfv\'en waves.
 
In a collisionless plasma, viewed (locally) in a frame with zero electric field,
a charged particle will spiral around a magnetic field line, 
executing cyclotron motion while the center of the transverse circular orbit is 
``guided" along the field line.  Ignoring the cyclotron motion
and the drift away from the field line, 
the particle is thus ``stuck'' on the field line \citeeg{northrop-teller1960}.  
The manifestly frame-invariant version of this statement is that the particle's 
worldline is stuck on the field sheet.  
That is, its possible motions are the timelike trajectories on the sheet.

When one can furthermore neglect radiation reaction from ``curvature radiation" due to the bending of the field sheet, then the motion of the particle is in fact \textit{geodesic} on the field sheet.  This follows simply from the fact that the Lorentz force $q F_{ab}U^b$ vanishes for a four-velocity $U^a$ tangent to the sheet.  This viewpoint makes it easy to exploit symmetries. For example, in a stationary, axisymmetric magnetosphere, each field sheet will have a helical symmetry under a combined time-translation and rotation.  The field sheet particle motion, being one-dimensional, will thus be integrable using the associated conserved quantity (see Sec.~\ref{sec:chiF}). 

Field sheet geometry also governs the propagation of Alfv\'en waves,  which
are transverse oscillations of the magnetic field lines embedded in a plasma \citep{alfven1942}. In a force-free plasma these are characterized by a wave 4-vector whose pullback to the field sheet is null with respect to the field sheet metric \citep{uchida1997linear2}, which implies that their group 4-velocity is null and tangent to the field sheet. Thus wavepackets propagate at the speed of light along the field sheets.

\subsubsection{Degenerate fields and differential forms}

The mathematical language of \textit{differential forms} is ideally suited to working with
degenerate fields, and we shall make extensive use of it in this paper.
The basic properties of differential forms are summarized in Appendix \ref{sec:forms},
to which we refer for all definitions. One of the reasons it is so convenient is that electromagnetism in general, and especially when fields are degenerate, has a rich differential and algebraic structure
that is in fact independent of the spacetime metric. By using the (metric-independent) exterior derivative, and (metric-independent) wedge products rather than covariant derivatives and inner products,  we avoid unnecessary appearance of the metric and thus keep the formalism as close as possible to the structure inherent in the field itself. The metric does of course play a role, but for the most part we can sequester that in the Hodge duality operator (which is especially simple to work with in stationary axisymmetric spacetimes).

The field strength tensor is a 2-form, denoted simply by $F$, and the source-free Maxwell 
equation (\ref{dF=0}) corresponds to the statement that the
2-form $F$ is ``closed", i.e.\ 
\beq\label{dF=0-form}
dF=0,
\eeq
where $d$ is the exterior derivative.
This equation, which we dub the \textit{covariant Faraday law}, 
encompasses both the absence of magnetic monopoles and 
the 3+1 Faraday law \eqref{Faraday}.
The degeneracy condition (\ref{FF=0}) and decomposition (\ref{simple}) are 
expressed using the wedge product $\w$ as 
\beq
F\wedge F=0
\eeq
and
\beq\label{simpleform}
F = \a\wedge \b,
\eeq
for some pair of 1-forms $\a$ and $\b$.
A 2-form $F$ with this property is sometimes called \textit{simple}.

To prove that the field sheets exist one can invoke a version 
of the Frobenius theorem: it follows from $dF=0$ and 
$F=\a\w\b$, together with the anti-derivation property of $d$ 
and antisymmetry of $\w$, 
that $d\a\wedge\a\wedge\b=d\b\wedge\a\wedge\b=0$.
This guarantees complete integrability of the Pfaff system $\a=\b=0$ \citep{choquetbruhat-dewitt1982}, which means that the vectors annihilating both $\a$ and $\b$ are tangent to submanifolds. 
A more intuitive argument for integrability will be given in the next subsection.   

\subsubsection{Frozen flux theorem}

If the electric field vanishes in the local rest frames 
defined by a timelike vector field $U$,
\beq\label{U.F=0}
U\cdot F = 0,
\eeq
then the magnetic flux is ``frozen in" along the flow of $U$.  
[The dot in \eqref{U.F=0} is our notation for the contraction of a vector with the
first slot of the adjacent form, here $U^a F_{ab}$.]
More precisely, the flux through a loop is conserved if the loop is flowed along $U$.  In ideal MHD the fluid four-velocity satisfies \eqref{U.F=0}.  The frozen flux theorem (also known as the frozen-in theorem or Alfv\'en's theorem) is the source of 
much insight into the behavior of such plasmas. 

To prove the theorem, consider a loop flowed along $U$ to create a timelike tube, and form a closed 2-surface by capping the ends of the tube with topological disks bounded by the initial and final loops. The integral of $F$ over any closed 2-surface vanishes since $dF=0$. 
The difference of the fluxes through the initial and final caps is therefore 
equal to the integral of $F$ on the tube wall, which vanishes because 
the vector $U$ that annihilates $F$ \eqref{U.F=0} is tangent to the wall.
Using the language of differential forms, Alfv\'en's theorem is thus recovered immediately, with no calculation, in an arbitrary curved spacetime.
It is interesting to contrast the simplicity of this completely general 
derivation with the usual one using electric and magnetic fields in flat spacetime.

A differential version of the statement of flux freezing may be obtained from the 
relation between the Lie derivative and the exterior derivative, sometimes called ``Cartan's magic formula", 
\beq\label{magic}
\cL_v\, \o = v\cdot d\o + d(v\cdot\o).
\eeq
Here $v$ is any vector field, $\o$ is any differential form, and $\cL$ is the Lie derivative. 
Applying the magic formula to $L_UF$, the $U\cdot dF$ term vanishes by the covariant Faraday law \eqref{dF=0}, and the $d(U\cdot F)$ term vanishes simply by the defining property (\ref{U.F=0}) of $U$.  Thus we obtain
\beq\label{LUF=0}
\cL_UF=0,
\eeq
stating that the field strength is preserved along the flow of $U$.  
In ideal MHD, the magnetic field is thus ``frozen into the fluid".

The frozen flux theorem is closely related to the integrability property 
that implies the existence of the field sheets. 
Recall that if $F$ is degenerate there is a two-dimensional space of 
vectors annihilating $F$ at each point. To prove that these  
are surface forming, let 
$X$ and $Y$ be two linearly independent
vector fields in the kernel of $F$, i.e., such that 
$X\cdot F=Y\cdot F =0$ everywhere.
As above, Cartan's magic formula implies $\cL_X F = 0$, 
so
$[X ,Y]\cdot F = {\cal L}_X(Y \cdot F) - Y \cdot \cL_X F=0$, 
hence 
$[X,Y]$ is also in the kernel. Since $X$ and $Y$  span the kernel, 
it follows that $[X,Y]$ is a linear combination of $X$ and $Y$, 
which by the Frobenius theorem implies that $X$ and $Y$ 
are tangent to 2-dimensional submanifolds. 
These integral surfaces 
are the field sheets. \textit{[Note: This paragraph replaces a less transparent argument given in the previous version; we thank John Friedman for suggesting this improvement.]}

\subsubsection{Euler potentials}

The covariant Faraday law (\ref{dF=0-form}) 
is equivalent, at least locally, to the statement that 
$F$ derives from a potential 1-form,
i.e.\ $F=dA$ for some 1-form $A$. 
For closed, simple 2-forms (such as degenerate EM fields), thanks to the existence of 
the field sheets, a much more 
restrictive statement holds: a pair of scalar ``Euler potentials" $\phi_{1,2}$ can be introduced 
such that \citep{carter1979,uchida1997general}
\beq\label{Euler}
F=d\phi_1\w d\phi_2.
\eeq
The  field sheets are the intersections of the hypersurfaces
of constant $\phi_1$ and $\phi_2$. 
To establish the (local) existence of the Euler potentials, 
note that coordinates $(x^A,y^i)$, $A,i=1,2$ can be chosen such that $y^i$ 
are constant on the field sheets, in which case we have $F=f(x^A,y^i) dy^1\w dy^2$, for some function $f$.
Then $dF=0$ implies that $f=f(y^i)$. Defining a new coordinate $\tilde{y}^1=\int f dy^1$,
we thus have $F=d\tilde{y}^1\w dy^2$.

The Euler potentials capture the freedom in a closed, simple 2-form, hence in any degenerate 
electromagnetic field. Rather than the four components of a (co)vector potential, 
there are just two scalar fields.
Even so, the potentials are not uniquely determined. 
$F$ defines an `area element' on the field sheets, which is preserved
under any replacement $(\phi_1,\phi_2)\rightarrow(\phi_1'(\phi_1,\phi_2), \phi_2'(\phi_1,\phi_2))$ with unit Jacobian determinant.  This is a field redefinition, not a dynamical gauge freedom.  In fact, the second time derivatives of both potentials are determined at each point by their value and first derivatives \citep{uchida1997general}.

\subsection{Euler-potential formulation of FFE}

Since all force-free fields are degenerate, we may formulate FFE as a theory of two scalar fields by plugging the Euler-potential form of a degenerate field strength \eqref{Euler} in to the force-free condition \eqref{FF}.  Rather than develop this technique in tensor language, we will instead discuss the differential forms version, which we find very useful in calculations.  We also discuss an action principle for the equations. 

\subsubsection{Force-free condition \& Euler currents}

The differential forms approach to Maxwell theory---using the current 3-form instead of the four-vector---is reviewed in Appendix \ref{sec:maxform}.
The force-free condition (\ref{FF}) can 
be expressed directly in terms of the current 3-form as 
\beq\label{FwJ}
F_{a[b}J_{cde]}=0.
\eeq
To see the equivalence with (\ref{FF}), contract
(\ref{FwJ}) with $\epsilon^{bcde}$ and use
$j^b = \frac{1}{3!}\epsilon^{bcde}J_{cde}$. 
In terms of the 1-form factors of $F$ (\ref{simpleform}),
this corresponds to the two conditions
\beq\label{FFabJ}
\a\w J = 0= \b\w J.
\eeq
The vanishing of these two 4-forms is an
extremely simple and convenient characterization of the force-free condition. 
Given Maxwell's equation $d*F=J$, it amounts to the conditions
\beq\label{FFFE}
\a\w d*F = 0=\b\w d*F.
\eeq
When the Euler potentials are used to express $\a$, $\b$ and 
$F$ as in (\ref{Euler}), Eqs. (\ref{FFFE}) become
\beq\label{EFFFE}
d\phi_{i}\w d*F = 0,\quad i=1,2
\eeq
which
comprise the full
content of force-free electrodynamics.

Note that these equations are equivalent to the statement that
two currents are conserved:
\beq\label{EulerCurrents}
d(d\phi_i\w *F)=0.
\eeq
The currents $d\phi_i\w *F$ 
deserve a name; we propose to call them \tit{Euler currents}.
That the force-free equations amount to the conservation of these two Euler currents
is a trivial but useful observation which does not appear to have been made previously.  In tensor notation, the Euler currents are given 
(up to a coefficient) by $F^{ab}\nabla_b \phi_i$.  Note that we could have alternatively defined the Euler currents to be $\phi_i J$, which differs from the previous definition by the identically conserved 3-form $d(\phi_i * \! F)$.

\subsubsection{Action}

One can arrive directly at the
force-free conditions (\ref{EulerCurrents}) starting from the usual Maxwell action
$-\tfrac{1}{2}\int F\w *F$, expressed as a functional of the potentials,
\beq\label{SFF}
S^{\rm FF} = -\frac12\int d\phi_1\w d\phi_2\w * (d\phi_1\w d\phi_2).
\eeq
Variation of this action with respect to 
$\phi_1$ and $\phi_2$ yields conservation of the 
Euler currents (\ref{EulerCurrents})
as a pair of Euler-Lagrange equations.\footnote{These Euler currents are the Noether currents associated with the global symmetries $\phi_1 \rightarrow \phi_1 + f_1(\phi_2)$ and $\phi_2 \rightarrow \phi_2 + f_2(\phi_1)$ of the action \eqref{SFF}.}
This action, and the  Hamiltonian formulation
derived from it, were
given by \citet{uchida1997general}.\footnote{An alternate approach \citep{thompson-blaes1998,buniy-kephart2014}
is to supplement the usual Maxwell action for the vector potential
with a Lagrange multiplier term enforcing the degeneracy condition,
$S= -\frac12\int F\w *F -\l F\w F$. The resulting Euler-Lagrange equations
are $d*F = d\l\w F$ and $F\w F=0$, another formulation of FFE.
We learn from this that $J=d\l\w F=d\l \w d\phi_1 \wedge d\phi_2$ for some scalar field $\l$, which immediately implies the force-free conditions \eqref{EFFFE}.  (Conversely, it is possible to show that $J$ has this form directly from the force-free conditions \eqref{FFabJ} and $dF=0$.)} Note that the Lagrangian is quadratic in time derivatives,
so the equations of motion are second order in time derivatives. 

The action is a scalar, 
so the stress-energy tensor is conserved when the equations of motion are
satisfied. This is to be expected, since our starting point was
the force-free condition which implies that the field transfers no energy or momentum 
to the charges. Moreover, the dynamics
shares the symmetries possessed by the $*$ operator on 2-forms,
namely symmetries and Weyl rescalings of the metric. 
This implies, for instance, that in a stationary axisymmetric spacetime there are 
conserved Killing energy and axial angular momentum currents, and that
force-free electrodynamics shares
with vacuum electrodynamics the property of depending only 
on the conformal structure of the spacetime. 
The potentials can also be restricted by a symmetry ansatz before variation,
to directly obtain the equations governing the symmetric solutions.

\subsubsection{Complex Euler potential}

Finally, it seems worth noting that the two Euler potentials
can be combined into one complex potential $\phi = (\phi_1+i\phi_2)/\sqrt{2}$. 
Then the field 2-form is given by $F=i\, d\phi\w d\phibar$, the 
force-free field equations correspond to the single complex equation
$d\phi\w d*F=0$, and the action is $\half \int  d\phi\w d\phibar \w *(d\phi\w d\phibar)$.  Whether this complex formulation is useful remains to be seen. 
 
\section{Poynting flux solutions}\label{sec:poynting}

In this section we recover and discuss a number of exact
solutions to the force-free field equations (\ref{FFFE}) using the method of exterior calculus.  In addition to introducing some important properties of force-free physics, we hope that this section will serve as a tutorial on computing with differential forms, for readers unfamiliar with that approach.  The most unfamiliar element is perhaps the use of the Hodge dual in place of the metric.  In Appendix \ref{Hodge} we review this operator and develop some computational techniques.  With the aid of these techniques, computations using forms can be remarkably simple, as we demonstrate below. We begin by discussing the magnetic monopole, then cover solutions describing purely outgoing (or ingoing)
Poynting flux, and finally superpose these to obtain the general solution used to construct monopole
magnetospheres in the following
section.

\subsection{Vacuum monopole}

To warm up, we begin with the magnetic monopole in the Schwarzschild background (which of course includes flat spacetime as a special case). It is a vacuum solution, and monopoles do not exist in nature, yet it has played an important role in the analytical modeling of force-free magnetospheres since the earliest years of the subject. The field strength 2-form is given by 
\beq\label{Fmon}
F^{\rm mon} = q \sin\theta\,  d\theta\w d\vphi.
\eeq
This is proportional to the area element on the sphere, and has the
same flux integral ($4\pi q$)  for any radius, so it is clearly the monopole
field.\footnote{Monopole charge is conventionally defined to equal the flux integral.  Our $q$ is thus $1/4\pi$ times the usual notion; we nevertheless refer to $q$ as the monopole charge.} But to illustrate the exterior calculus, let us check that the field equations
are satisfied. We have $dF^{\rm mon}=q\cos\theta\,  d\theta\w d\theta\w d\vphi$,
which vanishes because $d\theta\w d\theta=0$.
As for the other field equation, 
according to (\ref{*thetaphi}) the dual of the monopole 2-form 
is $*F^{\rm mon} = qr^{-2}dt\w dr$, so 
$d*F^{\rm mon} = -2qr^{-3}dr\w dt\w dr$. This too vanishes, 
because $dr\w dr = 0$.  The $3+1$ version of the magnetic monopole field in flat spacetime is $\vec{B}=(q/r^2) \hat{r}, \vec{E}=0$. 

The 2-form (\ref{Fmon}) is simple, i.e.\ the monopole field is degenerate. In particular this implies it
can be expressed in terms of Euler potentials, which can be taken as $\phi_1=-q\cos\theta$ and 
$\phi_2 = \vphi$. Note that the discontinuity of $\vphi$ at $2\pi$ means that the Euler potential is not globally
smooth. This presents no problem; moreover, were it not for this 
discontinuity, the field would be an ``exact form" $dA^{\rm mon}$, with $A^{\rm mon} = q\vphi \,d(\cos\theta)$,
so the total magnetic flux through the closed surface of the 2-sphere 
would necessarily vanish.\footnote{The discontinuity could be avoided by using 
instead  the potential $A^{\rm mon}=-q\cos\theta\,d\vphi$. However, the norm of the 1-form
$d\vphi$ is $(g^{\vphi\vphi})^{1/2}=1/(r\sin\theta)$, which blows up at the poles. This can be 
fixed at the north pole by using instead $A^{\rm mon,N}=-q(\cos\theta-1)\,d\vphi$ and at the
south pole by using $A^{\rm mon,S}=-q(\cos\theta+1)\,d\vphi$, which differs from the northern
potential by the pure gauge piece $-2q\, d\vphi$. The discontinuity of $\vphi$ implies that this
gauge transformation is not trival however, which accounts for the existence of a nonzero
magnetic flux through the sphere.}

\subsection{Outgoing Poynting flux}

The next solution we consider is genuinely force-free ($j^a \neq 0$) and remarkably simple and general.   The solution is on Schwarzschild (and flat) spacetime and has no symmetries at all, being given in terms of a free function of three variables as 
\beq\label{Fout}
F^{\rm out}=d\zeta \w du,
\eeq
where $\zeta=\zeta(\theta,\varphi,u)$ is a function of retarded time (outging Eddington-Finklestein time) $u$ and the sphere angles $(\theta,\varphi)$.  (In flat spacetime, $u=t-r$.)  This solution was first found in \cite{brennan-gralla-jacobson2013} using a Newman-Penrose formalism, but here we analyze it in the simpler language of differential forms.  Comparison with Eq.~\eqref{Euler} shows that $\zeta$ and $u$ are Euler potentials for this solution.  Since $du$ is null and orthogonal to $d\zeta$, $F^{\rm out}$ is a null 2-form. The flat spacetime electric and magnetic fields are given below in Eq.~\eqref{FoutEB}.
It is evident from Eq.~\eqref{Fout} that $d F^{\rm out}=0$.  To check the force-free condition we use Eq.~\eqref{nulldual} for the dual of a null 2-form, giving $*F^{\rm out}=*(d\zeta\w du) \sim \star d\zeta\w du$, where $\star$ indicates dual on the sphere.  The current is $J= d*F^{\rm out}\sim d\theta\w d\vphi\w du$, showing that $d\zeta\wedge J = du \wedge J=0$, i.e., the force-free equations (\ref{FFFE}) are satisfied.

The electromagnetic stress-energy tensor $T_{ab}^{\rm EM}$ (\ref{TEM}) associated with Eq.~\eqref{Fout} is given by 
\beq
T_{ab}^{\rm out}=|d\zeta|^2  (du)_a (du)_b
\eeq
where $|d\zeta|^2$ denotes $g^{ab}(d\zeta)_a (d\zeta)_b$. Thus the solution represents a flow of electromagnetic energy along the outgoing radial null direction $(du)^a$.  
Because of this flux, we refer to $F^{\rm out}$ as the outgoing flux solution.  
The net flux of Killing energy leaving the system at retarded time $u$, calculated at $r=\infty$, is given by
\beq\label{Prad}
\mathcal{P}^{\rm out}(u) \equiv \lim_{r \rightarrow \infty}\int T_{ab}^{\rm out} (\partial_t)^a (dr)^b d\Omega = \int |d\zeta|^2 d\Omega,
\eeq
where $d\Omega$ is the area element on the unit sphere.  Since the Killing energy is conserved as it propagates, this is also the Killing flux per Killing time through a sphere at any radius.\footnote{The concept of Killing time applies to an individual integral curve of the Killing field $\xi^a$, and is given by the lapse of $\lambda$ along the curve, where $\lambda$ is any function satisfying $\xi^a \nabla_a \lambda=1$ on the curve.  In Schwarzschild, possible choices for $\lambda$ include the usual time coordinate $t$ as well as the outgoing and ingoing Eddington-Finklestein coordinates $u$ and $v$.  The Killing time may equivalently be defined as the lapse of parameter along the curve, when parameterized so that the tangent vector equals the Killing vector.}

The energy flow in the field \eqref{Fout} is unlike ordinary electromagnetic radiation in that the flux persists for stationary fields, i.e., energy is carried away even if $\zeta$ is independent of $u$.  In this case the solution has more the character of a flow than a wave, and such flows are sometimes called ``electromagnetic winds'' or ``Poynting winds''.  For vacuum fields this situation is impossible with isolated sources, but it does occur in waveguides and in planar symmetry.  In fact, these scenarios admit vacuum solutions that are highly analogous to Eq.~\eqref{Fout}.  In Appendix \ref{sec:flux} we explore these examples as context for understanding the outgoing flux solution.

By itself, the outgoing flux solution is unphysical, since it describes energy emerging from the origin of coordinates in flat spacetime (where the solution is singular), or from the past horizon on the analytic extension of the Schwarzschild spacetime.\footnote{Note also that the solution is not regular on the future horizon unless $d \zeta$ vanishes as $u \rightarrow \infty$.} Additionally, as a null field, it lies on the threshold of the electrically dominated regime, and thus might be unstable to non-force-free processes.  However, as described below, the solution is physically realized as part of magnetically dominated field configurations associated with a rotating star or black hole, which sources the outflow of energy.

The current $J \sim d\theta\w d\vphi\w du$ of the outgoing flux solution is a null 3-form.  The dual of such a form is proportional to the null factor $du$ (see Appendix \ref{sec:null3dual}), so we have $j^a \sim (du)^a$.  That is, the current 4-vector is null and radial.  If the charges all have the same sign, they must be moving at the speed of light, but a null current can also be composed of charges of opposite sign moving such that the net charge density is equal to the magnitude of the net 3-current in any Lorentz frame.   The force-free equations are sensitive only to the \textit{net} charge-current.

Using the standard orientation $dt \wedge dr \wedge d\theta \wedge d\vphi$, 
the current for this solution is given explicitly by 
\begin{align}\label{Jout}
J&=(d\star d\zeta)\w du \nonumber \\
&=(\Delta_2\zeta) \sin\theta\,  d\theta\w d\vphi\w du,
\end{align}
where $\Delta_2$ is the Laplacian on the unit sphere. This expression reveals two important points. First, the integral of the current over angles vanishes, so there is no net current entering or leaving the system.  Since the current is null (equal magnitude of charge and current), this also indicates that there is no net charge.\footnote{The reason for this can be traced to force-free condition $\vec E\cdot\vec j =0$ (\ref{FFEB}). Since the current is radial, this condition implies that $\vec E$ has no radial component, which implies that the flux of $\vec E$ through a sphere vanishes, so there can be no net charge inside the sphere.} Second, there is no vacuum solution of this sort that is everywhere 
regular on the sphere. In vacuum the current vanishes, which would require that $\zeta$ be a harmonic function on the sphere, $\Delta_2\zeta=0$. Other than a constant (which yields zero field), no such functions exist.

It is rather curious that a purely outgoing solution exists on a Schwarzschild background.  One would expect that waves would backscatter from the effective potential caused by the spacetime curvature. The existence of non-scattering solutions like these was  discovered by \citet{robinson1961}. He showed that, associated with any shear-free null geodesic congruence there is a family of null, nonscattering \textit{vacuum} solutions to Maxwell's equations.  For the radial outgoing null congruence in the Schwarzschild spacetime, the Robinson solutions are exactly the fields \eqref{Fout} with $\Delta_2\zeta=0$. These are in some sense illusory solutions, since they are not globally regular on the sphere. However, they are resurrected
as \textit{bona fide}, regular solutions in the force-free context.

\subsection{Outgoing flux from an arbitrary worldline}
\label{Fout-worldline}

In flat spacetime, the energy flux of Eq.~\eqref{Fout} emerges from the origin of coordinates, which may be identified with a stationary worldline.  In fact, the solution generalizes readily to an \textit{arbitrary} timelike worldline, where $u$ is taken to be the associated retarded time.  That is, on the future lightcone of any point $p$ on the worldline, $u$ is the proper time at $p$.  Then precisely the same expression \eqref{Fout} is a solution, if $(r,\theta,\vphi)$ are any coordinates such that $d\theta$ and $d\vphi$ are orthogonal to $du$, for example, global inertial spherical angles.  This follows from the same computation used to check that \eqref{Fout} is a solution.  This solution was first found in  \citet{brennan-gralla2014} using the Newman-Penrose formalism.

\subsection{Outgoing flux in Kerr}\label{sec:kerrflux}

We next present the generalization of the outgoing flux solution \eqref{Fout}
to a rotating black hole background, i.e.\ the Kerr spacetime.  
The stationary axisymmetric version of this solution was found by 
\citet{menon-dermer2007,menon-dermer2011}, and it was generalized to the nonstationary, nonaxisymmetric case in \citet{brennan-gralla-jacobson2013} using the Newman-Penrose formalism. Here we recover that generalized solution 
using the exterior calculus. 

It is simple to describe the solution 
using outgoing Kerr coordinates $(u, \bar{\vphi}, \theta, r)$, which are
defined in Appendix \ref{sec:Kerr}. A first guess would be that 
the field \eqref{Fout} is a solution, with the substitution
$\vphi \rightarrow \bar{\vphi}$.  However, that is not correct,
because in Kerr the 1-form $du$ is timelike rather than null, and 
the null property of $du$ played a critical role in establishing
that \eqref{Fout} is a solution in Schwarzschild. To motivate a modification, and to proceed with the calculations, 
we need the following properties of the Kerr metric
in these coordinates (see Appendix \ref{sec:Kerr}): ($i$) the 1-form $du - a\sin^2\theta\, d\bar{\vphi}$ is null 
and orthogonal to the 1-forms $du$, $d\bar{\vphi}$, and $d\theta$; and 
($ii$) $d\theta$ and $d\bar{\vphi}$ are orthogonal to each other, and the
ratio of their norms is $\sin\theta$.  

The analogy with the case of the Schwarzschild metric 
now motivates the initial ansatz 
\beq \label{FoutKerr1}
F^{\rm out, Kerr}=d\zeta\w(du - a\sin^2\theta\, d\bar{\vphi}),
\eeq
where as before $\zeta =\zeta(\theta,\bar{\vphi},u)$. However, 
note that $dF^{\rm out, Kerr}$ will be nonzero if $\zeta$
depends on $u$. Hence let us assume that 
$\zeta = \zeta(\theta,\bar{\vphi})$ is independent of $u$, and check the
force-free equations.
(We will generalize this to non-stationary solutions momentarily.)
The dual of \eqref{FoutKerr1} is given by $*F^{\rm rad, Kerr}=-\star d\zeta\w(du - a\sin^2\theta\, d\bar{\vphi})$, where $\star d\zeta$ is an $r$-independent linear combination of $d\theta$ and $d\bar{\vphi}$.
The current is thus proportional to $d\theta\w d\bar{\vphi}\w du$, which 
has vanishing wedge product with the two factors of  \eqref{FoutKerr1},
so indeed the force-free field equations (\ref{FFabJ}) hold. 

Now to allow for $u$ dependence, we must generalize the 
ansatz to 
\beq \label{FoutKerr2}
F^{\rm rad, Kerr}=(A\, d\theta + B\, d\bar{\vphi})\w(du - a\sin^2\theta\, d\bar{\vphi}),
\eeq
where $A=A(\theta,\bar{\vphi},u)$ and $B=B(\theta,\bar{\vphi},u)$. This is not necessarily a closed 2-form so we must impose the covariant Faraday law
\beq
dF^{\rm out, Kerr}=[A_{,\bar{\vphi}}-B_{,\theta}+(a\sin^2\!\theta) A_{,u}]\, d\bar{\vphi}\w d\theta\w du = 0,
\eeq
which results in the differential equation
\beq\label{ABKerr}
A_{,\bar{\vphi}}-B_{,\theta}+(a\sin^2\!\theta) A_{,u}=0.
\eeq
In the non-spinning ($a=0$) or stationary cases, the last term
vanishes and we find $A=\zeta_{,\theta}$ and $B=\zeta_{,\bar{\vphi}}$
for some $\zeta$ 
as before. In the 
spinning, non-stationary case,  
we could for example choose any $A$, and define $B$ by integration
with respect to $\theta$ (although only for some $A$ will the solution 
be smooth at the poles). Once we have solved \eqref{ABKerr}, all
that remains is to impose the force-free conditions, but these hold 
by exactly the same reasoning just used for the stationary solutions. 

Note that, like in the Schwarzschild case, this solution has the remarkable property that the radiation has no backscattering. This is again directly linked to Robinson's theorem: the congruence tangent to the null vector obtained by contraction of $du - a\sin^2\theta\, d\bar{\vphi}$ with the inverse metric is geodesic and shearfree. (It is the outgoing principal null congruence of the Kerr metric \citeeg{poisson-book}.)  And again, there is no globally regular vacuum solution of this type, but in the presence of nonzero current there are regular force-free solutions.  These solutions were first found by assuming the current is along the principal null congruence \citep{brennan-gralla-jacobson2013}.  That analysis also shows that there are no other solutions with such a current.

\subsection{Ingoing flux}

By taking the time-reverse\footnote{In Kerr the time-reverse refers to sending $t\rightarrow-t$ and $\vphi \rightarrow -\vphi$.} of the outgoing flux solution, one obtains an ingoing flux solution.  This solution represents energy emerging from a distant region and converging on the origin of flat spacetime, or entering the horizon of a black hole.  In the black hole case
the ingoing flux is regular at the future horizon and totally absorbed by the black hole, with no backscattering.

\subsection{Superposed monopole and flux}\label{sec:superposed}

Since FFE is non-linear, in general the superposition of two solutions does not yield a third solution.  However, the vacuum monopole field (\ref{Fmon}) has no current, and exerts no force on the current of the radial flux solution (\ref{Fout}) 
(i.e., $F_{ab}^{\rm mon}j^{{\rm out} \ \! a}=0)$, so their superposition yields a solution,
\begin{equation}\label{Fsup}
F^{\rm sup} = q \sin \theta\, d \theta \wedge d\varphi + d \zeta \wedge du.
\end{equation} 
The field in Eq.~\eqref{Fsup} is magnetically dominated when $q\neq0$, and is 
otherwise null.  It is in fact the general force-free solution with radial, 
null current in Schwarzschild (and flat) spacetime \citep{brennan-gralla-jacobson2013}.

Unfortunately, this simple construction does not generalize to the Kerr
background. Although exact monopole\footnote{The exact magnetic monopole solution on Kerr is obtained by taking the dual of the solution generated by the vector potential $A_a=\xi_a$, where $\xi^a$ is the (asymptotic) time-translation Killing field.} 
and outgoing flux solutions on Kerr are known, the monopole field
exerts a force on the null current. [This obstruction is a special case of 
a general theorem: a solution with current along a null geodesic twisting congruence
cannot be magnetically dominated \citep{brennan-gralla-jacobson2013}.]

An interesting generalization applies however to a monopole 
moving along an arbitrary worldline in Minkowski space:
the dual of the Lienard-Wiechert vacuum field
can be superposed with the outgoing flux solution described 
in subsection \ref{Fout-worldline} \citep{brennan-gralla2014}.  
This yields a magnetically 
dominated solution, in which normal radiation 
(in the dual Lienard-Wiechert field) 
coexists with current-supported Poynting flux.

It is instructive to write the $3+1$ version of the superposed solution for a stationary worldline in flat spacetime.  The electric and magnetic fields associated with Eq.~\eqref{Fsup} in flat spacetime are given by the orthonormal frame components
\begin{subequations}\label{FoutEB}
\begin{align}
E_{\hat{\theta}} & = B_{\hat{\vphi}}    = \frac{1}{r} \partial_\theta \zeta \label{pen} \\
 E_{\hat{\vphi}} & =  - B_{\hat{\theta}} = \frac{1}{r \sin \theta}\partial_{\vphi} \zeta, \label{pencil} \\
 B_{\hat{r}} & = \frac{q}{r^2}. \label{marker}
\end{align}
\end{subequations}
The outgoing flux part of the solution [second term in \eqref{Fsup}] corresponds to Eqs.~\eqref{pen}-\eqref{pencil}, while the monopole corresponds to \eqref{marker}.  The radial Poynting flux is carried by orthogonal $E$ and $B$ fields tangent to the sphere and equal in magnitude, while the magnetic monopole gives the magnetic field lines a radial component and ensures magnetic domination.  The $3+1$ version of the statement that this solution has a null, radial four-current is that the 3-current density is radial and equal in magnitude to the charge density,
\beq
\rho = -\frac{\Delta_2\zeta}{r^2}, \quad \vec{\jmath} = \frac{\Delta_2\zeta}{r^2} \hat{r},
\eeq
where we remind the reader that $\Delta_2$ is the Laplacian on the unit sphere.  The monopole field is a vacuum solution and the charge and current come entirely from the outgoing flux solution. The fact that these solutions may be superposed can be understood by noting that the magnetic monopole field is in the same (radial) direction as the current of the outgoing flux solution, so that the addition of the monopole yields no Lorentz force.  As described below, different choices of $q$ and $\zeta$ give rise to different solutions relevant to the exterior of different rotating bodies.

\section{Monopole magnetospheres}\label{sec:magnetospheres}

In this section we apply the solutions discussed in the previous section 
to model magnetospheres external to rotating stars and black holes with monopole charge.  
These models present basic physical properties of force-free magnetospheres in a simple
setting, most importantly the conversion of rotational kinetic energy to Poynting flux.
 Using the same solutions, a closer approximation to real magnetospheres is obtained 
 by ``splitting'' the monopole, as discussed in Sec.~\ref{sec:sheets}.

\subsection{Rotating monopole (Michel solution)}

The \citet{michel1973} rotating monopole solution has served for decades as a starting point for analytical modeling of pulsar and black hole magnetospheres.  
Michel found his solution using an early version of the stationary, axisymmetric framework that we treat in Sec. \ref{sec:stataxi}.  Here we instead recover the solution as a special case of the monopole/flux solution \eqref{Fsup}.  Specifically, the Michel solution is given by specializing to flat spacetime and choosing $\zeta=q \Omega \cos \theta$  with constant $\Omega$,
\beq\label{FMichel}
F^{\rm Michel}= -q\, d(\cos\theta)\w(d\vphi - \O \, du).
\eeq
This solution can be terminated on the surface of a perfectly
conducting star rotating with angular velocity $\O$. The 1-forms
$d\theta$ and $d\vphi - \O \, du$ both vanish when contracted with the 4-velocity 
of any point co-moving with the surface (which is proportional to $\partial_t + \Omega \partial_\vphi$), so that the electric field vanishes in the conductor rest frame.  The conducting boundary conditions only require the tangential components to vanish; the fact that also the perpendicular component also vanishes is a consequence of 
the force-free magnetosphere outside, and would not hold for 
the (non-degenerate) exterior field of 
a rotating magnetized conductor in vacuum (see discussion at the end of Sec.~\ref{O=OF}).
For comparison with the more realistic cases of higher multipoles, 
it is conventional for a spherical star to rewrite the monopole charge 
$q$ in terms of the magnetic field strength $B_0$ at the surface 
of radius $R$, $q=B_0 R^2$.

The current 3-form for the Michel solution is
given by (\ref{Jout}), which evaluates to 
$J=-q\O  \sin2\theta\, d\theta\w d\vphi\w du$. Equivalently, the current 4-vector is equal to the radial null vector 
$j^a = -2q\O (\cos\theta/r^2)(\partial_r)^a$.\footnote{The vector $\partial_r$ is
defined in the $(u,r,\theta,\vphi)$ coordinate system, so it corresponds
to translation of $r$ at fixed $u,\theta,\vphi$ and hence is a future pointing,
outgoing null vector.}  In the northern hemisphere this is a radial in-going 3-current and a negative charge density of the same magnitude as the 3-current, while in the southern hemisphere it is a radial out-going 3-current and a positive charge density.

The energy flux away from the rotating monopole comes only from 
the radiation part of the field, and is given by Eq.~(\ref{Prad}), which evaluates
here to 
\beq\label{PMichel}
{\cal P}^{\rm Michel}= \frac{8\pi}{3}q^2\O^2= \frac{8\pi}{3}B^2R^4\O^2.
\eeq
This outflow of energy is transferred from the rotational kinetic energy
of the conductor, which is possible because the field is not force-free in the conductor.  The physics of the transfer can be understood as follows: Free charges in the conductor are carried by the rotational motion and hence feel a Lorentz force that drives a current in the surface from north to south.  This current in turn feels a Lorentz force opposite to the motion, producing a reaction torque on the conductor, which acts to slow the rotation. 

The Michel monopole illustrates a key physical effect of pulsar physics: 
a rotating, magnetized conductor generates an outgoing energy flux, 
even when stationary. 

\subsubsection{Field sheet geometry of the Michel monopole}\label{sec:michel-sheet}
\begin{figure}
\centering
\includegraphics[scale=.38]{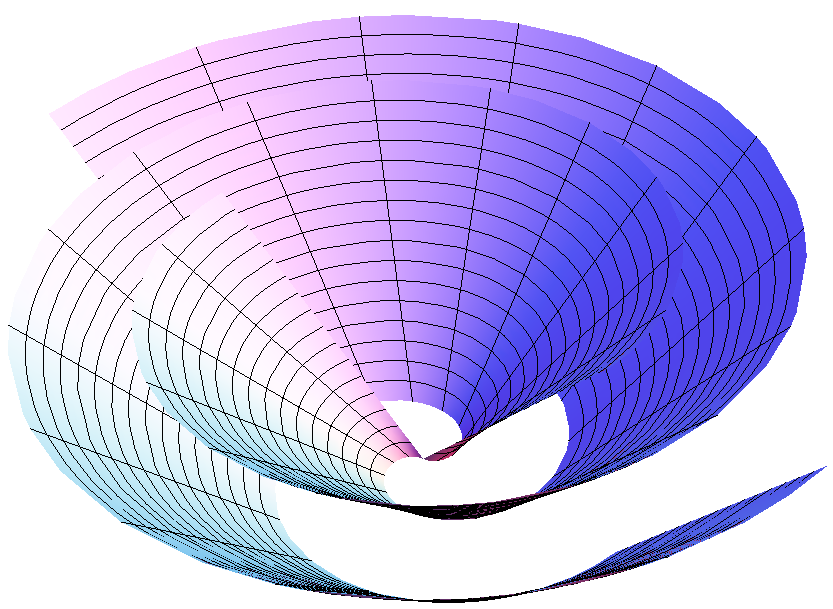}
\caption{Two equatorial field sheets of the Michel monopole.}\label{fig:michel_sheets}
\vspace{-3mm}
\end{figure}

The field of a monopole rotating in vacuum would of course be identical
to that of a static monopole, but the Michel solution is in a certain sense ``really rotating".  The structure of this
field can be elucidated via the geometry of its field sheets.
The Euler potentials can be taken as
$\phi_1=-q\cos\theta$ and $\phi_2=\vphi -\O(t-r)$, so the field
sheets are the surfaces where $\theta$ and 
$\vphi -\O(t-r)$ are constant.
Lab frame field lines (intersections of the sheets with constant $t$ planes) 
form Archimedean spirals in the equatorial plane, and conical helices for 
other values of $\theta$ (see Fig.~\ref{fig:michel}).
At successive times $t$ and $t+\Delta t$ these lines are rotated relative to each other by an angle $\O \Delta t$, 
so one may think of the lines as rotating with angular velocity $\O$. 
They are also related by $r \rightarrow r+\Delta t$, however, 
so one may equivalently think of them as expanding outward at the speed of light. The field sheet is independent of which way one thinks of field line evolution (and also of the choice of frame used to define field lines).  A spacetime plot of two equatorial field sheets is given in Fig.~\ref{fig:michel_sheets}. 
The field sheet metric is obtained by imposing
the conditions $d\theta=0$ and $d\vphi=\O\,du$ in the Minkowski
metric, which yields $ds^2 = -(1-r^2\O^2\sin^2\theta)du^2 - 2du\,dr$.
Amusingly, this is nothing but 1+1 dimensional de Sitter spacetime
in ``Eddington-Finkelstein" form. The de Sitter horizon corresponds to the light cylinder 
$r\sin\theta=1/\Omega$ where a co-rotating observer would move with the velocity of light. The ``Hubble constant" is $\O\sin\theta$, which is also the surface gravity of the horizon.  This Killing horizon interpretation of the light cylinder extends to general stationary axisymmetric magnetospheres, as we discuss in section \ref{sec:lightsurfaces}.

\subsubsection{Differential Rotation}

Eq.~\eqref{FMichel} remains a solution when $\Omega$ is promoted to an arbitrary function of $\theta$.  This corresponds to a conducting star that rotates at latitude-dependent speed.  This generalization of the Michel monopole was first noted by \citet{blandford-znajek1977} [see Eq.~(6.4) therein].  It corresponds to choosing $\zeta=- q \int \sin\theta\, \Omega(\theta) d\theta$ in the superposed solution Eq.~\eqref{Fsup}.  On account of the $\theta$-dependence of $\zeta$, the power radiated is modified from the Michel form \eqref{PMichel}.

\subsubsection{Variable Rotation Rate}

Eq.~\eqref{FMichel} remains a solution when $\Omega$ is promoted to an arbitrary function of $u$, or more generally of $u$ and $\theta$.  This corresponds to a conducting star whose rotational speed changes in time.  The changes propagate outward into the magnetosphere at the speed of light.  This generalization of the Michel monopole was first noted by \citet{lyutikov2011}.  It  corresponds to choosing $\zeta=-q\int  \sin\theta\, \Omega(\theta,u) d\theta$ in the superposed solution Eq.~\eqref{Fsup}.  The flux at each retarded time $u$ is given by the instantaneous value of the associated stationary solution.

\subsection{Whirling monopole}

As a final generalization of the Michel monopole Eq.~\eqref{FMichel} terminated on a conducting star, we take the star to be a sphere, and allow it 
to undergo \textit{arbitrary} time-dependent rigid rotation with fixed center.  
The Michel solution corresponds to the choice 
$\zeta(\theta,\vphi,u) = q\Omega\cos \theta$ in the superposed solution \eqref{Fsup}.  
To produce the whirling monopole we replace the constant $\Omega$ by $\Omega(u)$,
and we replace
$\theta$,  the angle between the field point and the fixed rotation axis of the rotating monopole, 
by $\Theta=\Theta(\theta,\vphi,u)$, the angle between the field point and the rotation axis at the retarded time.  
In terms of the angular velocity vector $\vec{\Omega}(t)$  and the radial unit vector $\hat{r}(\theta,\vphi)$ we have $\Omega(u)\cos \Theta=\vec{\Omega}(u)\cdot\hat{r}(\theta,\vphi)$.
This defines a suitable $\zeta(\theta,\vphi,u)$ for the monopole/flux solution \eqref{Fsup},
and yields the whirling solution
\beq\label{Fwhirl}
F^{\rm whirl} = q \sin \theta\, d\theta \wedge d\vphi + q\, d[\vec{\Omega}(u)\cdot\hat{r}(\theta,\vphi)]\wedge du.
\eeq
At any retarded time this agrees with the Michel solution corresponding to the instantaneous 
angular velocity vector, hence it satisfies the conducting boundary condition on the surface of
the whirling sphere. Furthermore, the flux of the whirling monopole at retarded time $u$ agrees with the flux \eqref{PMichel} of the corresponding Michel solution.  Thus, even if a pulsar undergoes a dramatic whirl (as could occur during a glitch), then the monopole model predicts no additional associated energy losses. 

\subsection{Black hole monopole (BZ solution)}\label{sec:BZ}

As described in section \ref{sec:superposed}, the procedure of superposing monopole and outgoing flux solutions fails to produce a solution in Kerr.  However, long ago Blandford and Znajek (BZ) found a perturbative monopolar solution describing a stationary, axisymmetric outgoing flux of energy from a Kerr black hole, to second order in the black hole spin parameter $a$.  This solution may be recovered to first order in $a$ simply by promoting the Michel monopole solution to Kerr, as we now explain.  Recovering the second-order perturbations is also straightforward, though more involved.  We focus on the first-order piece, which provides the leading outgoing energy flux.

Although we explicitly considered flat spacetime when discussing 
the Michel monopole (\ref{FMichel}), it is also a valid solution in Schwarzschild, with $u$ the Schwarzschild retarded time (outgoing Eddington-Finklestein time).  (Both follow from Eq.~ \eqref{Fsup} with $\zeta=q \Omega \cos \theta$.)   Through first order in $a$, the BZ solution is just the Michel monopole, exported from Schwarzschild to Kerr by identifying the Schwarzschild coordinates with the Boyer-Lindquist (BL) coordinates.  Thus, the ansatz for the solution is
\beq\label{FBZansatz}
F_{\rm ansatz}^{\rm BZ}= q \sin\theta\, d\theta\w (d\vphi - \O du).
\eeq
Expecting $\O$ to be controlled by the spin of the black hole, we regard this quantity as $O(a)$.  Note that the background solution for this perturbation analysis is then the vacuum monopole in Schwarzschild, Eq.~\eqref{Fmon}.\footnote{When perturbing a vacuum solution $F^{(0)}$ to a force-free solution $F^{(0)}+F^{(1)}$, the first-order force-free condition is simply $F^{(0)}\cdot j^{(1)}=0$.  Thus one may choose \textit{any} conserved current $j^{(1)}$ transverse to the background field and then construct an associated Maxwell field $d*F^{(1)}=*j^{(1)}$.  BZ eliminated this freedom by demanding that 
perturbative solutions approach a genuine non-linear force-free solution (in this case the Michel monopole) at large $r$.  Here we simply promote the Michel solution to Kerr and note that the field equations hold to $O(a)$.}  We may take $u$ to be the outgoing Kerr coordinate \eqref{u}, 
since that differs from the Schwarzschild one only at $O(a^2)$ when expressed in terms of $t$ and $r$. This ansatz obviously satisfies $d F=0$, so it remains only to check the force-free conditions.

Since there is no current in the monopole solution, the second factor in the force-free condition (\ref{FFabJ}) 
vanishes at $O(a^0)$, hence in the $O(a)$ equation the first factor ($\a$ or $\b$) may be taken to be the 
zeroth order parts $d(-q\cos\theta)$ and $d\vphi$. 
Thus, up through $O(a)$, the force-free conditions amount to $d\theta\w J = d\vphi\w J = 0$,
i.e.\ the statement that both $d\theta$ and $d\vphi$ are factors in the current 3-form $d*F$. 
The $O(a)$ terms in $d*F$  have two origins: the $\Omega$ term in (\ref{FBZansatz}) and the
$O(a)$ part of the action of $*$ on the zeroth order (monopole) solution. 
The contribution of the $\O$ term to the current is 
$\O d*(\sin\theta\, d\theta\w du) = \O d(\sin^2\!\theta\,  d\vphi\w du)\sim d\theta\w d\vphi\w du$, 
which has both $d\theta$ and $d\vphi$ as factors. The $O(a)$ part of 
$*(d\theta\w d\vphi)_{\m\n}=2\e^{\theta \vphi rt}g_{r[\m} g_{\n]t}$ 
comes from $g_{\vphi t}$, the only $O(a)$ part of the Kerr metric in BL coordinates. 
Since $g_{r\m}\propto\d_\m^r$, this $O(a)$ contribution 
has the form $C(r,\theta) dr\w d\vphi$ for some function $C$. It therefore
contributes to the current $d*F$ a 3-form $\sim d\theta\w dr\w d\vphi$, which also has both
$d\theta$ and $d\vphi$ as factors. Hence the force-free condition is satisfied at $O(a)$.

Up to this point, the derivation would have also worked beginning with general superposed solution \eqref{Fsup}, provided the outgoing flux part is treated as $O(a)$. However there is one more, crucial, consideration regarding this rotating black hole solution: it should be regular on the future event horizon. It is easy to see that this requirement can be met within this class \textit{only} by the Michel monopole solution, with a specific value of $\O$. The 1-forms $d\vphi$ and $du$ are singular on the Kerr horizon, but there is a value of $\O$ for which their singularities cancel in $d\vphi - \O du$.  To see this we use ingoing Kerr coordinates $v$ and $\vphit$, which are related to $u$ and $\vphi$ by 
\eqref{v}, \eqref{vphih}, and \eqref{u}.  
Using $a=\Omega_H(r_+^2 + a^2)$, it follows that
\beq\label{etavsetatilde}
d\vphi - \O du = d\vphit - \O dv + \frac{2\O(r^2 + a^2)-\O_H(r_+^2 + a^2)}{\D}dr.
\eeq
Thus the singularity at the horizon is avoided if and only if $\O$ is \textit{one-half} the horizon angular velocity,
\beq\label{OBH}
\O=\frac{1}{2}\O_H.
\eeq
Since $\O_H \sim a$, it was indeed consistent to treat $\O$ as an $O(a)$ quantity when verifying that the force-free conditions are met.  We may thus write the BZ solution, to $O(a)$, in the exceptionally simple form,
\beq\label{FBZ}
F^{\rm BZ}= q \sin\theta\, d\theta\w (d\vphi - \tfrac{1}{2}\O_H du).
\eeq
This may of course also be written in a way that is manifestly regular on the horizon. 
Using \eqref{etavsetatilde}, the second factor in \eqref{FBZ} may be replaced
by $d\vphit - \half\O_H dv +\O_H(1+2M/r) dr$, dropping terms of $O(a^2)$.

The net energy flux can be computed far from the black hole where the
metric is flat, hence the flux associated with (\ref{FBZ}) is 
given by the same expression \eqref{PMichel}
as for the Michel monopole, 
\beq\label{PBZ}
{\cal P}^{\rm BZ} = \frac{8\pi}{3}q^2\left(\frac{1}{2}\O_H\right)^2 \approx 
\frac{\pi}{24} q^2\frac{a^2}{M^4} = \frac{2\pi}{3}B_0^2 a^2.
\eeq
(Here $B_0=q/(2M)^2$ is defined as the magnetic flux 
through the horizon, divided by the horizon area.)
The energy-momentum tensor (\ref{TEM}) contains
cross-terms between the monopole and the $O(a^2)$ part of $F$,
which we have not computed. However, since the monopole field has only a $\theta\vphi$ component,
no $T^r{}_t$ component of the stress tensor arises in this way,
so (\ref{PBZ}) is the full flux at this order.
Note that, unlike with the rotating monopole terminated on a
star, the energy carried by this flux does not appear in the field by violation
of the force-free condition. Rather, the conserved Killing energy 
on the rotating black hole background is locally \textit{momentum}
in the ergosphere, hence can be negative there.
A flux of negative Killing energy crosses the horizon,
balancing the outward positive flux.  The nature of this process is discussed more fully at the beginning of section \ref{BHM}.

\subsubsection*{Rotating stars}

To first order in the spin, the exterior metric of a rotating star is given by the Kerr metric linearized in $a$ \citep{hartle-thorne1968}.  We may thus also use \eqref{FBZansatz} to model stellar magnetospheres, including the leading gravitational effects of spin. As in the previous subsection, imposition of conducting boundary conditions at the star will fix $\Omega$ to equal the rotational velocity of the star.  It is interesting to compare this with the black hole case \eqref{OBH}, where there is an additional factor of one-half. As will be seen in section \ref{BHM}, the energy flux from any axisymmetric black hole magnetosphere would vanish if the angular velocity of the field were equal to that of the black hole horizon.

\section{Current sheets and split monopoles}\label{sec:sheets}

We have seen that the superposition of monopole and outgoing radiation solutions provides a simple analytic solution describing energy flux from rotating stars and black holes.  The catch, of course, is that real stars and black holes do not have monopoles inside them!  A cheap trick for addressing this last point is to artificially split the monopole in two or more parts, reversing the sign of the monopole charge (and perhaps also rescaling the charge) when passing from one region to the next.  A crude model of a dipole can be constructed in this way, for example, while still using only the monopole solution.  However, this splitting of the field has a dramatic consequence that must be confronted: since the field changes direction discontinuously across the splitting surface, Maxwell's equations imply the presence of a surface current 
and surface charge. Fortuitously, rather than being an unphysical embarrassment, this current sheet actually \textit{enhances} the correspondence of the solution with a pulsar magnetosphere.  We discuss the general necessity of such current sheets in Sec.~\ref{sec:pulsar-magnetosphere}.

\subsection{Split monopole} 
To illustrate the basic idea of a split monopole, consider first the field of a point magnetic monopole in vacuum, $\vec B=(q/r^2)\hat r$, and modify it by reversing the sign of the charge across the equatorial plane, yielding $\vec B^{\rm split}=\sgn(\cos \theta)(q/r^2)\hat r$.  In order for this to remain a solution to Maxwell's equations, there must be a surface layer of azimuthal current on the equatorial plane, i.e., a  current sheet.   Taking this solution to extend inward only to some radius $r=R$, one may regard it as the exterior of a star that has been magnetized in a peculiar split-monopole pattern.  Since the magnetic flux through closed surfaces vanishes, no monopole is required and ordinary currents flowing in the star can generate the field.

\subsection{Generalized split field construction}

We may split a field configuration across more general surfaces as follows.
  Begin with any force-free solution $F$, and replace it with a new solution
\beq\label{Fsplit}
F^{\rm split} = \s F,
\eeq
where $\s$ is a ``step function" on spacetime: constant except where it 
has a jump across a timelike 3-volume $\cS$, the world-volume of the current sheet. In the case of the vacuum monopole discussed above, $\cS$ 
would be the equatorial plane, extended in time, but in general it is 
a dynamical sub-manifold whose motion must be determined.

The jump conditions implied by Maxwell's equations must 
hold at the current sheet. As explained in Appendix \ref{sec:maxform}, these are that 
($i$) the pullback to $\cS$ of the jump in $F$ must vanish (implying no monopole surface charge or 
current), and ($ii$) the pullback to $\cS$ of the jump in 
$*F$ is the surface current 2-form $K$ (which describes both charge and 2-current densities).
The jump in $F^{\rm split}$ is $[F^{\rm split}] = [\s] F$, so the jump conditions are 
\beq
\label{splitjumps}
F|_\cS = 0,\qquad *F|_\cS = K/[\s],
\eeq
where the bar notation $|_{\cS}$ denotes the pullback to $\cS$.
The first of these conditions implies that the current sheet fully contains 
any field sheet that intersects it at a point where $F\ne0$: 
at a point where a field sheet intersects but is not contained in $\cS$, 
there exists a basis consisting of three vectors tangent to $\cS$ and 
a fourth tangent to the field sheet, and $F$ vanishes when contracted
with any pair of these, so $F=0$. 
It follows that the three-dimensional current sheet world-volume 
must be foliated by field sheets. Equivalently, the current sheet must 
be given by an equation $f(\phi_1,\phi_2)=0$, where 
$f$ is some function depending only on the two Euler potentials.
This criterion is necessary and sufficient for a valid split of the form \eqref{Fsplit}.

In terms of a $3+1$ split, these considerations tell us the possible shapes of current sheets of the form \eqref{Fsplit} and specifies their unique time evolution: an initial configuration for a current sheet must be a two-dimensional surface tangent to magnetic field lines, and the time evolution is that of the field lines.  In a spacetime sense, the world-volume $\cS$ of a current sheet may be generated by selecting a single ``seed curve" $\g$, transverse to field sheets, and flowing to all points on the field sheets intersecting $\g$.  Put differently, it is just the bundle of field sheets over $\g$.

So far we have treated current sheets as infinitesimally thin regions where the field has a discontinuity.  A physical sheet would have a finite thickness determined by its internal structure and the forces confining it.  A simple model for a finite-thickness current sheet is obtained by using a smooth transition function $\s(x)$ instead of the step function of \eqref{Fsplit}, yielding a degenerate, but not force-free, field $\tilde{F} \equiv \s(x) F$.  Provided $F$ is magnetic, this leads to opposing, compressional Lorentz forces as follows.  Like all electromagnetic fields, $\tilde{F}$ must satisfy Faraday's law $d\tilde{F}=0$, which implies $d \sigma \wedge F=0$.  Thus $\sigma$ must be constant on the field sheets.  The divergence of the stress tensor $\tilde{T}_{ab}=\sigma^2 T_{ab}$ is equal to $T^{ab} \nabla_b \sigma^2$ since the original field was force-free ($\nabla_a T^{ab}=0$).  Using Eq.~\eqref{Tdeg} for the stress tensor of a degenerate field, only the $h^\perp$ term contributes (since $\sigma$ is constant on the field sheets) and we find that the Lorentz force $-\nabla^a \tilde{T}_{ab}$ is equal to $-\tfrac{1}{4} F^2 \nabla_b \s^2$.  For magnetically dominated fields this force is towards the center of the sheet on both sides, i.e., compressional.  A more complete model would account for the opposing force establishing equlibrium; for example, thermal pressure provides the support in a Harris current sheet \citep{harris1962}.

\subsection{Rotating split monopole}

We now apply the splitting procedure to the Michel monopole \eqref{FMichel} and discuss its application to the BZ black hole monopole \eqref{FBZ} and the whirling monopole \eqref{Fwhirl}.  

\subsubsection{Aligned split monopole}\label{sec:aligned-split}
We first perform the split in the equatorial plane.  Since all field lines in the equatorial plane remain in the plane [see Fig.~\eqref{fig:michel}], it is clear that this plane is a valid location for a current sheet.  This original Michel split monopole is simply
\beq\label{Faligned}
F^{\rm aligned} = \sgn (\cos\theta) F^{\rm Michel},
\eeq
where as before the solution should be terminated on a rotating, conducting star.  
We label this field as ``aligned'' because the magnetic axis is aligned with the spin axis.
The surface current for the resulting equatorial current sheet is given
by \eqref{splitjumps} with $[\s] = 2$, i.e.\ $K^{\rm split Michel}= 2*F^{\rm Michel}|_{\cS}$.
Taking the dual of \eqref{FMichel}, we thus have  
\beq
K^{\rm split Michel}=  \frac{2q}{r^2}dt\w dr+ 2q\O d\vphi\w du.
\eeq
The first term is an azimuthal current density that falls off like $r^{-2}$, 
while the second term is a radial null current density that falls
off like $r^{-1}$ (because $|d\vphi| = 1/r\sin\theta$). 
The latter is the ``return current" in the complete circuit: whereas the
non-split monopole has a current flowing in from infinity in the northern hemisphere
and out to infinity in the southern hemisphere, the split monopole has current flowing 
inward in both hemispheres, and outward in the current sheet.

\subsubsection{Inclined split monopole}
We may equally well consider the inclined case, with the star magnetized 
in a split monopole pattern with the split along an equator 
inclined at an
angle $\a$ to the rotation axis $\hat{z}$ and co-rotating with the star. 
This provides a model for a pulsar with inclined magnetic axis.

\begin{figure}
\centering
\subfigure{\label{fig:michellines}\includegraphics[width=40mm]{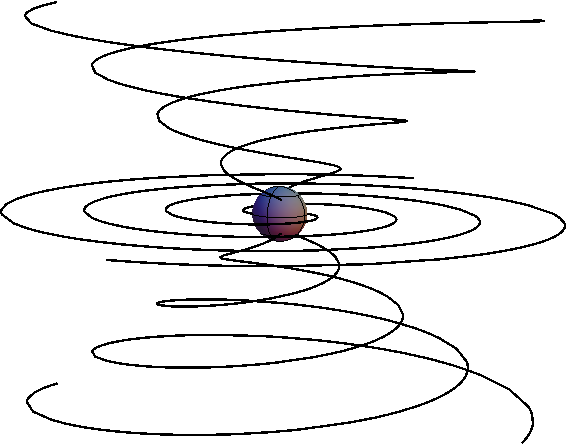}}
\subfigure{\label{fig:michelsheet}\includegraphics[width=40mm]{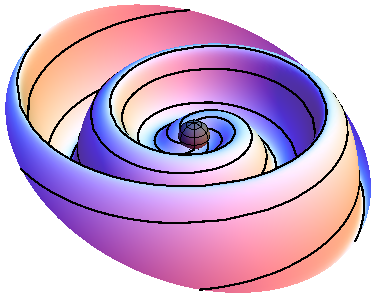}}
\caption{The Michel monopole, with central star drawn in. On the left, some representative lab frame magnetic field lines. On the right, the current sheet in the inclined case, with tangent field lines drawn in black.  The pattern of the sheet rotates rigidly with the star, or equivalently moves radially outward at the speed of light.}
\label{fig:michel}
\end{figure}

Recall that the Michel field sheets are 
specified by the values of the two Euler potentials,
$-q\cos\theta$ and $\vphi -\O u$, so that the current sheet must
be given by an equation of the form $f(\theta,\vphi-\O u)=0$.  To produce an inclined sheet we choose $f$ to vanish on the co-rotating inclined circle.
This circle at one time plays the role of the curve $\g$ that generates the current sheet.
A function satisfying this requirement is 
$f(\theta,\vphi-\O u)=\hat{m}(u)\cdot\hat{r}(\theta,\vphi)$,
where $\hat{m}(u)$ is the rotating split-magnetization 
axis inclined at the angle $\a$ to $\hat{z})$, and
$\hat{r}(\theta,\vphi)$ is the angle-dependent radial unit vector.
Since $\hat{m}$ uniformly rotates with angular velocity $\O$
about $\hat{z}$, $\hat{m}\cdot\hat{r}$ actually depends on
$u$ and $\vphi$ only through $\vphi-\O u$; explicitly, 
\beq
\hat{m}(u)\cdot\hat{r}(\theta,\vphi)=\cos\a\, \cos\theta +\sin\a\, \sin\theta\, \cos(\vphi-\O u).
\eeq
The inclined split monopole is thus given by
\beq\label{Finclined}
F^{\rm inclined} = \sgn[ \hat{m}(u)\cdot\hat{r}(\theta,\vphi)] F^{\rm Michel}.
\eeq
The rather intricate shape of this current sheet is shown in Fig.~\eqref{fig:michel}.  The complete field configuration, where the field changes sign on either side of the dynamical current sheet, is sometimes known as a ``striped wind''.  Eq.~\eqref{Finclined} was derived by \citet{bogovalov1999} in $3+1$ language. 
A current sheet of nearly identical shape and dynamics is observed outside the light cylinder in simulations of inclined dipolar magnetospheres \citep{spitkovsky2006,kalapotharakos-contopoulos-kazanas2012}.\footnote{Such sheets are supported in simulations by (non-force-free) prescriptions that enforce magnetic domination.} 

The dipolar split monopole is the most relevant split configuration for emulating a dipole pulsar, but a variety of other configurations are possible.  For example, 
one may split the solution on cones of fixed latitude, as is clear from the field lines shown in Fig.~\ref{fig:michel}.  Having two such cones, say at latitudes where $\cos\theta = \pm1/\sqrt{3}$, provides a rough imitation of a quadrupole pulsar. 
 In this aligned case the conical sheets are stationary, but it would be straightforward to determine the more complicated shapes and dynamics in the inclined case.   Most generally, one may use \textit{any} seed curve on the sphere
at one time to construct a sheet, since the monopolar (radial) component of the Michel monopole ensures that all such curves are transverse to field lines.  In this sense one may consider a sphere of arbitrary split-monopolar magnetization.

\subsubsection{Black hole split monopole}

One may also split the black hole version of the Michel monopole, as done by BZ in their original paper.  The procedure is precisely analogous to the case of flat spacetime  discussed above.  The BZ model involves splitting in the equatorial plane, Eq.~\eqref{Faligned} with $F^{\rm Michel}$ replaced with $F^{\rm BZ}$ \eqref{FBZ}.  The sheet extends all the way to the event horizon.  In nature, a magnetized accretion disc could source a field, and the current sheet becomes a crude model of such a disc.  However, Lyutikov has raised the interesting possibility that  the gravitational collapse of a pulsar could form a split monopole black hole magnetosphere, where the current sheet originally present outside the light cylinder (e.g., Fig~\ref{fig:aligned_pulsar}) meets the horizon.  If so, then the split BZ model would directly describe an astrophysical magnetosphere, if only for a brief time before magnetic reconnection destroys the sheet.

Although only the equatorial splitting has been explicitly considered in the black hole context, the more general splits discussed in the previous section are also possible.  In particular, the inclined equatorial split also yields Eq.~\eqref{Finclined}, with $F^{\rm Michel}$ replaced with $F^{\rm BZ}$ \eqref{FBZ}.  As argued by Lyutikov in the aligned case, it is conceivable that this solution could model a black hole newly formed from the gravitational collapse of an inclined pulsar.

\subsubsection{Whirling split monopole}

In the whirling case \eqref{Fwhirl}, as in the simple rotating case, any curve tangent to the sphere is transverse to field lines, and so is a valid seed curve for a splitting.  Thus, while we do not construct explicit examples, our results do cover the magnetosphere, including current sheet dynamics, of an arbitrarily whirling, arbitrarily split-monopole magnetized, conducting sphere.  Astrophysically, the whirling split monopole could be helpful for modeling emission (or lack thereof) associated with pulsar glitches, including the case where  
in addition to the magnitude the direction of angular velocity is modified. 

\subsection{Reflection split}

The preceding picture of current sheet behavior applies to sheets produced by
simple \textit{rescalings} of the field strength across the sheet \eqref{Fsplit}.  
While this type of sheet commonly appears (for example in the outer region of 
many pulsar magnetosphere models), 
Maxwell's equations admit other 
types of field discontinuities supported by a current sheet $\cS$,
provided only that the pullback to $\cS$ of the jump in $F$ vanishes.
For this type of discontinuity the magnetic field is not necessarily tangent to the sheet, and there is no simple story regarding the location and dynamics of the current sheets. 

A common example of such a discontinuity occurs in reflection-symmetric magnetospheres, where $F$ is reflected across the equatorial plane, entailing an equatorial current sheet. The field at $\cal S$, the
world volume of the equatorial plane, can be decomposed as 
$F = F_\parallel + F_\perp$,
where $F_\parallel$ is the projection of $F$ into $\cS$ and is invariant under reflection, 
and $F_\perp$ flips sign under reflection.
$F_\parallel$ comprises
the magnetic field normal to and the electric field tangent to the symmetry plane, 
and $F_\perp$ comprises the tangent magnetic field and the normal electric field. 
The jump in $F$ across $\cS$ is thus $[F]=2F_\perp$, and the pullback of this vanishes,
so the jump condition on $F$ is satisfied. For the other jump condition, note that 
the dual $*F_\perp$ is entirely ``parallel", so the jump in $*F$ is 
$[*F]=2*F_\perp =K$, which determines the surface current.

The aligned split monopole discussed in Sec.~\ref{sec:aligned-split} is a special case of this construction. In that example $F_\parallel$ vanishes, so the effect of the reflection is an overall sign change.  Examples with a nonzero normal component of the magnetic field (and tangential component of the electric field) are the pulsar magnetospheres considered in \citep{gruzinov2011,contopoulos-kalapotharakos-kazanas2014}, 
the black hole magnetospheres of \citep{uzdensky2005}, and the paraboloidal magnetospheres of \citep{blandford1976,blandford-znajek1977}.

\section{Stationary, axisymmetric magnetospheres}\label{sec:stataxi}

We turn now from specific analytical models to a general treatment of stationary, axisymmetric force-free magnetospheres, relevant both to spinning stars and to black holes.  This section consists primarily of a systematic review and derivation of the standard mathematical and physical results, but using new computational techniques and the conceptual framework developed in Sec.~\ref{sec:FFE}. It can be seen as a spacetime counterpart to the 3+1 presentation of \citet{macdonald-thorne1982}, using an extension to curved spacetimes of the Euler potential methods developed by \citet{uchida1997symmetry}.  With our systematic use of differential forms, the efficiency and elegance of Uchida's approach is fully realized. 
In the following sections we apply these results to pulsar and black hole
magnetospheres.

Our treatment is spacetime geometrical in the sense that we do not decompose tensors into spatial components and temporal components with respect  
to a time foliation. However, we make heavy use of the existence of a
coordinate system in which the metric components are block diagonal and
do not depend on the two ``symmetry coordinates". This hybrid technique of using
spacetime objects, specifically differential forms, in concert with special 
coordinates, is both remarkably efficient for computations and revealing about the structure of the theory. Another source of the efficiency and simplicity is the avoidance of unnecessary introduction of metric-dependence
into the calculations, and of confining what metric dependence there is 
to the action of the Hodge dual operator and metric determinants. 
This is achieved by using the exterior derivative rather than covariant derivatives, 
integrating $p$-forms on $p$-surfaces, and using the Hodge dual operator.

\subsection{2+2 decomposition of spacetime} 
In this section we set up the decomposition of spacetime that is central to our treatment.  We assume that the spacetime is stationary and axisymmetric, and that these two symmetries commute, so that there exist coordinates $t,\varphi$ such that $\partial_t$ and $\partial_\varphi$ are the time-translation and axial 
rotation Killing fields, respectively. Moreover, we assume that these Killing fields are orthogonal to two-dimensional surfaces. These assumptions 
should hold to a very good approximation in most astrophysically relevant settings.\footnote{There is no loss of generality in assuming the symmetries commute \citep{carter1970}, and 
for asymptotically flat solutions to Einstein's equation
in vacuum or with a circularly rotating fluid source, the 2-surface orthogonality property necessarily holds \citep{wald-book1984}. Non-circular spacetimes result from gravitational effects 
of meridional matter flow or toroidal magnetic fields \citep{gourgoulhon-bonazzola1993}. A fully geometrical treatment of ideal MHD in stationary, axisymmetric spacetimes, allowing for non-circularity, is given in \cite{gourgoulhon-etal2011}.} 
We refer to these surfaces as the ``poloidal subspaces", and to the surfaces generated by the Killing vectors $\partial_t,\partial_\varphi$ as the ``toroidal subspaces".  This is the standard usage of the word poloidal, while it generalizes toroidal to refer to the $t$-$\vphi$ sector, rather than just the spatial $\varphi$ direction.  We will label the poloidal subspaces with coordinates $(r,\theta)$ that are constant along the integral curves of the Killing fields, so that the metric components in these coordinates are block-diagonal,
\beq\label{gsymm}
g_{..}(r,\theta) = \left( \begin{array}{cc}
g_{..}^T & 0  \\
0 & g_{..}^P \end{array} \right).
\eeq
We refer to $g_{..}^T$ and $g_{..}^P$ as the toroidal and poloidal metrics, respectively.  Although it will not be necessary for all considerations, we further assume that $g_{..}^T$ is Lorentzian, while $g_{..}^P$ is Riemannian.  These metrics depend only on the point in the poloidal surface, i.e.\ their components are functions only of $r,\theta$.  We emphasize that here $r$ and $\theta$ are just names for arbitrary poloidal coordinates.

We adopt the orientation of $dt \wedge d\varphi \wedge dr \wedge d\theta$ for all integrals and dualization.   
The corresponding metric volume elements $\epsilon$, $\epsilon^T$ and $\epsilon^P$ on full, toroidal, and poloidal subspaces (respectively)
are given by 
\beq
\epsilon = \sqrt{-g}\, dt \wedge d\varphi \wedge dr \wedge d\theta,
\eeq
\beq\label{eTeP}
\epsilon^T = \sqrt{-g^T}\, dt \wedge d\varphi,\quad
\epsilon^P = \sqrt{g^P}\, dr \wedge d\theta,
\eeq
where $g$, $g^T$ and $g^P$ are the determinants of the corresponding metrics in these coordinates.
These satisfy the identities
\beq\label{epsilonTP}
\epsilon=\epsilon^T \wedge \epsilon^P, \quad *\epsilon^T = - \epsilon^P, \quad * \epsilon^P = \epsilon^T.
\eeq
We use $\e^T$ and $\e^P$ to define the Hodge dual operator 
on toroidal and poloidal forms, and we denote this operator by $\star$, reserving $*$ for the spacetime dual.  
Specifically, if $\omega^P$ is a poloidal differential form (a form made 
from poloidal cotangent vectors),
then $\star \omega^P$ denotes its Hodge dual on the 
poloidal space with respect to the poloidal metric, 
and similarly for $\star \omega^T$. 
On toroidal 1-forms, $\star\star=1$, while on poloidal 1-forms, $\star\star=-1$. The signs are opposite
to these on 0-forms and 2-forms. 
The dual operators satisfy
the following useful identities:
\beq\label{starstar}
*(\omega^T \wedge \omega^P) = - \left. \star \omega^T \wedge \star \omega^P \right.,
\eeq
\beq\label{star*}
*\o^P= \star\omega^P \wedge \e^T,
\eeq
where $\omega^T$ and $\omega^P$ are toroidal and poloidal 1-forms, respectively.  
More discussion of orthogonal subspaces and duality is given in Appendix \ref{sec:forms}.

\subsection{Degenerate, stationary, axisymmetric fields} 
\label{Dsaf}
A stationary, axisymmetric electromagnetic field satisfies
$\mathcal{L}_{\partial_t}F=\mathcal{L}_{\partial_\vphi}F=0$.  
In this subsection we assume the electromagnetic 
field is degenerate, $F\wedge F=0$, but not necessarily force-free.
Thus it can be expressed in terms of Euler potentials as 
$F=d\phi_1\w d\phi_2$. 
The Euler potentials need not share the symmetry of $F$, but their dependence on the ignorable coordinates $t$ and $\vphi$ is very restricted. Their form is worked out in Appendix \ref{sec:appEPsymm}, following \citet{uchida1997symmetry}.
Apart from the special case of purely toroidal magnetic field ($\partial_\vphi \cdot F=0$), 
one may always choose Euler potentials given by \eqref{EPsymmAppendix},
\beq\label{EPsymm}
\phi_1 = \psi(r,\theta), \quad \phi_2 = \psi_2(r,\theta) + \varphi - \Omega_F(\psi) t.
\eeq
We focus on this generic case in the following. The special case $\partial_\vphi \cdot F=0$ is treated briefly in section \ref{sec:special-case} below.

Field sheets are surfaces of constant $\phi_1$ and $\phi_2$, hence are labeled
by a value of $\psi$ and a value of $\phi_2$. If the field is 
magnetically dominated, as we assume from now on unless otherwise
stated, the field sheets are timelike. A magnetic field line, defined
with respect to $t$, is the intersection of a field sheet with a surface of 
constant $t$.  Besides the ``true" magnetic field lines one can also define 
\tit{poloidal field lines}, which are just the $\psi$ contours in the 
poloidal space. The bending of a true field line in the azimuthal direction, 
i.e.\ the variation of its $\vphi$ coordinate at fixed $t$, is determined by $\psi_2$.
As explained below, $\psi$ determines the polar magnetic flux,
and the function $\O_F(\psi)$ determines the angular velocity of the 
field lines.

The field strength corresponding to the general Euler potential of the form (\ref{EPsymm}) is
\beq\label{Fsymm}
F = d \psi \wedge d \psi_2 + d \psi \wedge \eta,
\eeq
where $\eta\equiv d\varphi - \Omega_F(\psi) dt$. (The properties of this useful 1-form are discussed below in Sec.~\ref{sec:eta}.)  Note that there is no term proportional to $d\vphi\w dt$, i.e.\ no ``toroidal electric field". This is a consequence of Faraday's law for stationary, axisymmetric fields, since the toroidal line integral of the electric field must be equal to minus the vanishing time derivative of the magnetic flux through the loop. It does not depend on the field being force-free or even degenerate.  Using Eq.~\eqref{starstar}, the dual of $F$ is given by
\beq\label{starFsymm}
*F =  {\frac{I}{2\pi}}\,  dt \wedge d \varphi - \star d\psi \wedge \star \eta,
\eeq
where, for the moment, $I$ is simply defined by 
\beq\label{Idef}
 *(d\psi \wedge d\psi_2)={\frac{I}{2\pi}}\, dt \wedge d\varphi,
\eeq
but it will be interpreted below as the polar current. 
We can express $F$ in terms of $I$ instead of $\psi_2$, 
by taking the dual of (\ref{Idef}), using Eq.~\eqref{epsilonTP} and $**=-1$ on 2-forms, as
\beq\label{FsymmI}
F = \frac{I}{2\pi(-g^T)^{1/2}} \epsilon_P + d \psi \wedge \eta.
\eeq
This displays the field as a sum of its poloidal and toroidal parts. 
Note the potentially confusing fact that because the magnetic field vector is defined via the \textit{dual} of the field strength 2-form $F$, the poloidal part
of $F$ [the first term in \eqref{FsymmI}] actually corresponds 
to the \textit{toroidal} magnetic field, i.e.\ the magnetic field component
in the $\partial_\vphi$ direction according to an observer at rest in the poloidal subspace. 
The notation commonly used for this
toroidal field is $B_T = I/2\pi$. Note that the proper magnitude of the 
toroidal field is thus not $B_T$ but rather $B_T/\sqrt{-g_T}$. 

The invariant $F^2=F_{ab}F^{ab}$ is the sum of the invariant squares 
of the toroidal and poloidal parts in \eqref{FsymmI},  
\beq\label{F^2}
F^2 = \frac{I^2}{2\pi^2 (-g^T)} + 2|d\psi|^2|\eta|^2.
\eeq
Here and below we use the notation $|\eta|^2$ to denote 
$g^{ab}\eta_a\eta_b$.
The first, poloidal term is always positive or zero, while the sign of the
second, toroidal term is that of $|\eta|^2$, which is negative when $\eta$ is timelike.

In the following subsections we expand on the 
interpretation and properties of the quantities introduced here.

\subsubsection{Magnetic flux function $\psi$}
It was noted above that $\psi$ labels magnetic field lines, but it is also directly related to the flux as follows.  Fix a poloidal point $(r,\theta)$ and time $t$, denote by $\mathcal{C}$ the loop obtained by flowing along $\partial_\varphi$, and let $\mathcal{S}$ be any topological disk bounded by $\mathcal{C}$.  The integral of $F$ over $\mathcal{S}$ is the magnetic flux through $\mathcal{C}$.  (Integration of differential forms is reviewed in Appendix \ref{Integration}.)  Writing $F$ as an exact differential $F=d(\psi\,  d\phi_2)$ and using Stokes' theorem we find  $\int_{\mathcal{S}} F = \psi \int_{\mathcal{C}}d\phi_2= 2\pi\psi$, so 
\beq\label{psiflux}
\psi(r,\theta) = \frac{1}{2\pi} \int_{\mathcal{S}(r,\theta)} F.
\eeq
That is, $2\pi \psi(r,\theta)$ is the magnetic flux through the loop of revolution defined by the poloidal point $(r,\theta)$.  
This is why $\psi$ is often called the magnetic flux function.  Another common name is the stream function. We will use both of these names, depending on the context. 

In deriving \eqref{psiflux} we have chosen the orientation $d\varphi$ on the loop $\mathcal{C}$, which by Stokes' theorem fixes the orientation for the 2-surface $\mathcal{S}$ with respect to which the flux is defined.  We will call this the flux in the ``upwards'' direction.  To understand the name, consider flat spacetime in cylindrical coordinates $(t,z,\r,\vphi)$, and let $\mathcal{S}$ be a disk of constant $z$.  Then the orientation $d\vphi$ on the boundary corresponds to the orientation $d\r\w d\vphi$ for the disk.  Given the spacetime orientation $dt\w dz\w d\r \w d\vphi$, this corresponds to the flux of the magnetic field pseudovector using the surface-normal $+\partial_z$.

The potential $\psi$ is also related to the electrostatic potential as follows.
A particle of mass $m$ and charge $e$ in stationary gravitational and 
electromagnetic fields has a conserved energy $\xi\cdot(m U + e A)$, where 
$\xi$ is the stationary Killing vector, $U$ is the particle 4-velocity and 
$A$ is a vector potential that is invariant under the symmetry, ${\cal L}_\xi A = 0$. 
Then it is natural to define $\xi\cdot A$ as the ``electrostatic potential".
Although not gauge-invariant, under a gauge transformation $A\rightarrow A'=A + d\lambda$
this changes by $\xi\cdot d\lambda$, which must be a constant if ${\cal L}_\xi A'$ is to vanish.
Hence the electrostatic potential \textit{difference} between two points is gauge-invariant.
For the degenerate fields discussed here we may  use $A=\psi \ \! d\phi_2$, so that 
the electrostatic potential is $-\O_F(\psi) \psi$. This determines the 
``potential drop" between magnetic field lines.

\subsubsection{Polar current $I$}\label{sec:I}
The integral of the charge-current 3-form $J=d*F$ over the 3-surface $\mathcal{S} \times \Delta t$, formed by flowing $\mathcal{S}$ along $\partial_t$ for a coordinate distance $\Delta t$, is 
(by Stokes' theorem) equal to the integral of $*F$ over the boundary.  
The contributions from the initial and final copies of $\mathcal{S}$ cancel out by stationarity, leaving $\int J = \int_{\mathcal{C}\times\D t} *F$.
Since this surface extends only in $t$ and $\vphi$, 
only the first term of Eq.~\eqref{starFsymm} contributes, and we have simply
\beq\label{polarcurrent}
I(r,\theta) = \frac{1}{\Delta t} \int_{\mathcal{S}(r,\theta) \times \Delta t} J,
\eeq
assuming the orientation $dt \wedge d\vphi$ on $\mathcal{C} \times \Delta t$, which by Stokes' theorem fixes the ``upwards'' orientation on $\mathcal{S} \times \Delta t$.\footnote{In cylindrical coordinates in flat spacetime for a disk of constant $z$, this corresponds to the orientation $dt \wedge d\vphi \wedge d \rho$ on $\mathcal{S} \times \Delta t$, which, given the spacetime orientation $dt\wedge d\vphi \wedge d\rho \wedge dz$, corresponds to flux along the $+\partial_z$ direction.}
Thus $I(r,\theta)$ is equal to the electric current, with respect to Killing time, flowing 
in the upward direction through the loop of revolution 
defined by the poloidal point $(r,\theta)$.  
We will call $I$ the \textit{polar current}.  Another common name is the poloidal current; however, we reserve that name for the current density flowing in the poloidal subspace, as distinguished from the net current through a loop.  Besides its interpretation as a current,
recall [\textit{cf}. discussion below Eq.~\eqref{FsymmI}] that $I/2\pi$ is  
equal to $B_T$, the toroidal magnetic field times $\sqrt{-g_T}$, which controls the bending of field lines in the 
$\vphi$ direction. (This relation between $B_T$ and $I$ is an instance of Amp\`{e}re's law.)  
In addition to these roles, $I$ gives the angular momentum flux per unit $\psi$, Eq.~\eqref{Lflux} below.

\subsubsection{Angular velocity of field lines $\O_F(\psi)$}
\label{sec:O_F}

The stationary axisymmetry implies that the field $F$ is unchanged by a shift in $\vphi$ and/or $t$;
however, the potential $\phi_2$ is in general unchanged only by a combined, helical shift $(\D t,\D \vphi = \O_F(\psi)\D t)$.
Under such a helical shift, the two Euler potentials are both unchanged, so a field sheet maps into itself.
We may therefore interpret $\O_F(\psi)$ as the angular velocity of the field line, the latter being defined by the intersection of the field sheet with a surface  of constant $t$.

\subsubsection{Co-rotation 1-form $\eta$}\label{sec:eta}

It is already apparent that the 1-form
\beq\label{etadef}
\eta = d\varphi - \Omega_F(\psi) dt
\eeq
plays an important role in characterizing stationary, axisymmetric magnetospheres.  
In light of $(\partial_t + \O \partial_\vphi)\cdot \eta=\O-\O_F$, $\eta$ measures
the extent to which a trajectory co-rotates with field lines.   
We refer to $\eta$ as the \textit{co-rotation 1-form}.  
Defining the \textit{co-rotation vector} $\chi_F=\partial_t + \O_F \partial_\vphi$, 
we have $\chi_F\cdot\eta=0$, so $\eta$ and $\chi_F$ are 
orthogonal as vectors (using the inverse metric to convert $\eta$ to a vector).
Both vectors lie in the two-dimensional, timelike toroidal subspace 
so, being orthogonal, they evidently have opposite timelike/spacelike 
causal character.  Explicitly, 
\beq
|\chi_F|^2 = g^T|\eta|^2 ,
\eeq
where the determinant $g^T$ of the toroidal metric is negative 
(since that metric is Lorentzian).  
Observers co-rotating with the magnetic field therefore exist 
only where $\chi_F$ is timelike and $\eta$ is \textit{spacelike}.

\subsubsection{Light surfaces}
\label{sec:lightsurfaces}

At a point where $\eta$ and $\chi_F$ are null, the co-rotating observer would need to travel at the speed of light. For this reason a surface composed of such points is generally called a \textit{light surface}, 
other names being critical surface, singular surface, velocity-of-light surface, 
or light cylinder.\footnote{We caution the reader that some authors reserve the term 
``light surface'' for a place where $F^2$ vanishes. 
These two notions of light surface agree only when $I=0$ (see Eq.~\eqref{F^2}).}   
The latter name stems from the fact that in flat spacetime,
with $\Omega_F=\textrm{const}$, there is one light surface 
located where the cylindrical radius is equal to $1/\Omega_F$.

Light surfaces in magnetospheres 
play a significant role for two reasons. 
One is that the equation satisfied by the magnetic flux function 
(the so-called stream equation, cf.\ Sec.~\ref{sec:stream})
has a critical point at a light surface. 
The implications of this for solutions of the equation 
are described briefly in Sec.~\ref{sec:solvestream}.

The other role of light surfaces is that they determine 
causal boundaries of propagation of charged particle winds
and Alfv\'en waves. As explained in Sec.~\ref{fieldsheets}, the field sheet metric 
governs such transport. Where the co-rotation vector $\chi_F$ is null, it coincides with one of the two field sheet lightrays delineating the light cone on the sheet. Since $\chi_F$ is strictly toroidal, the light surface is evidently a causal boundary (at least locally) for either ingoing or outgoing motion on the sheet.
In the case of the Michel monopole solution \eqref{FMichel}, for example,
outside the light cylinder particles can propagate only to larger radii. 
For field sheet modes, the light cylinder is thus a horizon,
beyond which influences cannot affect the interior. 

In a general stationary, axisymmetric magnetosphere, the allowed direction of particle flow across the light surface, i.e. the direction of the other future pointing light ray on the 
field sheet, is the same as the direction of positive angular momentum 
flow in the field if $\O_F$ is greater than $\O_Z$, the angular 
velocity of the local zero angular momentum observer (ZAMO).
If instead $\O_F<\O_Z$, these directions are opposite. 
This is demonstrated in Sec.~\eqref{sec:tedchamp}.\\

\subsubsection{Field sheet Killing vector}
\label{sec:chiF}

As noted in Sec.~\ref{sec:O_F}, the Euler potentials are unchanged 
under a combined time translation and rotation $(\D t,\D \vphi = \O_F(\psi)\D t)$.
The field sheets and the field strength are preserved under this 
transformation, which is generated by the flow of the co-rotation vector field 
\beq\label{chiF}
\chi_F=\partial_t +\O_F(\psi)\partial_\vphi,
\eeq
introduced in Sec.~\ref{sec:eta}.\footnote{The existence of this symmetry of the Euler 
potentials is an example of a general property, discussed in Appendix \ref{sec:appEPsymm}, which holds for degenerate fields with two commuting symmetry vectors $X$ and $Y$, provided the (constant) quantity $X{\cdot} Y{\cdot}  F$ is non-vanishing.}
This is not only a symmetry of the electromagnetic field; it is also a symmetry
of the intrinsic geometry of the field sheets.
That is, although $\chi_F$ is not a spacetime Killing vector if
$\O_F(\psi)$ is not constant, it is always a Killing vector of the induced
metric on the field sheets, because $\psi$ is constant on a field sheet.
We therefore refer to $\chi_F$ also as the
\textit{field sheet Killing vector}.

As explained in Sec.~\ref{fieldsheets}, the field sheet metric governs the
propagation of collisionless charged particles and Alf\'en waves in a certain
approximation.
The field sheet Killing vector thus provides conservation laws
for these sorts of transport. In particular, there is a conserved 
quantity $\chi\cdot p=p_t +\O_F p_\vphi$ associated with each particle or wavepacket 
trajectory, where $p$ is the four-momentum or wave four-vector respectively.  
Since the field sheet metric is two-dimensional, the single conserved quantity is enough to completely determine the motion from a choice of initial position and velocity.  In applications, such initial conditions may be provided, e.g., 
by particle injection velocities at non-degenerate gaps in an otherwise force-free magnetosphere. The 4-velocity $u$ of a particle at any point is then determined by the equations $u^2 = -1$, $u\cdot F=0$, and $u_a \chi_F^a={\rm const}$.

In flat spacetime,  the conserved quantity for particles moving along field lines is $u_a\chi_F^a=\gamma(-1+ \Omega_F \rho \,  v_\vphi)$, where $\gamma$ is the Lorentz factor of the trajectory (in the rest frame defined by $\partial_t$),
$\r$ is the cylindrical radius, and $v_\vphi=\r\, d\vphi/dt$ is the azimuthal three-velocity.  This quantity is sometimes used to determine outflow velocities from force-free solutions [e.g., \citealp{contopoulos-kazanas-fendt1999} Eq.~(16)].  We have obtained the conserved quantity as a simple consequence of the existence of a Killing vector on the field sheets, a formulation that generalizes to arbitrary circular spacetimes.

We note that the intersection of a light surface with a given field sheet
is a \textit{Killing horizon} for the field sheet Killing vector. That is, it is 
a null curve to which the Killing vector is tangent.  As mentioned in section \ref{sec:michel-sheet}, 
in the case of the Michel monopole the field sheets are isometric to two-dimensional de Sitter space, 
and the light cylinder horizon is a de Sitter horizon. 

\subsection{Energy and angular momentum currents}
\label{EandL}

A physical system governed by a Lagrangian on a spacetime with a symmetry generated by a Killing field $\xi^a$ has an associated conserved Noether current, $\mathcal{J}_\xi$.  In Appendix \ref{Noether} we review how this comes about using the language of differential forms. The electromagnetic field contribution to the Noether current 3-form (neglecting couplings) is given by 
\beq\label{Jxi-main}
\cJ_\xi   =-(\xi\cdot F)\w*F + \fourth F^2 \xi\cdot\e.
\eeq
This is the dual of $-T^a{}_b\xi^b$, where $T^{ab}$ is the Maxwell
stress-energy tensor \eqref{TEM}. The current is conserved if and only if $F_{ab}J^b\xi^a=0$, i.e., 
when the component of four-force in the $\xi$-direction vanishes.  
As explained in Appendix \ref{Noether}, the second term of \eqref{Jxi-main} 
is conserved automatically when the electromagnetic field also shares 
the symmetry, $\mathcal{L}_\xi F=0$.

In terms of the Euler potentials for a degenerate field we have
\beq\label{xidotF}
\xi\cdot F =\xi\cdot(d\phi_1\w d\phi_2) =  (\xi\cdot d\phi_1)d\phi_2-(\xi\cdot d\phi_2)d\phi_1.
\eeq
The first term on the right vanishes for stationary, axisymmetric fields characterized 
by the Uchida potentials (\ref{EPsymm}), while the second term is simply $-d\psi$
for the angular Killing field $\partial_\vphi$ and $+\Omega_F\, d\psi$ 
for the time translation Killing field $\partial_t$. 
Thus the angular momentum and energy currents are given by
\begin{align}
\cJ_L & = -d\psi\w*F - \fourth F^2\,  \partial_\vphi\cdot\e. \label{JL} \\
\cJ_E &= -\Omega_F(\psi)\,  d\psi\w*F + \fourth F^2\, \partial_t\cdot\e \label{JE}
\end{align}
The angular momentum current is \textit{minus} the Noether current 
\eqref{Jxi-main}.\footnote{The total 4-momentum vector $P^a$ is defined by
$\int \cJ_\xi = - \eta_{ab} P^a \xi_\infty^b$, so that the time and 
space components of $P^a$, which define the energy and translational momentum, 
have opposite signs in relation to the corresponding ``Hamiltonian" $\int \cJ_\xi$.}
The conserved quantity associated with the asymptotic time translation symmetry 
is sometimes called \textit{Killing energy}, or \textit{energy at infinity}, 
to distinguish it from energy as defined by local observers.
We will often simply call it ``energy", when the meaning is clear from the context.

When the electromagnetic field is coupled to charges,  
the energy and angular momentum currents \eqref{JL} and \eqref{JE} are not conserved,
unless the four-force vanishes along $\partial_\vphi$ and  $\partial_t$ respectively.
Since the second terms in Eqs.~\eqref{JL} and \eqref{JE} are automatically conserved
for stationary axisymmetric fields (see note below \eqref{Jxi-main}), conservation of energy and angular momentum amounts to the condition $d\psi\w d*F=0$.  (In particular, if a stationary, axisymmetric, degenerate field conserves one of these, it also conserves the other.)  This is equivalent to the first of the two force-free equations (\ref{EFFFE}) or, equivalently, 
conservation of the first Euler current \eqref{EulerCurrents}.\footnote{This property also holds for configurations with a single symmetry: for a degenerate EM field that is Lie derived by a Killing field $\xi$ of the background spacetime, conservation of the current conjugate to $\xi$ is equivalent the 
force-free condition involving the potential that is invariant under the symmetry.  
This follows from \eqref{xidotF} and the 
the analysis of Appendix \ref{sec:onesymmetry}.}

Suppose that energy and angular momentum are conserved, either because the field is fully force-free, or because the dissipation vanishes in symmetry directions, $F_{ab}J^b \xi^b=0$.  We have shown above that this is equivalent to $d\psi\w d*F=0$.  Using Eq.~\eqref{starFsymm} for $*F$, we have
\begin{align}
0 &= d\psi \wedge d*F \nonumber \\
&= \frac{1}{2\pi}d \psi \wedge d I \wedge d\varphi \wedge dt - d\psi \wedge d(\star d\psi \wedge \star \eta).
\end{align}
The second term vanishes because when factored it contains three poloidal 1-forms, while the poloidal space is only two-dimensional. It follows that $d\psi \wedge dI=0$, which implies that 
\beq\label{Iinv}
I=I(\psi).
\eeq
Thus, for stationary, axisymmetric, degenerate, energy and angular momentum conserving fields, the polar current $I$, like the angular velocity of field lines $\O_F$, is a function of the stream function alone.  The physical interpretation is that the poloidal current flows along poloidal magnetic field lines,
so that the Lorentz force along  $\partial_\vphi$ vanishes.\footnote{In the electrically dominated case, we instead have that poloidal current flows along poloidal equipotentials, i.e.\ perpendicular to electric field lines.}

The angular momentum and energy currents (\ref{JL},\ref{JE}) both contain a $d\psi$ factor and therefore vanish when integrated on a surface of constant $\psi$. This means that there is no flux of angular momentum or energy across such a surface; put differently, these quantities flow along the poloidal field lines, as well as in toroidal directions.  (The vectorial characterization of this property is that poloidal part of the vector current $(*\mathcal{J})^a$ is tangent to poloidal field lines.) To evaluate the flux, let $\cal P$ be a poloidal curve, and consider the 3-surface $\cS={\cal P}\times S^{1}\times \Delta t$ generated by rotating $\cal P$ all the way around the axis, and extended in Killing time by an amount $\Delta t$.  The total flux of angular momentum across $\cal S$ is the integral of $\cJ_L$ over that surface.  The $\partial_\vphi\cdot \e$ term does not contribute, since its pullback to a surface including the $\partial_\vphi$ direction vanishes.  The $\partial_t\cdot \e$ term similarly vanishes for the total energy flux, since the surface also includes the $\partial_t$ direction. The total fluxes are therefore given by 
\begin{align}
\int_{\cal S} \mathcal{J}_L &= - \int_{\cal S} d\psi\w *F, \label{cucumber} \\
\int_{\cal S} \mathcal{J}_E &= -\int_{\cal S} \O_F\, d\psi\w *F.
\end{align}
Since the surface $\cal S$ extends in both $\vphi$ and $t$, the
integral vanishes unless the integrand contains a toroidal 2-form.
Since $d\psi$ is poloidal, only the pure toroidal part of $*F$ {\eqref{starFsymm},
i.e.\ $(I/2\pi)dt\w d\vphi$, contributes, and so the flux rates 
through ${\cal P}\times S^1$ are given by\footnote{With the outward orientation for $\mathcal{P}\times S^1$, Eqs.~\eqref{Lflux} and \eqref{Eflux} give the \textit{outward} flux of angular momentum and energy respectively, so they give \textit{minus} the angular momentum and energy change respectively of the system inside the surface.}
\begin{align}
d\cL/dt &= -\int_{\cal P} I\, d\psi,\label{Lflux}\\
d{\cal E}/dt &= -\int_{\cal P} \O_F I\, d\psi.\label{Eflux}
\end{align}
If the poloidal curve $\cal P$ is a line of constant $\psi$, i.e.\ a poloidal field line, these integrals obviously vanish, illustrating the point made above that these currents ``flow along the poloidal field lines".  When energy and angular momentum are conserved (such as when the fields are force-free), then we have $I=I(\psi)$ [Eq.~\eqref{Iinv}] and  Eqs.~\eqref{Lflux} and \eqref{Eflux} become ordinary one-dimensional integrals over a coordinate $\psi$, with limits corresponding to the value of $\psi$ at the start and end of the curve $\mathcal{P}$.

\subsubsection{Direction of particle flow at a light surface}\label{sec:tedchamp}

We now establish the result mentioned in Sec.~\ref{sec:lightsurfaces},
that the direction of particle flow across a light surface is
the same or opposite to the direction of positive angular momentum
flow, according to whether $\O_F -\O_Z$ is positive or negative.
Here $\O_Z$ is the ZAMO angular velocity discussed
beginning with Eq.~\eqref{gnice} below.

In the physical setting discussed in Sec.~\ref{fieldsheets}, charged particle motion is effectively tangent to the field sheets. The 4-velocity $u$ of such a particle thus satisfies
\beq
\label{udotF}
0=u\cdot F =  \frac{I}{2\pi(-g^T)^{1/2}}\, u\cdot \e^P  - (u\cdot\eta)\, d\psi + (u\cdot d\psi)\, \eta,
\eeq
using \eqref{FsymmI}.  The last term is toroidal, and vanishes identically since $\psi$ is constant on the field sheet. The poloidal angular momentum current 3-form (i.e., the part of $\mathcal{J}_L$ containing $\e^T$ as a factor) is given by
\begin{align}
{\cal J}^P_L & =[-d\psi\w *F]^P = - \frac{I}{2\pi(-g^T)^{1/2}} d\psi\w \e^T
\\ & = \frac{-1}{u\cdot\eta}\frac{I^2}{4\pi^2(-g^T)} (u\cdot\e^P)\w \e^T,
\end{align}
using \eqref{starFsymm} in the second step and \eqref{udotF} in the final step.  The particle current is a positive number times $u\cdot\e=(u\cdot\e^P)\w \e^T + \e^P\w u\cdot \e^T$, whose poloidal part is the first term, $(u\cdot \e^P) \w \e^T$. The relative sign of the poloidal angular momentum current and particle current
in the direction $u$ is thus equal to the sign of
$-u\cdot\eta$. Since $\eta$ is null at the light surface, the sign of
its contraction with all future pointing vectors is the same. In particular  the
ZAMO 4-velocity $u_Z=\partial_t + \O_Z\partial_\vphi$ is a future pointing
timelike vector everywhere,\footnote{$u_Z$ is future pointing timelike at infinity,
is timelike everywhere (since it is orthogonal to the spacelike vector $\partial_\vphi$
and lies in the toroidal plane), and is nowhere zero.  Hence it is future timelike everywhere.}
so $\sgn(u\cdot\eta)=\sgn(u_Z\cdot\eta)=\sgn(\O_Z-\O_F)$.
We conclude that particle flow and positive angular momentum flow have the
same direction if $\O_F>\O_Z$, and opposite direction if $\O_F<\O_Z$.
In flat or Schwarzschild spacetime, $\O_Z=0$, so all that matters is the sign of
the angular velocity of the field line, and the direction of particle flow agrees with the
direction of energy flow.

\subsection{Stream Equation}
\label{sec:stream}

Up to now our discussion of stationary, axisymmetric fields has assumed degeneracy of the field, but has not assumed it is force-free.  For force-free fields the stream function $\psi$ satisfies a non-linear partial differential equation which is known by many names: stream equation, Grad-Shafranov equation, trans-field equation, and, in flat spacetime, pulsar equation \citep{michel1973,scharlemann-wagoner1973,okamoto1974,blandford-znajek1977}.  We will call this equation the \textit{stream equation}, and we now derive it in the case of a general stationary, axisymmetric metric of the block diagonal form \eqref{gsymm}. 
A similar equation can be derived in the presence of other sorts of symmetries. 

The stream equation follows directly from the two force-free equations (\ref{EFFFE}). 
As already demonstrated above Eq.~\eqref{Iinv}, the first force-free condition implies that $I=I(\psi)$, which is equivalent to conservation of energy and angular momentum.   
The second force-free condition yields
\begin{align}\label{enchilada}
0 &= d\phi_2\w d*F\nonumber\\
&=(d\psi_2 + \eta -\O_F' t \, d\psi ) \wedge \left(\frac{I'}{2\pi} d\psi \wedge dt \wedge d\varphi - d(\star d\psi \wedge \star \eta)\right) \nonumber \\
&=\frac{I'}{2\pi} d\psi_2 \wedge d\psi \wedge dt \wedge d\varphi - \eta \wedge d(\star d \psi \wedge \star \eta) \nonumber \\
& = \frac{I I'}{4\pi^2 g^T} \e + d(\eta \wedge \star d\psi \wedge \star \eta) - d \eta \wedge \star d \psi \wedge \star \eta \nonumber \\
& = \frac{ I I'}{4\pi^2 g^T}\epsilon - d(|\eta|^2 \star d\psi \wedge \epsilon^T  ) + \O_F' d\psi \wedge dt \wedge \star d\psi \wedge \star \eta \nonumber \\
& = \frac{ I I'}{4\pi^2 g^T}\,\e - d(|\eta|^2 * d\psi) - \O_F' |d\psi|^2 \la dt,\eta\ra\, \e.
\end{align}
In the second line prime denotes a $\psi$ derivative, and we use Eqs.~\eqref{EPsymm}, \eqref{etadef}, \eqref{starFsymm} and \eqref{Iinv}. Of the six cross terms, only two survive in the third line; two vanish because they contain three poloidal 1-forms, one vanishes because it contains three toroidal 1-forms, and one vanishes because it contains the same 1-form twice.
In the fourth line, in the first term we use the dual of (\ref{Idef}), together with (\ref{eTeP}) and (\ref{epsilonTP})
[alternatively, (\ref{aw*b})]. The other two terms arise from ``integration by parts" of the second term in the previous line,
using the anti-derivation property of $d$. To obtain the fifth line we use (\ref{aw*b}) in the second term, 
and the definition of $\eta$ (\ref{etadef}) in the third term. In the last line we use (\ref{star*}) in the second term,
and (\ref{aw*b})  and (\ref{epsilonTP}) in the third term. 

Finally, since $d*\o=\nabla_a\o^a\,\e$ for any 1-form $\o$, the last line of (\ref{enchilada}) yields
\beq\label{streameqn}
\nabla_a ( |\eta|^2 \nabla^a \psi ) + \Omega_F' \langle dt,\eta \rangle |d\psi|^2 - \frac{I I'}{4\pi^2 g^T}  = 0,
\eeq
where $\nabla_a$ is the covariant derivative determined by the spacetime metric.
This is the stream equation, in a form that holds for any metric of the form \eqref{gsymm}. 
If $\Omega_F$ and $I$ are specified as given functions of $\psi$, then 
Eq.~\eqref{streameqn} becomes a quasilinear elliptic equation for $\psi$, with critical 
points where the 1-form $\eta$ is null, i.e.\ at light surfaces (see Sec.~\ref{Dsaf}).  

The stream equation (\ref{streameqn}) involves the quantities $|\eta|^2$ and $\la dt,\eta\ra$,
which depend on $\O_F$ and the toroidal metric.
Without loss of generality we may write this metric in the common form
\beq\label{gnice}
(ds^T)^2 = - \alpha^2 dt^2 + \rho^2 (d\varphi - \Omega_Z dt)^2,
\eeq
where $\alpha$, $\rho$, and $\O_Z$ are functions of the poloidal coordinates $(r,\theta)$. The quantity $\Omega_Z$ is the angular velocity of zero angular momentum observers (ZAMOs), who follow the (non-gedoesic) toroidal curves orthogonal to the angular Killing field  $\partial_\vphi$,  while $\alpha$ is the rate of ZAMO time with respect to $t$, sometimes called the redshift factor \citep{macdonald-thorne1982}.
In terms of these quantities, those appearing in the stream equation are given by
\begin{align}
|\eta|^2 & = \rho^{-2} - \alpha^{-2}(\Omega_F-\Omega_Z)^2 \label{etasquared} \\
\langle dt,\eta \rangle & = \alpha^{-2}(\Omega_F-\Omega_Z) \label{dteta} \\
-g^T & = \alpha^2 \rho^2. \label{gT}
\end{align}
In particular, the light surfaces are located 
where $\rho=\pm \alpha/(\Omega_F-\Omega_Z)$, and $\langle dt,\eta \rangle$ 
vanishes where $\Omega_Z=\Omega_F$.  
 
For comparison with other treatments, note that the four-dimensional determinant 
$g$ can also be expressed as $-\a \rho g^P$ or as $g^Tg^P$.  We may thus write the
first term in \eqref{streameqn} using the 
covariant derivative $\bD_a$ on the three-dimensional surfaces of constant $t$
or the two-dimensional poloidal covariant derivative $D_a$, giving
\begin{align}
\nabla_a ( |\eta|^2 \nabla^a \psi ) & = \a^{-1/2}\bD_a[\a^{1/2}|\eta|^2 \bD^a\psi] \label{3Dstream} \\ 
& = (-g^T)^{-1/2}D_a[(-g^T)^{1/2}|\eta|^2 D^a\psi]. \label{2Dstream}
\end{align}
The RHS of Eq.~\eqref{3Dstream} is the standard $3+1$ form \citep{macdonald-thorne1982}, while Eq.~\eqref{2Dstream} gives a $2+2$ form.

It is worth mentioning that the stream equation can apply more generally than
in the stationary axisymmetric case. In particular,  for any $2+2$
metric, if the field is symmetric under one
of the factors of the $2+2$, and falls into
Uchida's case 1 (Appendix \ref{sec:twosymmetries})
then the same manipulations above will give rise to a stream equation that differs only in minor details.  For example, a stream equation applies to the case where the field is plane symmetric, i.e.\ $x$ and $y$ are the ignorable coordinates, while the fields depend on $z$ and $t$, in flat spacetime.

\subsubsection{Action derivation of stream equation}

The stream equation can also be efficiently derived directly
from the action \eqref{SFF}, with the symmetric form \eqref{EPsymm}
for the potentials. \citet{uchida1997general} worked this out and explained the 
relation to the Scharlemann-Wagoner action \citep{scharlemann-wagoner1973} from which the
derivation is even simpler. Here we will briefly summarize Uchida's 
analysis using our methods. 

The action \eqref{SFF} takes the form 
\beq
S^{\rm sym}=-\half\int (d\psi\w d\psi_2)\w *(d\psi\w d\psi_2) + |\eta|^2 |d\psi|^2 \e
\eeq
The quantities to be varied are $\psi$ and $\psi_2$,
while $\O_F(\psi)$ in $\eta$ is treated as an fixed function. The variation
of $\psi_2$ yields the equation $d\psi\w d*(d\psi\w d\psi_2)=0$,
which using \eqref{starFsymm} implies $d\psi\w dI=0$, and hence $I=I(\psi)$.
[This is basically the same as the derivation of \eqref{Iinv}.]
The variation of $\psi$ in the second term of the action 
yields minus the first two terms of the stream equation 
\eqref{streameqn} times $\e$, while variation in
the first term yields 
\beq
d\psi_2\w d*(d\psi\w d\psi_2)=\frac1{2\pi}d\psi_2\w dI\w dt\w d\vphi=
\frac{II'}{4\pi^2 g^T}\e,
\eeq
where in the last step we used the conclusion $I=I(\psi)$ from the 
$\psi_2$ variation, together with 
\beq
d\psi_2\w dI = I'd\psi_2\w d\psi = - \frac{II'}{2\pi(-g^T)^{1/2}}\e^P.
\eeq
Hence we recover the stream equation \eqref{streameqn}.

It is tempting, after having found that $I=I(\psi)$, to substitute
$d\psi\w d\psi_2 = (I/2\pi\sqrt{-g^T})\e^P$ back into the
action, eliminating $\psi_2$ and 
yielding $(-I^2/4\pi^2 g^T)\e$ for the first term in the 
integrand, and then treating $I$ as a fixed function.
This is not correct:
it would be like solving for a velocity component $\dot q(p,q^i,\dot{q}^i)$ in mechanics 
in terms of a conserved conjugate momentum $p$ and the other 
coordinates and velocities, and substituting that
back into the action. The resulting action would yield
invalid equations of motion, because in the original action the
conserved quantity was not held fixed. However, if 
at the same time one modifies the Lagrangian by 
addition of $-p\dot q(p,q^i,\dot{q}^i)$, the procedure is then correct.
(This amounts to using the Hamiltonian formalism for $q$, and
the Lagrangian formalism for the remaining coordinates.)
Following an analogous procedure to trade the $\psi_2$ dependence
of the action in favor of $I(\psi)$, Uchida shows that
the net result is simply to flip the sign of the $I^2$ term, yielding the action
\beq
S^{\rm SW}=-\half \int  \left(\frac{I^2}{4\pi^2 g^T} + |\eta|^2|d\psi|^2\right)\e.
\eeq
This is the \textit{Scharlemann-Wagoner action} \citep{scharlemann-wagoner1973}, from which the stream 
equation \eqref{streameqn} follows immediately as the $\psi$ stationarity condition
when treating $I(\psi)$ as a fixed function.

\subsubsection{Solution of the Stream Equation}\label{sec:solvestream}

The stream equation \eqref{streameqn} for the stream function $\psi$ has the peculiar feature that it contains unknown functions $\Omega_F(\psi)$ and $I(\psi)$ which must also be somehow determined.  In this subsection we briefly discuss the nature of this equation and mention several approaches to finding solutions.

If $\Omega_F(\psi)$ and $I(\psi)$ are specified then the stream equation \eqref{streameqn} becomes a quasilinear equation for $\psi$.  Where $|\eta|\neq0$ (i.e., away from any light surfaces) the equation is second order, with elliptic principal part.  Thus on a domain not containing light surfaces one expects unique solutions given suitable boundary data for $\psi$.\footnote{For Dirichlet data, choosing $I(\psi)$ and $\O_F(\psi)$ is equivalent to specifying $I$ and $\O_F$ on the boundary.  Thus the total boundary data is a component of the poloidal magnetic field (derived from $\psi$), a component of the poloidal electric field (obtained from $\O_F$ and $\psi$), and the toroidal magnetic field (proportional to $I$).}  At a light surface (where $|\eta|=0$) the stream equation becomes first order,\footnote{If $|\eta|^2$ vanishes quadratically or
faster, the equation would actually be zeroth order, i.e., algebraic.
However, this situation does not arise in practice.}
\beq\label{streambc}
\nabla_a (|\eta|^2) \nabla^a \psi + \Omega_F' \langle dt,\eta \rangle
|d\psi|^2 + \frac{I I'}{4 \pi^2 g^T}=0.
\eeq
When $I(\psi)$ and $\Omega_F(\psi)$ are specified, this may be viewed
as a Robin-type boundary condition for $\psi$ at a new boundary, the light surface. If a single light surface cuts a domain in two, one expects a unique solution on either side, but the solutions will generally not match smoothly at  the light surface.  It is thus plausible that the requirement of smooth matching restricts the choice of $I(\psi)$ and $\Omega_F(\psi)$ to a single free function on field lines, at least on field lines (values of $\psi$) that cross the light surface. If a second light surface is crossed by the same field line, one expects both $I(\psi)$ and $\Omega_F(\psi)$
to be determined.

These expectations are borne out in numerical calculations that iteratively update guesses for the free functions until a sufficiently smooth match is achieved across all light surfaces.  This approach to solving the stream equation in the presence 
of light surfaces was introduced by \citet{contopoulos-kazanas-fendt1999} and later used by several other authors \citep{uzdensky2005,timokhin2006,gruzinov2006,contopoulos-kazanas-papadopoulos2013,nathanail-contopoulos2014}.  For a pulsar magnetosphere, $\Omega_F(\psi)$ may be fixed in advance to be 
the (constant) angular velocity of the star (cf. Sec.~\ref{O=OF}), and the single free function $I(\psi)$ may be determined by matching across the single light surface.  Black hole magnetospheres 
are qualitatively different in three respects: 
(i) the location of the light surfaces generally depends on $\O_F(\psi)$ and $\psi$,
(ii) if the black hole is spinning there can be two light surfaces (cf. Sec.~\ref{sec:bhls}), 
and
(iii) at the horizon there is a fixed relation between 
$\psi$, $I$, and $\O_F$, the Znajek condition (cf. Sec.~\ref{sec:znajek}).
The Znajek condition can be viewed as determining $\psi$ on the horizon, 
given $I(\psi)$ and $\Omega_F(\psi)$.  
On field lines that cross both light surfaces the latter two functions would also be determined.

In order to find analytic solutions to the stream equation one approach is to restrict the dependence of $\psi$ to a one-dimensional subspace of the two-dimensional poloidal space,  converting the stream equation into an ordinary differential equation (ODE). Then, for example if $\O_F$ is a fixed constant, the boundary condition on $\psi$ can determine $I(\psi)$ locally, leaving just an ODE to be solved.  This kind of tactic was used for example by \citet{menon-dermer2007}, who found a family of solutions in the Kerr spacetime where $\psi$, $\Omega_F$ and $I$ are independent of Boyer-Lindquist radial coordinate $r$.

Finally, stationary, axisymmetric force-free solutions can be
generated by time-dependent evolution from non-force-free initial
data, using numerical devices that short out electric fields and
dissipate energy.  Thus one effectively solves the stream equation
through time-dependent evolution
\citeeg{komissarov2001,mckinney2006,spitkovsky2006,komissarov-mckinney2007}.

\subsection{Field line topology}

In this subsection we establish two restrictions on the possible topology 
of magnetic field lines in stationary, axisymmetric force-free magnetospheres. 
In sections \ref{sec:pulsar-magnetosphere} and \ref{BHM} we apply the second of these results to pulsar and black holes magnetospheres.

\subsubsection{No closed loops} 
We begin with the simpler of the two restrictions:
\begin{quote}
\textit{A stationary, axisymmetric, force-free, magnetically dominated 
field configuration cannot possess a closed loop of poloidal field line.}
\end{quote}
By a closed loop of poloidal field line we mean a level set of $\psi$ that forms a smooth closed curve, i.e., a closed set $\psi=\textrm{const}$ on which $d\psi \neq 0$.  Note that such loops do not in general correspond to closed loops of ``true'' field line, since those lines bend in the $\partial_\vphi$ direction. To establish the result we employ the expression 
\eqref{EFFFE} of the force-free condition in terms of the conservation of the Euler currents.  In particular, we use the fact that the Euler current $J_2 = d \phi_2 \wedge *F$ is a closed 3-form. By Stokes' theorem, this implies the 
vanishing of the integral of $J_2$ over any closed 3-surface bounding a force-free region of spacetime.

\begin{figure}
\centering
\subfigure[]{\label{fig:noloops}\includegraphics[width=20mm]{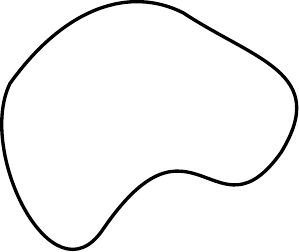}}
\hspace{1.5cm}
\subfigure[]{\label{fig:nolsloops}\includegraphics[width=20mm]{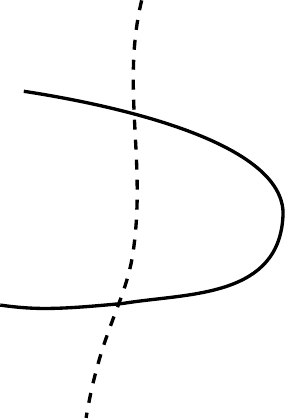}}
\caption{Disallowed topologies of force-free poloidal magnetic field lines.  (a) No closed loops when magnetically dominated.  (b) No light surface loops when $\Omega_F'=I'=0$.}
\label{fig:topology}
\end{figure}

Suppose for contradiction that a closed loop of poloidal field line exists, i.e., that a smooth level curve $\cC$ of $\psi$ is closed.  Flowing this loop along $\partial_\vphi$ one obtains a torus, and flowing that along $\partial_t$ (by an amount $\Delta t$), one obtains a closed 3-surface, consisting of a timelike tube $\cS={\cal C}\times S^{1}\times \Delta t$ and initial and final spacelike, solid torus caps ${\cal C}\times S^{1}$.  Integrating $J_2$ on this surface, the contributions from the caps cancel. The timelike tube is a surface of constant $\psi$ on which $d\psi\ne0$, so we can express it using (\ref{v.dyintegral}), with 
$v$ any vector field such that $v\cdot d\psi=1$ on $\cS$:
\begin{align}\label{notalooptobefound}
0&=\int_\cS d\phi_2\w *F \nonumber\\
&= \int_\cS v\cdot(d\psi\w d\phi_2\w *F)\nonumber\\
&=\int_\cS v\cdot(F\w *F)\nonumber\\
&=\frac{1}{2}\int_\cS  F^2 \, v\cdot \epsilon \nonumber \\
&=\pi \D t \oint_\cC F^2 \sqrt{-g^T} v\cdot \epsilon^P
\end{align}
In the third line we used $F=d\phi_1 \wedge d\phi_2$ (\ref{Euler}) and $\phi_1=\psi$ (\ref{EPsymm}), in the fourth line we used \eqref{aw*b}, and in the last line we used (\ref{epsilonTP}) and carried out the toroidal part of the integral. The contraction of the poloidal 1-form $v\cdot \epsilon^P$ with a tangent vector to $\cC$ vanishes only if $v$ is also tangent to $\mathcal{C}$, which is excluded by $v \cdot d\psi=1$.
Thus, if $F$ is magnetically dominated ($F^2>0$) everywhere on $\cC$, the integral
\eqref{notalooptobefound} cannot vanish, so we have a contradiction. This establishes our ``no closed loops" result.

\subsubsection{Light surface loop lemma}
\label{sec:light-surface-lemma}

In this subsection we prove a light surface loop lemma that will be useful when we treat pulsar and black hole magnetospheres.  This lemma was part of the   ``no ingrown hair'' argument of \citet{macdonald-thorne1982}, but here we present it on its own and also discuss two related results.  The lemma states that for stationary, axisymmetric, force-free fields (not necessarily magnetically dominated),
\begin{description}
\item {\it No poloidal field line may pierce a light surface twice in a contractible region where} $\Omega_F'=I'=0$.
\end{description}
The condition $\Omega_F'=0$ indicates that all field lines rotate with the same angular velocity, while $I'=0$ implies that no poloidal current flows in this region (otherwise $I$, being the total current through the cap with boundary at $\psi$, would depend on $\psi$).  Our hypotheses allow non-zero toroidal magnetic field $I$, supported by poloidal current flowing elsewhere in the magnetosphere.

The proof is based on the fact that when $\Omega_F'=I'=0$, conservation of the second Euler current is equivalent to conservation $|\eta|^2 *d\psi$,
\beq\label{newcurrent}
0 = d(d\phi_2\wedge *F) = d(|\eta|^2*d\psi).
\eeq
Eq.~\eqref{newcurrent} result follows from the derivation of the stream equation \eqref{enchilada}, and the vectorial version $\nabla_a(|\eta|^2 \nabla^a\psi)=0$ can be seen directly from the final form of the stream equation, Eq.~\eqref{streameqn}.  Suppose for contradiction that a smooth ($d\psi\neq0)$ poloidal field line intersects a light surface (where $|\eta|=0$) twice, without intersecting another light surface, as depicted in Fig.~\ref{fig:nolsloops}. We may construct a closed 3-surface by considering the portions of the field line and light surface that form a closed poloidal loop, and flowing this loop along $\partial_t$ (for time $\Delta t$) and along $\partial_\varphi$.  Integrating $|\eta|^2 *d \psi$ over this 3-surface, the initial and final spacelike caps cancel by stationarity, while the timelike portion involving the light surface vanishes since $|\eta|=0$ there.  The remaining portion is that generated by the field line segment, which is a surface of constant $\psi$ on which $d\psi\neq0$. With the same notation as in the previous subsection, that integral is given by 
\begin{align}
0&=\int |\eta|^2 v\cdot (d\psi\w *d \psi)\nonumber\\
&=\int |\eta|^2 |d\psi|^2 v\cdot \e.\label{notalightsurfacelooptobefound}
\end{align}
By assumption $\eta$ is non-null everywhere on the field line segment, so cannot change sign.  Since the poloidal subspace is Riemannian, we have $|d\psi|^2>0$.  Thus $|\eta|^2 |d\psi|^2$ does not change sign, and the reasoning used below Eq.~\eqref{notalooptobefound} implies that the integral \eqref{notalightsurfacelooptobefound} cannot vanish, a contradiction.

If more than one light surface is present, then, for any field line segment that pierces a single light surface twice (or more), there will also be a subsegment that pierces a (possibly different) light surface twice without encountering any other light surface.  We may then run the above argument on that subsegment, again concluding that the field configuration is impossible.  This establishes the lemma.

The conclusion of the light surface lemma also holds in two additional cases.  First, since it is the product $\Omega_F'\langle dt,\eta \rangle$ that appears in the stream equation \eqref{enchilada} or \eqref{streameqn}, we may replace the assumption $\Omega_F'=0$ with $\langle dt,\eta \rangle=0$.  The interpretation of $\langle dt,\eta \rangle=0$ is that field lines co-rotate with zero angular momentum observers, cf.\ Eq.~\eqref{dteta}.  

Second, we may drop the force-free assumption and instead assume that i) angular momentum is conserved (first force-free condition is satisfied and hence $I=I(\psi)$) and ii) there is a reflection isometry\footnote{A definition of a reflection isometry is given in Sec.~\ref{sec:openclosed}} (of the spacetime and the fields) about a spacelike 2-surface (poloidal curve flowed in toroidal directions) intersecting the light surface loop. The reflection isometry implies that $I$ is odd under reflection (see Eq.\eqref{FsymmI}, noting that $F$ is even while $\epsilon_P$ is odd), but $I=I(\psi)$ and the evenness of $\psi$ implies that $I$ is even.  Thus we in fact have $I=0$, and the assumption $I'=0$ is superfluous.  We must still include $\Omega_F'=0$ as an assumption, and in this case we still have the last equality in \eqref{newcurrent}, $d\phi_2 \wedge d*F=d(|\eta|^2*d\psi)$.  These quantities are now \textit{not} known to vanish, but we may use the reflection isometry to argue that their integrals do vanish.  To do so note that $d\phi_2$ is even (first use the evenness of $\eta$ and $F$ to establish evenness of $d\psi$, and then $F=d\psi \wedge d\phi_2$ implies $d\phi_2$ is even), while $d*F$ is odd on account of the duality.  Thus the form $d\phi_2 \wedge d*F$ is odd.  Since the shape of the light surface loop is symmetric, the integral of this form over the interior of the loop (flowed in $\vphi$ and $t$ to form a four-volume) is vanishing.  This establishes the first line of \eqref{notalightsurfacelooptobefound}, and the same arguments establish a contradiction.  To summarize, we have shown that for stationary, axisymmetric, reflection-symmetric, degenerate, energy and angular momentum conserving fields with $\Omega_F'=0$ (or $\langle dt,\eta \rangle=0$), no light surface loops straddling the reflection surface may exist.

\subsection{Special case: no poloidal field}\label{sec:special-case}
As mentioned above, the form \eqref{EPsymm} of the Euler potentials of a stationary, axisymmetric solution is valid only when $F\cdot \partial_\vphi\neq 0$. The case $F\cdot \partial_\vphi = 0$ corresponds to a purely toroidal magnetic field, and may be useful in situations where the poloidal field is small \citeeg{contopoulos1995}.
When $F\cdot \partial_\vphi = 0$ one may choose the form \eqref{EPsymmAppendixEx},
\beq\label{EPsymm2}
\phi_1 = \chi(r,\theta), \quad \phi_2 = \chi_2(r,\theta) + t.
\eeq
We may then write $F$ as
\beq
F = \frac{I}{2\pi (-g^T)^{1/2}} \epsilon_P + d\chi \wedge dt,
\eeq
where $I$ satisfies $*(d\chi \wedge d\chi_2)=(I/2\pi) dt \wedge d\vphi$, the analog of \eqref{Idef}. It is evident that $\chi$ is an electric potential for the (purely poloidal) electric field, while $I$ is again equal to the electric current through a toroidal loop (Sec.~\ref{sec:I}). As in the generic case, the first force-free equation implies $I=I(\chi)$ (see discussion above Eq.~\eqref{Iinv}).  To derive the associated stream equation we may follow the same steps of equations \eqref{enchilada}, finding 
\beq\label{streameqn2}
\nabla_a ( |dt|^2 \nabla^a \chi ) - \frac{I I'}{4\pi^2 g^T} = 0,
\eeq
where $'$ is a $\chi$ derivative. Note that \eqref{EPsymm2} and \eqref{streameqn2} 
can be obtained from the generic versions \eqref{EPsymm} and \eqref{streameqn} (respectively) by the replacements $\psi \rightarrow \chi$, $\vphi \rightarrow t$, and $\Omega_F \rightarrow 0$. 
However, since $\psi$ determines the poloidal magnetic field (which vanishes in the special case), it is more physical to obtain them from the limit $\psi \rightarrow 0$ and $\Omega_F \rightarrow \infty$ with the product 
$\Omega_F \psi \rightarrow -\chi$ finite. 

An example of this limit is
provided by the Michel monopole solution $-qd(\cos\theta)\w (d\vphi - \O_F du)$.
In the limit $q\rightarrow 0$, $\O_F\rightarrow\infty$, with $q\O_F$ held fixed.  The
vacuum monopole term (which provides the poloidal magnetic field) vanishes, leaving just the stationary, axisymmetric outgoing Poynting flux solution \eqref{Fout}, $F=q\O_F d(\cos\theta)\w du$, which satisfies the stream equation \eqref{streameqn2} rather than the generic stationary axisymmetric stream equation.  Euler potentials for this solution in the above notation are
specified by $\chi = q\O_F \cos\theta$ and $\chi_2 = -r$.}

Another interesting example arises in the Menon-Dermer solution, i.e.\ the 
stationary axisymmetric case of the Poynting flux solution \eqref{FoutKerr1} in Kerr spacetime \eqref{FoutKerr2}, $A(\theta)d\theta\w (du - a\sin^2\theta\, d\bar{\vphi})$. That solution falls in the generic class on account of the $d\bar{\vphi}$ term, but since that term vanishes along the axis, the axis limit lands on this special case. The angular velocity of the field lines in this solution is $\O_F = 1/(a\sin^2\theta)$, whose limit indeed diverges as the axis is approached.

The case of no poloidal field does not appear to have been previously considered in the force-free context.  However, \cite{gourgoulhon-etal2011} have given a completely general treatment of stationary, axisymmetric equilibria in the context of ideal magnetohydrodynamics, which includes the magnetically dominated force-free case as a limit.

\section{Pulsar Magnetosphere}\label{sec:pulsar-magnetosphere}

This section addresses general features of magnetospheres around conducting, magnetized stars in the case of aligned rotation and magnetic axes.  Such a configuration is stationary and axisymmetric, so does not pulse; however, it serves as a simple example of key properties of pulsar magnetospheres, and as an approximation for a nearly aligned pulsar. Specifically, we discuss the boundary condition at the stellar surface which 
determines the angular velocity $\O_F(\psi)$ of the field, and the roles of the light cylinder and current sheet in delimiting the region of closed field lines.

The pulsar magnetosphere has mainly been studied in flat spacetime.  We will discuss general features of pulsar magnetospheres based on the general metric \eqref{gsymm}, so our comments will hold when gravity is included.   Our analysis also serves to identify precise circumstances under which each particular feature must hold.

\subsection{Angular velocity of field lines}
\label{O=OF}

The angular velocity of field lines $\Omega_F$ may be determined by the assumption of a perfectly conducting stellar surface, which should be a good approximation for neutron stars.
If $U$ is the 4-velocity field of a perfectly conducting surface, then the contraction of $U\cdot F$ with any vector tangent to the surface vanishes. That is, the electric field in the rest frame of the conductor must have no component tangent to the surface.  If the surface is that of an axisymmetric star with
four-velocity $U\propto \partial_t + \O\partial_\vphi$, then for a stationary, axisymmetric degenerate field (\ref{Fsymm})
we have $U\cdot F =-(U\cdot\eta)d\psi\propto(\O_F-\O)d\psi$.
Provided the poloidal magnetic field is not tangent to the stellar surface (i.e., provided 
there is a surface tangent vector $v$ with $v\cdot d\psi\neq0$), it follows that 
$\Omega=\Omega_F$.  
We have thus shown that for stationary, axisymmetric, degenerate fields,
\begin{quote}
\textit{Poloidal field lines that non-tangentially intersect a perfectly conducting star must have $\Omega_F=\Omega$.}
\end{quote}
Thus the field lines co-rotate with the star.  Note that when $\Omega=\Omega_F$ we have $U\cdot F=0$ (see expression in text above), implying that also the \tit{normal} component of the electric field in the rotating frame must vanish at the surface of the conducting star.  Thus there is no induced charge on the stellar surface, according to co-rotating observers.  Static observers, on the other hand, will generically measure induced charge, depending on the assumptions for the field configuration within the star.
 
The lack of induced surface charge in the rotating frame is a direct consequence of degeneracy and the conducting boundary condition: Since the tangential components must vanish on the conducting surface, the electric field is purely normal.  But if the magnetic field has a normal component, $\vec{E}\cdot\vec{B}=0$ implies that the electric field vanishes entirely, and there is no induced charge.  If the star were instead surrounded by vacuum, the field would not be degenerate, and generically there would be a surface charge and a normal component of the electric field in the rotating frame. 

\subsection{Open and closed zones}\label{sec:openclosed}

\begin{figure}
\centering
\includegraphics[scale=.8]{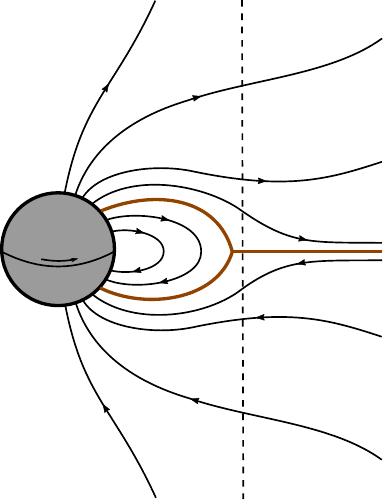}
\caption{Diagram illustrating the poloidal structure of the standard aligned pulsar magnetosphere.  A current sheet (thick brown line) separates poloidal field lines (black) into three zones, one closed and two open.  The closed zone terminates at, or just within, the light cylinder (dashed line, shown artificially close to the star).  }\label{fig:aligned_pulsar}
\vspace{-3mm}
\end{figure}

Closed field lines are defined to be field lines that intersect the star twice, while open field lines intersect it once.  In vacuum, the field lines of a monopole star would all be open, while those of a dipole are all closed.  The standard aligned force-free pulsar magnetosphere (Fig.~\ref{fig:aligned_pulsar}), on the other hand, is a mix: the field lines form a dipole pattern at the star, but only some of them return to close, with the rest opening up to infinity.  This basic structure of closed and open zones was postulated in the earliest work on the subject \citep{goldreich-julian1969}, and later work has confirmed that such solutions do exist.  

An important feature of all configurations previously considered is that the closed field lines remain within the light cylinder,\footnote{To model pulsars we focus on spacetimes for which degenerate field configurations will have a single light surface with the topology of a cylinder, outside of which co-rotating trajectories are spacelike.  We refer to this surface as the light cylinder.}  
unless they pass through a non-force free region such as a current sheet.
While it is commonly asserted that closed force-free field lines \textit{must} remain within the light cylinder, we are unaware of any explicit demonstration in the literature.  In this section we will critique the reasoning that one often hears or reads, and then demonstrate several related results based on various specific assumptions. We will conclude by explaining why, despite these results, 
the possibility that closed force-free field lines could venture outside the light cylinder has not (yet) been ruled out.

An argument one often hears or reads for the impossibility of closed magnetic field lines in a degenerate field outside the light cylinder is based on the notion that particles stuck on such field lines (in cyclotron motion) would have to be moving faster than the speed of light.  It seems unsatisfactory to invoke ``particles'' to determine something about force-free fields, however, since the matter plays no role in the dynamics of those fields (other than to carry the current).  To the extent that such an argument is valid, it should be possible to reformulate it without reference to particles.

To identify such a reformulation, we note that  
(i) particles are stuck on field lines only if the field is magnetically dominated, 
and (ii) such field lines 
\textit{always} admit sub-luminal particles stuck on them,  
because magnetically dominated, degenerate fields 
always have timelike field sheets. Hence 
 to rule out some behavior of a degenerate field 
 ``because particles stuck on field lines cannot 
go faster than light" must be logically equivalent 
to ruling it out simply by the assumption that 
the field is magnetically dominated. 
In what follows we will base our arguments on magnetic domination 
rather than considerations of particles.  
We will also give one argument that does not require magnetic domination.

A stationary, axisymmetric field cannot remain magnetically
dominated outside the light cylinder if 
the field line has no toroidal component,
because such a line sweeps out a spacelike field sheet outside the light cylinder. 
To reach this conclusion computationally, note that 
the absence of a toroidal magnetic
field is equivalent to the condition $I=0$ [cf. \eqref{FsymmI}],
which implies $F^2=2|d\psi|^2 |\eta|^2$ \eqref{F^2}.
The factor $|d\psi|^2$ is non-negative, since the poloidal subspace is Riemannian, 
and outside the light cylinder $|\eta|^2<0$ by definition, so if $I=0$ then
$F^2$ must be negative (or zero) outside the light cylinder, 
violating magnetic domination. This establishes a link between the 
absence of polar current and the confinement of magnetic field lines:
\begin{quote}
\textit{In a stationary, axisymmetric, degenerate, magnetically dominated field configuration, field lines with $I=0$ must lie within the light cylinder.}
\end{quote}

The idea that \textit{closed} field lines must remain within the light cylinder may in part be due to an association between closed field lines and $I=0$.  The original Goldreich-Julian model postulated a co-rotating portion of the magnetosphere, where one sign of charge rotates rigidly with the star.  
This portion indeed has $I=0$, since there is no poloidal current.  
In most self-consistent models constructed by solving the stream equation \citeeg{contopoulos-kazanas-fendt1999}, $I=0$ is assumed in the closed zone.  One possible reason for this assumption is the fact that
\begin{quote}
\textit{Reflection symmetry and the force-free condition imply that closed field lines crossing the equatorial plane must have $I=0$.}
\end{quote}
By ``reflection symmetry'' we mean an isometry of the spacetime and fields under which the volume element $\epsilon$ changes sign, and which leaves fixed a 3-surface composed of a spacelike 2-surface flowed along the timelike Killing field.  In flat spacetime this corresponds to the usual reflection symmetry about the plane $z=0$, and in the general case we refer to the spacelike 2-surface as the equatorial ``plane'' although it would generally have intrinsic curvature.  To establish the result we note that $I(\psi)$ is constant on the field line, hence is even under reflection, but it is also required to be odd since $F$ (\ref{FsymmI}) is even while the poloidal area element $\e^P$ is odd.  Since this result relies only on $I=I(\psi)$, it suffices to assume just that energy and angular momentum are conserved, rather than the full force-free condition (see discussion in Sec.~\ref{EandL} above).

Combining the previous two indented italicized statements, we have the following theorem on closed field lines:
\begin{quote}
\textit{In a stationary, axisymmetric, reflection symmetric, force-free, magnetically dominated field configuration, field lines crossing the equatorial plane must have $I=0$ and remain within the light cylinder.}
\end{quote}
Note that reflection symmetry implies that field lines which come from the star and cross the equator are closed,
so this result provides a precise set of assumptions under which the basic topology of the standard pulsar magnetosphere is to be expected. Again, the force-free assumption may be replaced by that of conservation of energy and angular momentum. 
The assumption 
of magnetic domination can be replaced by the condition that 
$\Omega_F'=0$ which, as shown in Sec.~\ref{O=OF}, necessarily 
holds if the field terminates on a rigidly rotating, perfectly conducting
star with non-tangential poloidal surface field.
Then, together with the other assumptions, 
the light surface loop lemma  of Sec.~\ref{sec:light-surface-lemma} applies
and establishes that closed lines cannot extend outside the light cylinder.  
We thus have the following alternate result:
\begin{quote}
\textit{In a stationary, axisymmetric, reflection symmetric, force-free, field configuration
with $\O_F'=0$, 
field lines crossing the equatorial plane must have $I=0$ and remain within the light cylinder.}
\end{quote}

Given the above considerations about closed field lines, a natural (and so-far standard) configuration for the dipole pulsar magnetosphere is that shown in Fig.~\ref{fig:aligned_pulsar}.  Field lines near the equator of the star would, in vacuum, have returned more quickly to the star, so it stands to reason that the closed zone of the configuration will be near the equator.  On the other hand, field lines near the poles would have extended far from the star in vacuum, so it makes sense that in the force-free case these field lines will open up to infinity.  Outside the light cylinder, there are no closed field lines, and the reversal of the sign of the field across the equator implies the presence of a current sheet (see Sec.~\ref{sec:sheets}).  A current sheet is also present at the boundary between closed and open zones, connecting to the star.  This general structure of the aligned dipole pulsar force-free magnetosphere has become standard, but alternative models do exist \citep{gruzinov2011,contopoulos-kalapotharakos-kazanas2014}. 

We conclude this section with a discussion of situations in which 
closed field lines could be present outside the light cylinder in a magnetically dominated field configuration.  We first discuss fields that are assumed to be degenerate but are otherwise arbitrary.  In this case it is straightforward to construct closed zones that proceed outside the light cylinder, since $\psi$, $\Omega_F(\psi)$, and $I$ may be chosen freely.  In particular, we may take $\Omega_F={\rm const}$, choose $\psi$ corresponding to the standard closed/open structure of the pulsar magnetosphere except with the closed zone penetrating the light cylinder, and then choose $I$ large enough that $F^2$ is positive everywhere in the closed zone (see Eq.~\eqref{F^2}).  This configuration must violate reflection symmetry, but by a similar construction we may form closed loops outside the light cylinder in a reflection-symmetric magnetosphere by confining those loops to a hemisphere. 

A more interesting example is provided by the new pulsar magnetosphere of \citet{contopoulos-kalapotharakos-kazanas2014}.  There the field is magnetically dominated, reflection symmetric, satisfies $\O_F'=0$, and is force-free everywhere except in the equatorial plane where lies a current sheet.  There are closed field lines with $I(\psi)\ne0$ that extend outside the light cylinder and have a kink at the current sheet. This configuration escapes our no-go theorem because of the infinitesimally thin current sheet violating the force-free condition.  Specifically, a nonzero current is consistent with reflection symmetry because $I(\psi)$ flips sign across the current sheet. That is, 
$I_{\rm below}(\psi)=-I_{\rm above}(\psi)$. This reversal of the sign of $I$ on a field line would not be allowed if the field were force-free everywhere.

Adopting the additional constraint that the fields be everywhere force-free has the potential to eliminate 
the possibility of closed poloidal field lines venturing outside the light cylinder,
but we have not found an argument that does so.  \citet{gruzinov2006} has found force-free solutions with closed lines having $I\neq0$, showing that there is no difficulty of the lines closing.  He has chosen to study configurations where the closed lines remain within the light cylinder, but we see nothing preventing alternative configurations for which they do not.  As far as we are aware, it is currently an open question whether magnetically dominated force-free fields can have closed field lines outside the light cylinder.

\section{Black Hole Magnetosphere}
\label{BHM}

Force-free magnetospheres of spinning black holes differ qualitatively from those of stars because of the presence of the event horizon and ergosphere.  Within a star, the force-free condition does not hold, so energy (and angular momentum) can be transferred from the star to the electromagnetic field. In the case of a stationary black hole, by contrast, the force-free condition may in principle hold up to and across the horizon, and conditions behind the horizon cannot affect the exterior, so there no analog of the star transferring energy to the field. On the other hand, the meaning of this conserved energy is modified
because of the spacetime curvature: as discussed in Sec.~\ref{EandL}, it is the integral of the Noether current associated with the ``time translation" Killing vector.  On a stationary 
spinning black hole spacetime, this Killing vector is timelike far from the black hole, but spacelike
near the black hole in the \tit{ergosphere}, due to the extreme ``dragging of inertial frames". In the ergosphere,
therefore, Killing energy is actually spatial momentum as defined by local observers. Hence the electromagnetic
field can have a local negative Killing energy density. Killing energy can therefore be extracted from the 
black hole, despite its conservation, because a corresponding \tit{negative} Killing energy can flow 
across the horizon into the black hole. A process in which rotational energy is extracted from a spinning black hole,
with a negative Killing energy flux across the horizon balancing a positive Killing 
energy flux at infinity, is (or should be\footnote{The term is sometimes reserved for a particular class of such processes. However, while Penrose introduced the idea
with the example of lowering a mass into the ergosphere at the end of a rope, 
he concluded: ``Thus, in a sense, we have found away of extracting rotational energy from the `black hole'. Of course, this is hardly a practical method! Certain improvements may be possible, e.g., using a ballistic method.$^7$ But the real significance is to find out what can and what cannot be done in principle since this may have some indirect relevance to astrophysical situations.'' (His footnote 7 states ``Calculations show that this can indeed be done,'' and goes on to describe the particle splitting method (now usually called \textit{the} Penrose process).})
called a \tit{Penrose process} \citep{penrose1969,penrose19692002}. 
When the mechanism involves a force-free magnetosphere 
it is generally known as the \textit{Blandford-Znajek process} or mechanism.  For further discussion of the relationship between the Penrose and BZ processes see e.g. \citep{komissarov2009,lasota-etal2014}.

Another qualitative difference between stellar and black hole magnetospheres lies in the topology of their field lines. The main qualitative feature of the pulsar magnetosphere relative to vacuum is that some field lines are open, although some loop back on to the star.  In the case of a spinning black hole, it turns out that \textit{all} field lines extending from the horizon must be open, unless they enter or loop around a non-force-free region, such as an accretion disk.  In reference to the ``no hair'' theorems on black hole uniqueness in vacuum, we name this result the \textit{no ingrown hair} theorem.  This illustrates the fact that while the magnetic field lines (hairs) may indeed emerge from the horizon of a black hole with a force-free magnetosphere, they may \textit{not} return unless they encounter a non-force-free region.  The no ingrown hair result was first derived by \citet{macdonald-thorne1982}; our proof is similar but uses some different arguments.

Finally, there is a tricky technical point that arises only in the treatment of black hole magnetospheres. The $2+2$ poloidal/toroidal decomposition of the spacetime, which is so useful in handling the stationary axisymmetric force-free equations, breaks down at the horizon. One issue is that the 1-form $dr$ becomes null (it is normal to the horizon, which lies at constant $r$), so that the poloidal subspace becomes null, rather than spacelike. Another is that the 1-forms $dt$ and $d\vphi$ diverge, both in a manner proportional to $dr$, so that the toroidal subspace also becomes null, with the same null direction as the poloidal subspace. One way to handle this is to instead use coordinates that are regular at the horizon. Alternatively, one can continue to use the $2+2$ decomposition, being careful to determine the appropriate conditions that must be imposed to ensure regularity at the horizon. Here we will do some of both.

In this section we adopt the Kerr metric for the black hole; however, the main important property is the presence of a horizon-generating Killing vector
$\partial_t+\Omega_H \partial_\vphi$, so analogous results could be easily derived for other spinning black hole metrics of the form \eqref{gsymm}.

\subsection{Znajek horizon condition}\label{sec:znajek}

On the future horizon of
the Kerr spacetime the quantities
$\O_F$, $I$, and $\psi$ 
are not independent, but instead obey the \textit{Znajek condition} \citep{znajek1977}, 
\beq\label{regularity-condition}
I = 2\pi(\O_F-\O_H)\psi_{,\theta}\sqrt{\frac{g_{\vphi\vphi}}{g_{\theta\theta}}}.
\eeq
This holds for any stationary axisymmetric degenerate solution of 
Maxwell's equations with $\partial_\vphi\cdot F\ne0$.
Here $\Omega_H$ is the angular velocity of the horizon, defined by the condition that the Killing field
\beq\label{chi}
\chi= \partial_t+\O_H\partial_\vphi,
\eeq
is tangent to the null horizon generators.  
Znajek obtained \eqref{regularity-condition}
by demanding that the corresponding electromagnetic field strength $F$ 
(Eq.~\eqref{FsymmI} in our notation) is regular on the 
horizon, and the condition is often employed to guarantee regularity in calculations involving irregular coordinates.  However, as elaborated below, all quantities appearing in \eqref{regularity-condition} have invariant geometrical status, making the condition independent of any coordinate concerns, including their regularity.  To emphasize this point we begin by providing a 
tensorial
derivation of \eqref{regularity-condition}.  We then present a
derivation expressing $F$ in regular coordinates, 
which shows how 
\eqref{regularity-condition} guarantees regularity of $F$ on the future horizon for non-extremal black holes.  We find an additional condition required for regularity in the extremal case $a=M$.  Finally, we reproduce the surprising fact, first pointed out by \cite{macdonald-thorne1982}, that the stream equation \eqref{streameqn} in fact \textit{implies} the Znajek condition up to sign, corresponding to regularity on either the future or past horizon.

To see that \eqref{regularity-condition} is a relationship between invariants,
recall that $\O_F$ may be defined by the condition $(\partial_t +\O_F\partial_\vphi)\cdot F=0$ for degenerate fields $F$ with $\partial_\vphi \cdot F \neq 0$, the polar current $I$ is the upward flow of charge per unit Killing time through a polar cap bounded by an axial loop, $2\pi \psi$ is the upward flux of $F$ through such a loop, and $2\pi \sqrt{g_{\vphi\vphi}}$ is its circumference.  These notions are valid on the horizon, where the remaining ingredient $\psi_{,\theta}/\sqrt{g_{\theta\theta}}$ is the derivative with respect to proper length along the horizon in the direction orthogonal to the two Killing vectors, away from the upward pole.

The tensorial derivation makes use of the horizon-generating Killing field
\eqref{chi} which, in view of the defining property of $\O_F$ and the structure
of the field strength \eqref{Fsymm}, has the useful property
\beq\label{chidotF}
\chi\cdot F = (\O_F-\O_H)d\psi.
\eeq
The derivation proceeds by evaluating the polar current on the horizon and performing 
a few manipulations:
\begin{align}
I/2\pi &=*F_{ab}\partial_t^a \partial_\vphi^b\\
&=*F_{ab}\chi^a \partial_\vphi^b\\
&=F_{ab}*\! (\chi^a \partial_\vphi^b)\\
&=F_{ab}\chi^a \partial_\theta^b\sqrt{g_{\vphi\vphi}/g_{\theta\theta}}\label{*Fline4}\\
&= (\O_F-\O_H)\psi_{,\theta}\sqrt{g_{\vphi\vphi}/g_{\theta\theta}}.
\end{align}
In the first line we used \eqref{Idef}, in the second line we used antisymmetry of $F_{ab}$ to replace $\partial_t$ by $\chi$, and in the third line we shifted the duality operation from $F_{ab}$ to $\chi^a \partial_\vphi^b$.  In the fourth line we used the fact that on the horizon $\chi$ is null and orthogonal to $\partial_\vphi$, so that the  2-form $\chi_{[a}(\partial_{\vphi})_{b]}$ is null and therefore can be dualized as explained in Appendix \ref{sec:null2dual}. In this step we also use the fact that $\partial_\theta$ is orthogonal to both $\partial_\vphi$ and $\chi$. We have chosen the sign appropriate for the future horizon.\footnote{To determine the sign of $\star\chi$, note that on the future horizon $\chi_a \sim (du)_a$ (both 1-forms are null and normal to the horizon).  As shown at the end of Appendix \ref{sec:null2dual} in the Schwarzschild case, $\star du=du$, hence $\star \chi = \chi$. The Kerr case is related by a continuous deformation to Schwarzschild, so the sign is the same. On the past horizon, $\chi_a \sim (dv)_a$, and $\star dv=-dv$, hence $\star \chi = -\chi$.} 
Finally, in the last line we use Eq.~\eqref{chidotF}.
Note that, other than stationary axisymmetry with commuting Killing fields,
the only special property of the spacetime used in this derivation
is the existence of a Killing horizon generated by $\chi$.

When thinking of the intrinsic quantities $I$, $\O_F$ and $\psi$ individually, we are unaware of any reason to expect them to be related on the horizon. However, since the poloidal and toroidal subspaces become null, and with 
non-trivial intersection in the limit at the horizon, the corresponding parts 
of the 2-form $F$ are not independent.  This makes the existence of the Znajek condition less surprising.

To show that \eqref{regularity-condition} guarantees regularity on the future horizon of Kerr
(in the non-extremal case) we begin with Eq.~\eqref{FsymmI}, $F=I/(2\pi\sqrt{-g^T})\epsilon_P+d\psi \wedge \eta$. Using $\sqrt{-g^P/g^T}=(r^2 + a^2\cos^2\theta)/(\D\sin\theta)$, the first term can be written as 
\beq\label{Iterm}
\frac{I}{2\pi(-g^T)^{1/2}} \epsilon_P = \frac{I}{2\pi}\frac{r^2 + a^2\cos^2\theta}{\D\sin\theta} \,dr\w d\theta,
\eeq 
where we have used relations from Appendix \ref{sec:Kerr}. 
Notice that this term is singular on the horizon $r=r_+$,  where $\Delta\equiv(r-r_+)(r-r_-)=0$. 
The second term $d\psi \wedge \eta$ is also singular because 
$\eta$ \eqref{etadef} is composed of the singular 1-forms $d\vphi$ and $dt$.  

To isolate the divergent behavior, we define a ``regularized"  co-rotation 1-form, 
\beq
\tilde\eta=d\vphit -\O_F\, dv,
\eeq
where $v$ and $\vphit$ are the  regular ingoing Kerr coordinates 
\eqref{v}, \eqref{vphih}. Then 
$\tilde\eta$ is regular on the future horizon, and differs from 
$\eta$ by a singular form proportional to $dr$,
\begin{align}
\eta &= \tilde\eta +[\O_F(r^2+a^2)-a]\frac{dr}{\D}\nonumber\\
& = \tilde\eta+[\O_F(r^2+a^2)-\O_H(r_+^2+a^2)]\frac{dr}{\D}.
\end{align}
The $\eta$ term in $F$ can then be written as
\begin{align}\label{etaterm}
d\psi&\w\eta = d\psi\w \tilde\eta \nonumber\\
& - \psi_{,\theta}\frac{(\O_F(r^2+a^2)-\O_H(r_+^2+a^2)}{\D}\,dr\w d\theta.
\end{align}
The field strength is the sum of \eqref{Iterm} and \eqref{etaterm}, 
\beq\label{nice}
F=d\psi\w\tilde\eta + \frac{f(r)}{(r-r_+)(r-r_-)}\, dr\w d\theta,
\eeq
with 
\begin{align}
f(r)= &\frac{I(r^2 + a^2\cos^2\theta)}{2\pi\sin\theta}\nonumber\\
& -\psi_{,\theta}[(\O_F(r^2 + a^2)-\O_H(r_+^2 + a^2)].
\end{align}
Regularity of the field at the horizon requires $f(r_+)=0$,
i.e.\
\beq\label{reg-Kerr}
I = 2\pi (\O_F-\O_H)\psi_{,\theta}\frac{(r_+^2 + a^2)\sin\theta}{r_+^2 + a^2\cos^2\theta},
\eeq
evaluated at $r_+$.
The rightmost factor agrees with $\sqrt{g_{\vphi\vphi}/g_{\theta\theta}}$ 
on the horizon, so we have recovered the Znajek condition \eqref{regularity-condition}
on the future horizon. 

In the non-extremal case $r_+\ne r_-$, this condition is also sufficient for regularity, 
and the horizon value of the field takes the form
\beq
F=d\psi\w\tilde\eta + \frac{f'(r_+)}{r_+-r_-}\, dr\w d\theta.
\eeq
In the extremal case, $f'(r_+)=0$ is a \textit{second} necessary condition, 
and together the two are sufficient.
Using $r_+=M=a$, which holds in the extremal case, this
second condition becomes 
\begin{align}
0=\frac{f'(r_+)}{2M^2}=&\frac{I'\psi_{,r}(1+\cos^2\theta)}{4\pi\sin\theta}
+\frac1M\left(\frac{I}{2\pi\sin\theta}-\psi_{,\theta}\O_F\right)
\nonumber\\
&
-\psi_{,\theta r} (\O_F-\O_H)- \psi_{,\theta}\psi_{,r}\O_F',
\label{reg-extreme}
\end{align}
where now  $\O_H=1/2M$, and the expression is evaluated at $r=r_+$. 
We are not aware of a previous derivation of this condition, although 
the analogous condition is known in the Reissner-Nordstrom case \citep{takamori-etal2011}.
In the \textit{near} extremal case, $F$ would become large if $f'(r_+)$ 
does not go to zero as $r_+-r_-=2M\sqrt{1-(a/M)^2}$.  It seems reasonable to expect that $F$ does not blow up as extremality is approached (for physically reasonable boundary conditions), so we expect 
\eqref{reg-extreme} to be approximately satisfied for near-extremal black holes.

\cite{blandford-znajek1977} regarded the Znajek condition \eqref{regularity-condition} as a boundary condition at $r=r_+$ for the stream equation.  However, \cite{macdonald-thorne1982} pointed out that 
it in fact \textit{follows} from the stream equation, up to a sign. 
To see this, first write the stream equation \eqref{streameqn}
near the horizon as
\begin{align}\label{streamI}
\frac{I I'}{4\pi^2} & = \frac{A}{\Sigma} \sin\theta\, \partial_\theta\left[\frac{\sin \theta}{\Sigma} \left( \O_F-\O_H \right)^2 \partial_\theta \psi \right] \nonumber \\ &  \qquad - \frac{A}{\Sigma^2}\sin^2\!\theta\, \O_F' \left( \O_F-\O_H \right) (\partial_\theta \psi)^2 + O(\Delta),
\end{align}
where $A$, $\Sigma$, and $\Delta$ are defined in Appendix \ref{sec:Kerr},
and we have made use of Eqs.~\eqref{etasquared}-\eqref{gT}.
Here we assume $\psi$, $\O_F$ and $I$ are smooth in $r$ and $\theta$.  The fact that only $\theta$-derivatives appear is related to the vanishing of $g^{rr}$ on the horizon.  On the horizon $r=r_+=\textrm{const}$ we may relate $\psi$ and $\theta$ derivatives using the ordinary chain rule, $f' = \partial_\theta f / \partial_\theta \psi$ for functions $f$.  Then we have
\begin{align}
\partial_\theta \left(\frac{I^2}{8\pi^2} \right) & = \frac{I I'}{4\pi^2}\partial_\theta \psi \\
&= \partial_\theta \left[\frac{A\sin^2\!\theta}{2 \Sigma^2} \left(\O_F-\O_H\right)^2 (\partial_\theta \psi)^2 \right],
\end{align}
where the second line follows using \eqref{streamI} and the Leibniz rule.  We may now integrate in $\theta$ along the horizon, yielding
\begin{equation}
I^2 = 4\pi^2 \frac{A\sin^2\! \theta}{\Sigma^2} \left(\O_F-\O_H\right)^2 (\partial_\theta \psi)^2 + C.
\end{equation}
The integration constant $C$ must vanish for the fields to be regular on the axis (otherwise a line current will be present there), 
so we conclude that the stream equation implies
\begin{equation}\label{regularity-condition-pm}
I =  \pm 2\pi \frac{\sqrt{A}\sin \theta}{\Sigma} \left(\O_F-\O_H\right) \partial_\theta \psi.
\end{equation}
This is the Znajek condition \eqref{regularity-condition}, with an additional $\pm$ on the right hand side.

If the minus sign is chosen then the fields are regular on the past horizon instead of the future horizon and thus represent a white hole magnetosphere 
rather than a black hole magnetosphere.  Note that it is impossible for the fields to be regular on both horizons, since all quantities appearing in \eqref{regularity-condition-pm} take the same value on both horizons.  The fact that Eq.~\eqref{regularity-condition-pm} is always satisfied by smooth solutions to the stream equation indicates that, when solving the stream equation on $r\geq r_+$, the only boundary condition at $r=r_+$ that one need impose is the sign choice corresponding to a black hole.

The Znajek condition has a number of practical applications. First, it can be used as a horizon boundary condition for the stream equation, 
as in the original BZ paper.  (The BZ solution can equally well be derived 
without the use of the condition; cf. \cite{mckinney-gammie2004} and our Sec.~\ref{sec:BZ}.) 
The condition is also 
helpful in derivations of theoretical results.  We use it to obtain the illustrative flux formulae \eqref{Lhorizonflux} and \eqref{Ehorizonflux} [following \cite{blandford-znajek1977}] and to prove the no ingrown hair theorem 
[following \cite{macdonald-thorne1982}].

\subsection{Energy and angular momentum flux}

The general expressions \eqref{Lflux} and \eqref{Eflux} give the outward energy and angular momentum flux in terms of the invariants $I$ and $\Omega_F$.  In the Kerr spacetime it is instructive to push these integrals to the horizon.
That is, let the poloidal curve $\mathcal{P}$ be the horizon $r=r_+$.  Using the Znajek condition \eqref{regularity-condition} and the fact that $dr=0$ on the horizon, we have 
\begin{align}
d\mathcal{L}/dt & = 2\pi \int_0^\pi (\Omega_H-\Omega_F) (\psi,{}_{\theta})^2 \sqrt{\frac{g_{\vphi\vphi}}{g_{\theta\theta}}}
\, d\theta \label{Lhorizonflux}\\
d\mathcal{E}/dt & =  2\pi \int_0^\pi \Omega_F (\Omega_H-\Omega_F) (\psi,{}_{\theta})^2 \sqrt{\frac{g_{\vphi\vphi}}{g_{\theta\theta}}}\, d\theta.\label{Ehorizonflux}
\end{align}
It follows that positive Killing energy flows outward 
if and only if $\Omega_F$ is between $0$ and $\Omega_H$. 
Since no influence can emerge from behind the horizon, 
however, it is more natural to say
 that negative Killing energy flows inward across the horizon,
as explained at the beginning of this section. Note that the BZ solution \eqref{FBZ} has $\O_F=\O_H/2$, the value that maximizes the energy flux at fixed magnetic flux through the horizon.

It also follows from \eqref{Lhorizonflux} and  \eqref{Ehorizonflux} that any 
outflow of energy is accompanied by an outflow of angular momentum. 
This is consistent with the fact that the source of the 
energy is the rotation of the black hole. 
A universal upper limit to the energy extracted
for a given angular momentum 
extracted can be found using the null energy condition, 
$T_{ab} \ell^a \ell^b>0$ for null vectors $\ell^a$, which in particular is 
satisfied by the electromagnetic field stress tensor.  
Since the horizon-generating Killing field $\chi$ is null on the horizon, 
we have $T_{ab}\chi^a \chi^b\ge0$ there.  This expression is 
equal to the inward flux of energy $T_{ab}\partial_t^a \chi^b$ minus $\Omega_H$ 
times the inward flux of angular momentum $-T_{ab}\partial_\vphi^a \chi^b$.
It follows that the \textit{outward} flux of energy is bounded by 
$\delta \mathcal{E} \le \Omega_H \delta \mathcal{J}$.  The BZ process satisfies $\d \mathcal{E} = \O_F \d \mathcal{J}$, 
so its efficiency is governed by the ratio $\O_F/\O_H$.

As noted in the original paper, the process can be characterized in thermodynamic terms. The BZ process is stationary, but when the back reaction on the geometry is taken into account it becomes a quasi-stationary process in which the first law of black hole mechanics \citep{bekenstein1973,bardeen-carter-hawking1973} should apply, 
$\delta M - \Omega_H \delta J = (\kappa/8\pi) \delta A$,
where $M$ and $J$ are the mass and angular momentum of the black hole,
$\kappa$ is the surface gravity and $A$ is the horizon area. 
The second law of black hole mechanics (which follows from the 
null energy condition and cosmic censorship) states that the area
cannot decrease, $\delta A \ge 0$ \citep{penrose-floyd1971,hawking1972,hawking-ellis1973}.
In the BZ process, $\d M=-\d  \mathcal{E}$ and $\d J=-\d \mathcal{J}$, so 
the first and second laws imply the same upper bound obtained above using 
the null energy condition directly. Perfect efficiency corresponds to the case in
which the area of the horizon is unchanged. According to the 
second law of black hole mechanics, only in that limit is the process
reversible.

\subsection{Light surfaces in a black hole magnetosphere}\label{sec:bhls}

Recall that a light surface is a hypersurface in spacetime where the
field sheet Killing vector $\chi_F =\partial_t+\O_F(\psi)\partial_\vphi$ is null or, equivalently, where the co-rotation 1-form $\eta$ is null.  Light surfaces play a practical role in finding force-free solutions, since they correspond to singular points of the stream equation (see Sec.~\ref{sec:solvestream}).  They also act as horizons for the propagation of particles and Alfven waves through the magnetosphere (see Sec.~\ref{sec:lightsurfaces}).  In the Kerr spacetime there are in general \textit{two} light surfaces, an outer one qualitatively similar to the ordinary light cylinder, and an inner one within the ergosphere. Outside the outer light surface, 
a co-rotating curve with angular velocity $\O_F$ is rotating too fast to be timelike, whereas
inside the inner light surface, it is rotating too \textit{slowly} to be timelike.  
The existence of the inner surface follows from
the fact that within the ergosphere observers (i.e.\ timelike curves) 
must rotate with a minimum angular velocity,  
which approaches $\Omega_H$ at the event horizon. 
Any field line with $\O_F<\O_H$ will therefore cross an
inner light surface at some point sufficiently close to the horizon. The inner light surface meets the 
horizon at the poles.

\begin{figure}
\centering
\includegraphics[scale=.65]{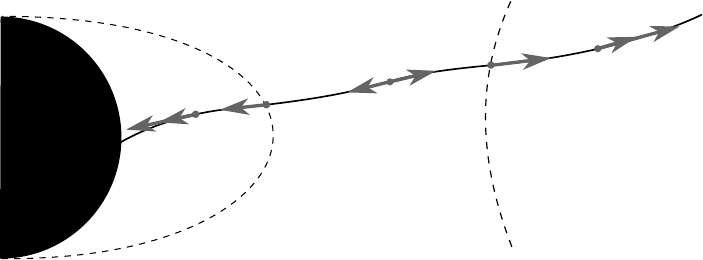}
\caption{Diagram of black hole light surfaces (dashed lines) and wind propagation along a field line.  The inner light surface is drawn exaggeratedly far from the black hole.  Arrows indicate the projection of the field sheet null vectors onto the poloidal plane.  These bound the possible poloidal velocities of particles moving on the field line.  Particles may only move inwards inside of the inner surface, and only outwards outside of the outer surface.}\label{fig:BH_wind}
\vspace{-3mm}
\end{figure}

The field sheet Killing vector is spacelike inside the inner light surface and outside the outer one, and timelike in between.  For $\Omega_F<\Omega_H$, wind particles and Alfven waves can travel only inward across the inner light surface and only outward across the outer light surface, as indicated in Fig.~\eqref{fig:BH_wind}.  This follows from the analysis of Sec.~\eqref{sec:tedchamp}, which established that the particle wind direction relative to that of positive angular momentum flow \eqref{Lhorizonflux} is determined by the sign of  $\O_F-\O_Z$, which is positive at the outer light surface and negative at the inner one.  This was shown using a different method by \citet{komissarov2004}, who has given a detailed discussion of the properties of the light surfaces of Kerr.

\subsection{No ingrown hair}\label{sec:no-ingrown-hair}

We now prove the no ingrown hair theorem, which forbids regions of closed field lines for black hole magnetospheres \citep{macdonald-thorne1982}.  By a closed field line we mean a smooth poloidal field line that non-tangentially\footnote{In earlier discussion we did not include this additional proviso as part of the definition of ``closed field line''.  Note, however, that non-tangential intersection is necessary for establishing that conductors determine the rotation frequency of their field lines (see Sec.~\ref{O=OF}).
} intersects the horizon twice, i.e., a level set of $\psi$ on which $d\psi\neq 0$, which intersects $r=r_+$ twice, each time with $\psi,{}_\theta\neq0$.  We first establish that $I=0$ and $\Omega_F=\Omega_H$ for a a closed field line in a force-free region.  (We use only the first force-free condition $I=I(\psi)$ for this part of the argument.)  The regularity condition \eqref{regularity-condition} implies that the sign of $I$ is determined by the product of $\O_F(\psi)-\O_H$, which is the same at the two ends (and everywhere on the line), with $\psi,{}_{\theta}$, which has opposite sign at the two ends (since loops with different values of $\psi$ are nested and $\theta$ is monotonic along the horizon).  Thus $I$ has opposite sign at opposite ends; however, $I$ is also constant on the line, and hence must vanish.  The regularity condition (together with $\psi,{}_{\theta}\neq0$) then implies that $\O_F=\O_H$ for the line.  

In a force-free region of closed field lines we thus have $\Omega_F'=I'=0$, so the conditions for the light surface loop lemma hold.  Furthermore, since $\O_F=\O_H$ the horizon itself is a light surface [recall that the Killing vector \eqref{chi} is null at the horizon], and the lemma applies to it.  This proves the no ingrown hair theorem: 
\begin{quote}
\textit{A contractible force-free region of closed poloidal field lines cannot exist in a stationary, axisymmetric, force-free Kerr black hole magnetosphere.}
\end{quote}
Thus a black hole cannot have a ``closed zone'' like a dipole pulsar does.  Notice that the theorem does not rely on reflection symmetry or magnetic domination.

\begin{figure}
\centering
\includegraphics[scale=.55]{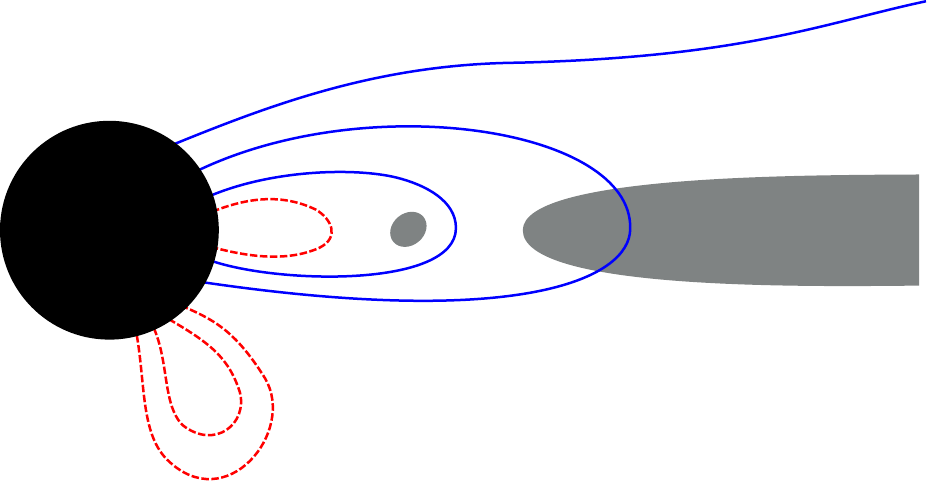}
\caption{Allowed (blue solid) and disallowed (red dashed) topologies of poloidal field lines in a force-free black hole magnetosphere.  Open lines are allowed.  Closed lines must pass through, or loop around, a non-force-free region (gray).}\label{fig:BH}
\vspace{-3mm}
\end{figure}

Closed field lines may exist if they pass through, or loop around, non-force-free regions of the magnetosphere.  For example, closed field lines may connect the black hole to an accretion disk.  Or, if a torus of material orbits the black hole, field lines may loop around the torus before returning to the horizon.  (These looping field lines must still have $\Omega_F=\Omega_H$ and $I=0$.)  The no-ingrown hair theorem is a natural generalization to the force-free setting of the no-hair idea that an astrophysical black hole cannot have its ``own'' magnetic field.  In order for closed field lines to exist, non-force-free currents must flow to support them.  These ideas are illustrated in fig.~\ref{fig:BH}.

Like its classical counterpart, the no ingrown hair theorem deals only with stationary situations, giving no insight into how any closed loops will be destroyed during the approach to stationarity.  It seems likely that loops will either be absorbed by the black hole \cite{membrane-book} or opened up by non-force-free processes \cite{lyutikov-mckinney2011}.  Qualitative discussions of different field line types in black hole magnetospheres can be found in \citep{membrane-book,blandford-lighthouse,hirose-etal2004,mckinney2005}.

\section*{Acknowledgments} We are grateful to Julian Krolik, Luis Lehner, Jon McKinney and Cole Miller for many helpful conversations.  T.J. is grateful for hospitality at the Institut d'Astrophysique de Paris, where some of this research was conducted.  This research was supported in part by Perimeter Institute for Theoretical Physics.  Research at Perimeter Institute is supported by the Government of Canada through Industry Canada and by the Province of Ontario through the Ministry of Research \& Innovation.  S.G. acknowledges support from NASA through the Einstein Fellowship Program, Grant PF1-120082.  T.J. was supported in part by the NSF under grant No. PHY-0903572.

\appendix

\section{Differential forms}\label{sec:forms}

A differential form is an antisymmetric, covariant tensor. Under the operations of
scalar multiplication, addition, and wedge product $\wedge$, differential forms comprise a graded algebra. The correspondence with tensor index notation is given by 
\beq
(\a\wedge \b)_{a_1\dots a_p b_1\dots b_q}=\frac{(p+q)!}{p!q!}\a_{[a_1\dots a_p}\b_{b_1\dots b_q]},
\eeq
where the square brackets denote antisymmetrization. Note that for any 1-form $\a$ we thus
have $\a\w\a=0$. 
The exterior 
(antisymmetrized) derivative $d$ is
a graded derivation on the algebra, and satisfies
$dd=0$. The graded derivation property is that for any $p$-form $\a$ and $q$-form $\b$, we have 
\beq
d(\a\w\b)=(d\a)\w\b + (-1)^p\a\w d\b.
\eeq
In tensor notation, 
\beq
(d\a)_{aa_1\dots a_p}=(p+1)\nabla_{[a}\a_{a_1\dots a_p]},
\eeq
where $\nabla_a$ is any torsion-free derivative operator, e.g.\
coordinate partial derivatives.  A form $\a$  is 
\textit{closed}, if $d\a=0$ and \textit{exact} if $\a=d\g$ for some $\g$.
An exact form is always closed, since $dd=0$.  In a contractible 
region, a closed form is always exact.

A $p$-form can be contracted with any number of vectors up to a maximum of $p$. 
Given a $p$-form $\a$ and a vector $v$, the contraction on the first ``slot"
or first ``index" of $\a$ will be denoted here by a dot:
\beq
(v\cdot\a)_{a\dots b}= v^{m}\a_{ma\dots b}.
\eeq
A more common notation for this operation is 
$i_v\a$. The \textit{pullback} of a form $\b$ to a $p$-dimensional submanifold 
$\cS$ is just $\b$, considered as a form on $\cS$. That is, the contraction of the pullback of $\b$ with any set of $p$ vectors tangent to $\cS$ is, by definition, the contraction 
of those vectors with $\b$.

\subsection{Integration of forms}
\label{Integration}
A $p$-form $\a$ can be integrated on a $p$-dimensional surface, or 
submanifold $\cS$.
Intuitively, one chops up $\cS$ into infinitesimal parallelopipeds 
each generated by $p$ infinitesimal vectors, evaluates $\a$
on these $p$ vectors, and adds the resulting numbers. Because 
of the multi-linearity and antisymmetry of $\a$, the result is independent
of how the chopping is done. Note, however, that the \tit{sign}
of the result depends on the order taken for the edge vectors of the parallelopipeds. 
Thus the integral is well-defined only once an \tit{orientation}
for $\cS$ is specified, i.e.\ an equivalence class of continuous, nowhere vanishing $p$-forms
on $\cS$ related by positive multiples. The edge vectors are ordered so that the result is positive on members of this class. (For brevity we will sometimes refer to a continuous, nowhere vanishing $p$-form $\o$ as an ``orientation" for a $p$-surface, meaning actually that $\o$ determines the orientation.)  An ordered coordinate system $(y^1,\dots,y^p)$ on $\cS$
determines the orientation $dy^1 \w \cdots \w dy^p$, with respect to which the integral 
of $\a=\a_{1\dots p}dy^1 \w \cdots \w dy^p$  is given by an ordinary multiple integral:
\beq
\int_\cS \a = \int \a_{1\dots p}\,  dy^1\cdots dy^p.
\eeq

In this paper we make use of two properties of 
$p$-form integrals. One is Stokes' theorem, which  
relates the integral of $d\o$ on $\cS$ to the
integral of $\o$ on the boundary $\partial\cS$:
\beq\label{Stokes}
\int_\cS d\o = \int_{\partial \cS} \o 
\eeq
where the orientation on $\partial\cS$ is the one
induced by contracting an outward-pointing vector 
on the first slot of the orientation form on $\cS$.
In particular, note that if $d\o=0$ then $ \int_{\partial \cS} \o =0$.

The other property pertains to the integral of a 3-form (more generally,
to the integral of an $(n-1)$-form over a hypersurface 
in an $n$-dimensional space).
Suppose $\cal S$ is 
a level hypersurface of the function $y$, i.e.\ it is defined 
by the equation $y=y_0$ for some constant $y_0$, and 
let $v$ be any vector field such that $v\cdot dy=1$ on $\cal S$.
Then the pullback of a 3-form $\o$ to $\cal S$ is equal to the pullback of  
$v\cdot(dy\w\o)$. (To show this just contract with any 
three vectors tangent to $\cal S$.) The integral of $\o$ on
$\cal S$ can thus be expressed as 
\beq\label{v.dyintegral}
\int_{\cal S} \o = \int_{\cal S} v\cdot(dy\w\o)
\eeq
This is useful when the 4-form $dy\w\o$ has 
properties that allow efficient computation of the integral.

\subsection{Hodge dual operator}
\label{Hodge}

To begin with an intuitive definition, 
the Hodge dual $*$ of a decomposable $p$-form is
simply the orthogonal decomposable form with the same squared norm,
up  to a sign. The sign of the squared norm of the dual is opposite in four dimensional spacetime, 
because the dual of a spacelike form is timelike, and vice versa. This definition is extended
linearly to linear combinations of such forms, and it defines the dual of a form up to a sign. 
A more precise definition fixes all of these signs, up to one overall sign that depends on the choice 
of an orientation. Keeping track of the signs can be tedious, but for many purposes they
need not be determined. The dual of the dual $*^2$ brings one back to the same form, up to a sign
that depends on the dimension of the space, the rank of the form, and the signature of the metric.
In two or four dimensional Lorentzian spacetime,  $*^2=\pm1$, with the $+$ sign for odd rank forms and
the $-$ sign for even rank forms, while for Euclidean signature the signs are opposite to these.

An explicit definition of the dual in an $n$-dimensional space 
can be given in terms of the metric-compatible volume element $\e_{a_1\cdots a_n}$,
which is the unique, up to a sign, 
totally antisymmetric tensor normalized by $\e_{a_1\cdots a_n}\e^{a_1\cdots a_n}=\pm n!$,
with the $+$ sign for Euclidean and the $-$ sign for Lorentzian signature.
(The indices are raised by inverse metrics as usual.) The choice between the two volume
elements is a choice of orientation. The dual of a $p$-form 
with respect to this orientation is defined by 
\beq
*\b_{b_1\cdots b_{n-p}}=\frac{1}{p!}\b^{a_1\cdots a_{p}}\e_{a_1\cdots a_p b_1\cdots b_{n-p}}
\eeq
For any pair of $p$-forms $\a$ and $\b$, one has the useful relation
\beq\label{aw*b}
\a\w *\b = \la \a,\b\ra_g *1,
\eeq
where 
\beq
\la \a,\b\ra=\frac{1}{p!}\a_{m_1\cdots m_{p}}\b^{m_1\cdots m_{p}},
\eeq
and $*1$ is the volume element (with a choice of orientation). In fact, the
dual is defined implicitly by the relation \eqref{aw*b}.

\subsubsection{Diagonal metrics}
If the line element is written in diagonal form, it is particularly easy to 
find the action of the dual. 
For example, consider the Schwarzschild line element:
\beq
ds^2 = -A\, dt^2 + A^{-1} dr^2 + r^2 d\theta^2 + r^2 \sin^2\theta\,  d\varphi^2,
\eeq
with $A=1-2M/r$ (in units with $G=c=1)$. We can read off that the four 1-forms
$A^{\frac12}dt$, $A^{-\frac12}dr$, $rd\theta$, and $r\sin\theta\, d\varphi$ are orthonormal,
the first being timelike and the others spacelike. The dual
$*(d\theta\w d\vphi)$ is therefore proportional to $dt\w dr$. To determine the
coefficient function, we can simply scale all the forms so they have unit 
norm. Thus, 
\begin{align}
*(d\theta\w d\vphi)&=\frac{1}{r^2\sin\theta}*\bigl((rd\theta)\w(r\sin\theta\, d\varphi)\bigr)\nonumber\\
 &=\pm\frac{1}{r^2\sin\theta}\, (A^{\frac12}dt)\w (A^{-\frac12}dr)\nonumber\\
 &=\pm\frac{1}{r^2\sin\theta}\, dt\w dr \label{*thetaphi}
\end{align}
Since $**=-1$ on spacetime 2-forms, it follows that also $*(dt\w dr) = \mp r^2\sin\theta\, d\theta\w d\vphi$.
As explained above, the overall sign depends upon the orientation. 
According \eqref{aw*b} we have
$d\theta\w d\vphi\w*(d\theta\w d\vphi)=\la d\theta\w d\vphi,d\theta\w d\vphi\ra *1$,
and since the metric on the angular subspace is positive definite this is a positive
number times the volume element $*1$. Hence the sign in \eqref{*thetaphi} 
is $+$ for the orientation of $d\theta\w d\vphi \w dt\w dr$ and $-$ for the opposite 
orientation.

\subsubsection{Orthogonal subspaces}
\label{A+B}

Suppose the metric space $V$ of dimension $n$ 
is the direct sum of two orthogonal subspaces,
$V=A\oplus B$, of dimensions $n_A$ and $n_B$,
and let the orientations be chosen so that the 
volume elements are related by $\e =\e^A\w \e^B$. 
Then the dual of a wedge product $\a\w\b$ of an $A$-$p$-form with a $B$-$q$-form 
is given by 
\beq
*(\a\w\b)=(-1)^{q(n_{\!A}-p)}\star\!\a\w\star\b, 
\eeq
where the symbol $\star$ denotes the Hodge dual on the subspaces $A$ or $B$, defined with respect to $\e^A$ and $\e^B$. We will find this very useful for simple 2-forms in stationary, axisymmetric spacetimes, in which case $p=q=1$ and $n_A=2$, so the sign is $-$.

\subsubsection{Dual of a null 2-form}
\label{sec:null2dual}
A null 2-form has the composition $\a\w n$, with $n$ a null
1-form and $\a$ a spacelike 1-form orthogonal to $n$. 
This is orthogonal to itself, and has zero norm, so 
the intuitive definition we began with does not specify the dual. However, 
we can decompose the space as in the previous subsection,
with $\a$ in $A$ and $n$ in $B$, so that 
$*(\a\w n)= -\star\a\w \star n$. Now the dual of a null 1-form
$n$ in a two-dimensional space satisfies $n\w\star n=0$, so
$\star n\propto n$. Also  
$\star^2 n = n$, so $\star n =\pm n$. One of the two null directions
has the plus sign and the other has the minus sign, but which is 
which depends on the orientation of $\e^B$.
We thus have
\beq\label{nulldual}
*(\a\w n)= \pm \star\a\w n,\\
\eeq
where the $\pm$ corresponds to $\star n=\mp n$.
Note that the 1-form $\star\a$ is not unique, because any multiple of $n$ can
be added to it without changing the wedge product in (\ref{nulldual}).
Thus, the particular $2+2$ decomposition of the space plays no role:
$\star \a$ can be defined as any 1-form with the 
same norm as $\a$ and orthogonal to both $\a$ and $n$. 

An example to be used in the text involves the retarded time coordinate 
\beq\label{uschw}
u = t - r^*
\eeq
on the Schwarzschild spacetime, where $r^*$ is the radial ``tortoise coordinate"
defined by $dr^*= A^{-1} dr$. In terms of $u$, the Schwarzschild line element 
takes the Eddington-Finkelstein form, 
\beq
ds^2 = -A\, du^2 - 2\, du dr + r^2 d\theta^2 + r^2 \sin^2\theta\,  d\varphi^2.
\eeq
A surface of constant $u$ is an outgoing 
lightlike, spherical surface in spacetime, so $du$ is a null 1-form. The dual of null 2-forms
involving $du$ is given by 
\begin{align}
*(d\theta\w du) &=\pm\sin\theta\, d\vphi \w du,\label{*dthetadu}\\ 
*(d\vphi\w du) &=\mp(\sin\theta)^{-1}\, d\theta \w du.\label{*dphidu}
\end{align}
The relative sign of these two duals is fixed by the fact that $**=-1$ on 2-forms.
To fix the overall sign we adopt the orientation of $d\theta\w d\vphi \w dt\w dr$
for $\e$, $d\theta\w d\vphi $ for $\e^A$, and $dt\w dr=du\w dr$ for $\e^B$.
It is then simple to see that $\star d\theta= \sin\theta\, d\vphi$. 
To find the sign of
$\star du=\pm du$ we compute: 
$(\star du)_a = g^{bc}(\e^B)_{ca}(du)_b=g^{ur}(\e^B)_{ra}=(-1)(-du)_a$.
Hence $\star du = du$, so the \textit{upper} signs in (\ref{*dthetadu}) 
and (\ref{*dphidu})
apply. Had we used instead the advanced time coordinate, 
a similar calculation would have yielded $\star dv = -dv$, since
$g^{vr}=1$.

\subsubsection{Dual of a null 3-form}\label{sec:null3dual}

The electric current density is a 3-form, and we shall be interested in the case in which 
this 3-form is null. A null 3-form has the composition $\a\w\b\w n$, where 
$n$ is null and orthogonal to both $\a$ and $\b$. The dual of this is a 1-form that 
is orthogonal to all three of the forms in this triple wedge product, hence is a multiple
of $n$. We can understand this, and the coefficient, using the method of orthogonal subspaces
described in Sec.~\ref{A+B}, with the
$2+2$ decomposition into the subspace spanned by $\a\w\b$ and the 
orthogonal subspace. Then we have $*(\a\w\b\w n)=\star(\a\w\b)\w \star n= 
|\a\w\b| \star n$, 
so the proportionality factor is just the norm of the 2-form $\a\w\b$:
\beq
*(\a\w\b\w n)=\pm |\a\w\b| n.
\eeq
For example,  $d\theta\w d\vphi\w du$ is null, and 
\beq
*(d\theta\w d\vphi\w du)=-(r^2 \sin\theta)^{-1}\, du. 
\eeq
\subsection{Electromagnetism and differential forms}\label{sec:maxform}

Maxwell's equations \eqref{dF=0} and \eqref{sources} take an elegant form in the language of differential forms,
\begin{align}
dF &= 0 \label{maxform1} \\ 
d\!*\! F &= J, \label{maxform2}
\end{align}
where $F$ is the electromagnetic field $F_{ab}$ and
$J$ is the current 3-form, related to the current vector $j^a$ by $J_{abc}=j^m\e_{mabc}$.
Current conservation $dJ=0$ is implied by $dd=0$.  The charge that flows (in spacetime) through a patch of oriented 3-surface $\Sigma$ is the integral $\int_\Sigma J$. If $\Sigma$ is spacelike with future orientation this is the total charge in a spatial 3-volume, 
whereas if $\Sigma$ is timelike it is the net charge that flows in the orientation direction 
across a spatial 2-surface over a lapse of time.\footnote{Given a spacetime orientation $\e$, a direction of flow across a 3-surface $\Sigma$ corresponds to an orientation $v \cdot \e$ on $\Sigma$, where $v$ is any vector transverse to $\Sigma$ sharing the flow direction.  The integral $\int_\Sigma J$ with respect to this orientation gives the current flowing across $\Sigma$ in the sense of $v$.}

The integral of the 2-form $F$ on a spacelike 2-surface is the magnetic flux through that surface. 
The choice of orientation for the surface integral corresponds in 3+1 terms to the 
choice of sign for the normal to the surface when defining the flux of the magnetic field 
pseudovector.\footnote{The surface orientation $\e_2$ is related to a vector $a$ defining
the ``outward" direction for the flux by $\e_2=a\cdot( u\cdot\e)$ (up to positive rescalings),
where $u$ is a future timelike vector and $\e$ is the spacetime orientation.}
 This integral vanishes if the 2-surface is the (closed) boundary of a 3-ball and $dF=0$ holds everywhere in the interior.  The magnetic flux through two homologous surfaces bounded by the same loop must therefore be the same, so the flux through a loop is well-defined.\footnote{An eternal black hole provides an example with nontrivial homology.  If the black hole carries a magnetic monopole charge, then the fluxes through two surfaces spanning a loop will not be the same if the two surfaces together enclose the horizon.}

A surface layer between two spatial regions can support a discontinuity in the field $F$
by carrying a surface charge and/or current density. The \textit{jump conditions} restricting 
such discontinuities are naturally formulated in terms of the three-dimensional world-volume $\cS$ of the 
surface, which allows for arbitrary motion of the surface in time. These conditions are that 
the pullback to $\cS$ of the jump of $F$ vanishes, and the pullback to $\cS$ of the jump of $*F$ is equal to the current 2-form $K$
on $\cS$,
\begin{equation}\label{jumps}
\left[ F \right]_{\cS} = 0, \qquad \left[*F \right]_{\cS} = K.
\end{equation}
The jump is defined as the discontinuous change when crossing $\cS$ 
in a given arbitrary ``jump direction".
The 2-form $K$ is defined so that when integrated on a patch of two-dimensional surface contained in $\cS$ it yields the same result as $J$ integrated on an infinitesimal 
thickening of that patch transverse to $\cS$.  
The orientation $\e_3$ of the thickened patch should 
satisfy $v \cdot \epsilon_3=\epsilon_2$ (up to positive rescalings), where $\e_2$ is the orientation of the patch and $v$ is a vector pointing in the jump direction. The 2-form $K$ may also be expressed in terms of a distributional current 3-form $J_{\rm surf}$, related to $K$ via
\beq
J_{\rm surf} = \delta(s)\, ds\w K,
\eeq
where $s$ is any function that is constant on $\cS$ and increasing in the jump direction. The surface current $J_{\rm surf}$ does not depend on the jump direction. 

The jump conditions \eqref{jumps} are established by integrating Maxwell's equations \eqref{maxform1} and \eqref{maxform2} over the thickened patch and using Stokes' theorem, 
in the limit that the width of the thickening goes to zero. 
It is easily checked that for a surface at rest in an inertial frame in flat spacetime, these conditions
agree with the familiar ones: the tangential electric field and normal magnetic field must be continuous,
the jump in the normal electric field is the surface charge density, and the jump in the tangential
magnetic field is the cross product of the surface current density with the unit normal to the
surface in the direction the jump is defined.

\section{Poynting flux examples}\label{sec:flux}

The fact that stationary field configurations can carry energy away from a source in force-free electrodynamics is counter to intuition from the vacuum case, where this role is normally reserved for time-dependent fields sourced by accelerated charges.  In order to have finite, non-vanishing net flux from a central source the Poynting vector (or at least its angular integral) must fall off as $1/r^2$. 
This indicates that $E$ and $B$ should fall off as $1/r$, which in vacuum occurs only for time-dependent, radiative behavior. In the stationary, vacuum case, the $E$ and $B$ fields fall off as $1/r^2$ and $1/r^3$ (or $1/r^2$, if we allow monopoles), respectively.  The Poynting flux is thus at best $1/r^5$ and so there is no net flux through a large sphere.  (By Poynting's theorem there is then no net flux through any closed surface surrounding the source.)

The situation is different, however, if charge current extends from the source out into the surroundings.   A helpful example
is an electric circuit with a battery and a resistor. 
In a steady state,
power flows from the battery into the resistor, and the energy is carried by a Poynting flux largely in vacuum
between battery and resistor \citeeg{galili-goihbarg2005}. A force-free plasma is, in effect, a distributed circuit, in which a
similar effect can take place. In particular, considering for example a rotating conductor as a localized
energy source, if charge-current extends to infinity, then $1/r$ behavior for the fields is possible in the stationary case, provided the charge and current fall off as $1/r^2$.  This entails an unphysical infinite total charge on some (most) conical wedges of space, even though the total charge may be zero. However, in reality the force-free magnetosphere extends only a finite distance, and quantities that are finite in the infinite-$r$ limit (such as the net flux) should adequately represent the physics of a real configuration that extends to large but finite $r$.

In the remainder of this Appendix we consider three simple, quantitative examples that help to elucidate the role of the current in allowing for energy transport by Poynting flux in stationary fields. The examples are a plane symmetric vacuum solution, a 
coaxial cable, and a cylindrical plasma-filled waveguide. The last two examples 
were used by \cite{punsly2008} to illustrate features and provide intuition about the 
physics of MHD magnetospheres, and we adapt them here to the force-free setting. 

\subsection{Planar symmetry in vacuum}
A simple vacuum solution with planar symmetry is given by
\beq\label{Fplane}
F^{\rm plane} = f(u)\, dx \wedge du,
\eeq
where $(t,x,y,z)$ are Minkowski coordinates, $u=t-z$, and $f$ is an arbitrary function.  (To check that 
Maxwell's equations are satisfied, note that $dF^{\rm plane}=0$ follows immediately from $du\w du=0$,
and $F^{\rm plane}$ is a null 2-form so (cf.\ Appendix \ref{sec:null2dual}) 
$*F^{\rm plane}= -f(u) dy \wedge du$, which is similarly closed,  $d*F^{\rm plane}=0$.)  
Notice the similarity to the non-vacuum force-free solution \eqref{Fout}.  If the function $f(u)$ is sinusoidal, $f \sim \sin(\omega u)$, Eq.~\eqref{Fplane} would typically be described as a plane wave polarized in the $x$ direction,
but any function $f(u)$ gives a solution.  The energy flux (Poynting vector) is proportional to $f(u)^2$ and persists in the stationary case $f={\rm const}$, for which Eq.~\eqref{Fplane} represents static crossed electric and magnetic fields filling all of space.  

This stationary case is evidently an ``energy-transporting field",  but it has no physical source. 
Nevertheless, a solution with global planar symmetry reveals a possible \textit{local} behavior of electromagnetic fields.  Vacuum electrodynamics does not allow this local behavior to be extended globally
with a localized source.  On the other hand, the force-free case does allow such a global extension.
The correspondence can be made precise by noting that Eq.~\eqref{Fplane} arises in a planar limit of 
Eq.~\eqref{Fout}, where $z$ and $x,y$ are identified with the normal and tangential directions (respectively) to the sphere about a point. The charge-current vanishes in this limit.

\subsection{Coaxial cable}

A coaxial cable consists of a pair of concentric, cylindrical conductors, and
supports transverse electromagnetic (TEM) modes whose behavior is closely analogous 
to the planar case \eqref{Fplane}. The relevant solution is
\beq\label{Fcoax}
F^{\rm coax} = \frac{f(u)}{\r}\, d\r \wedge du,
\eeq
where $\rho$ is the cylindrical radius of the $x,y$ plane.
The demonstration that it satisfies the vacuum Maxwell equations
is essentially the same as for the planar case \eqref{Fplane}.  
This field tensor corresponds to a radial electric field
and a circumferential magnetic field, both of which are transverse to the 
propagation direction.  Assuming that there is no radial magnetic field in the conductor, the boundary condition at the vacuum/conductor interface is that the pullback of $F^{\rm coax}$ to the world-volume of the conducting walls vanishes (see discussion of boundary conditions in Sec.~\ref{sec:maxform}).
The world-volume of a cylinder contains no radial vector, while $F^{\rm coax}$ has a $d\rho$ factor, so this is satisfied.  Thus $F^{\rm coax}$ is indeed a TEM mode, which propagates at the speed of light,
and is terminated at the cylindrical conductors 
on which it induces charge and current.  If there were no central conductor, the
field would be singular on the axis (at $\r=0$).

As in the planar case, oscillatory solutions $f(u)\sim \sin u$ are usually considered, 
viewed as transmission modes in a coaxial cable.  However, also as in the planar case, 
the local energy flux (Poynting vector) is proportional to $f(u)^2$, 
and persists for any $f(u)$. The stationary solution 
is just static crossed electric and magnetic fields, sourced by an infinite line charge and current 
in the conductors.  

While energy transport  is not physically realizable in the strictly planar configuration \eqref{Fplane}, 
it \textit{is} realizable in the coaxial case, even for static fields.  
Imagine embedding a finite-length coaxial waveguide in a longitudinal magnetic field, and attaching a conducting disc to one end and a resistor connecting the
inner and outer cylinders at the other.  If the conducting disc is set spinning (while the walls are fixed), 
it becomes a ``Faraday disk'' electric generator, driving current in the $z$ direction along the  
inner cylinder and in the opposite direction along the outer cylinder.  
Far from either end, and after initial transients, field takes the form of a uniform
magnetic field in the $z$-direction, $B_0 \r \, d\r\w d\vphi$, plus a transverse part, 
\beq\label{Fresistor}
F^{\rm coax \& res}=  \frac{1}{2\pi\r}\,d\r \w (\l\, dt +I\, dz).
\eeq
The constant $\l$ is the linear charge density, which determines the strength
of the radial electric field, and the constant $I$ is the current along the $z$ direction
in the central conductor, which determines the azimuthal magnetic field strength. 
The linear charge density is fixed by the voltage drop $V$ between the walls 
(which is in turn fixed by the disk rotation rate and magnetic field strength $B_0$), 
while the current is given by Ohm's law $V=IR$ in terms of the resistance $R$ of the resistor.  
The static Poynting vector points from the disc to the resistor, and we may regard the Poynting flux as delivering the energy from the agent spinning the wheel to the resistor on the other end. 

The solution \eqref{Fcoax} with $f(u)$ constant arises when we take the special case of \eqref{Fresistor}
with $I=-\l$. This case is selected by some unremarkable particular value for the resistance, but it can
also be selected by a sort of ``no outer boundary'' condition. 
Suppose the waveguide is infinitely long, and that the Faraday disk starts turning at time 
$t=0$. Then at any time the fields should remain zero
beyond some distance from the wheel, and we may model this by 
$F^{\rm propagating} = \theta(vt - z) F^{\rm coax \& res}$, where
$\theta$ is the Heaviside step function and $v$ is a constant speed.
The Maxwell equations then imply $d\theta\w F^{\rm coax \& res}=0$
and $d\theta\w *F^{\rm coax \& res}=0$, which in turn imply $v=1$ and $I=-\l$. 
(Note that this implies the current is null.) 
That is, the wavefront propagates at the speed of light, and behind it we are left
with the solution \eqref{Fcoax} in the static case.  

\subsection{Plasma-filled waveguide}

Now suppose we take out the central cylinder in the coaxial waveguide and 
fill the cylinder with force-free plasma. Instead of the current being
carried on the central cylinder, it is distributed throughout the plasma. 
The field must satisfy the perfect conductor boundary condition on the outer cylinder 
and on the Faraday disk rotating with angular velocity $\O$,
and we suppose it has a uniform magnetic field of magnitude  $B_0$ in the $z$ direction. 
In a stationary, axisymmetric configuration, the total field must then have the form
\beq\label{waveguide0}
F^{\rm waveguide}=B_0 \r \, d\r\w \bigl(d\vphi - \O dt + d\psi_2(\r,z)\bigr),
\eeq
as explained in Appendix \ref{sec:appEPsymm}. The function $\psi_2$ can be determined by the requirement that the field satisfies the two force-free conditions.  The first one is $d\rho \wedge d*F=0$, which implies $\psi_2 = f(\rho) z$, and the second one (or the stream equation) then implies $f(\rho)=\pm \Omega$.  The result (after rejecting a solution that has a line current singularity on the axis) is $d\psi_2 =\pm \O\, dz$, hence
\beq\label{waveguide}
F^{\rm waveguide}=B_0 \r \, d\r\w \bigl(d\vphi - \O d(t \mp z)\bigr).
\eeq
One thus has a uniform magnetic field superposed with a Poynting flux in either the positive or negative $z$ direction.  This is a precise 
 cylindrical analog of the Michel monopole solution
(\ref{FMichel}) for a rotating, conducting sphere
 generating a spherical outgoing flux.
 As in that solution, the current associated with (\ref{waveguide}) is null, i.e.\,
the charge density and 3-current vector have equal magnitude. 
The 3-current is uniform and in the $z$-direction. The force-free
condition is satisfied by virtue of a balance between the radial
inward Lorentz force acting on the 3-current and the repulsive
radial electric force acting on the charge density. 
Note that, in contrast to the coaxial vacuum waveguide (\ref{Fresistor}), the current and charge density in the force-free plasma filled waveguide are not independently specificable. 

\section{Kerr metric}
\label{sec:Kerr}

In this appendix we present a number of useful formulae related to the Kerr spacetime.  For a more detailed treatment, see (e.g.) \citet{poisson-book}. 
The Kerr metric for a black hole of mass $M$ and angular momentum $aM$ is given in Boyer-Lindquist coordinates by
\begin{align}
ds^2 & = - \left(1-\frac{2Mr}{\Sigma}\right)dt^2 - \frac{4Mar \sin^2 \! \theta}{\Sigma} dt d\vphi \nonumber \\ & \qquad + \frac{A}{\Sigma} \sin^2 \! \theta \ d\vphi^2 + \frac{\Sigma}{\Delta} dr^2 + \Sigma d\theta^2 \\
& = - \frac{\Sigma \Delta}{A} dt^2 + \frac{A}{\Sigma} \sin^2 \! \theta \left( d\vphi - \Omega_Z dt \right)^2 + \frac{\Sigma}{\Delta} dr^2 + \Sigma d\theta^2,
\end{align}
where
\begin{align}
\Sigma=r^2+a^2 \cos^2 \! \theta, \quad \Delta = &\  r^2-2Mr+a^2 \\ A=(r^2+a^2)^2-a^2 \Delta \sin^2 \! \theta, \quad &\Omega_Z=2Mar/A.
\end{align}
The inner/outer horizons $r_\pm$ are the roots of $\D=(r-r_+)(r-r_-)$, $r_\pm=M\pm\sqrt{M^2-a^2}$. The Killing field $\partial_t + \Omega_H \partial_\vphi$ generates the horizon, where $\Omega_H=a/(r_+^2 + a^2)$ is called the horizon angular velocity. The relevant metric determinants are given by
\begin{align}
\sqrt{-g^T} = \sqrt{\D}\sin\theta, \qquad &\sqrt{g^P} = \Sigma/\sqrt{\D}, \\
\sqrt{-g} = \sqrt{-g^T g^P} &= \Sigma \sin \theta,
\end{align}
where $T$ and $P$ refer to toroidal ($t\vphi$) and poloidal ($r\theta$), respectively. The inverse metric components are
\begin{align}
g^{tt}=-A/(\Sigma \Delta), \ & g^{t\vphi}=-2Mar/(\Sigma \Delta), \nonumber \\ g^{\vphi\vphi}= (\Delta -a^2 \sin^2\!\theta)/(\Sigma & \Delta \sin^2 \!\theta), \ g^{rr} =\Delta/\Sigma, \ g^{\theta \theta}=1/\Sigma. 
\end{align}

The BL coordinates are singular on the future and past event horizons. The ingoing Kerr coordinates $v$ and $\tilde{\vphi}$ are regular on the future horizon (but not the past horizon), and are related to $t$ and $\vphi$ by
\begin{align}
dt &= dv -[(r^2 + a^2)/\D]dr, \label{v} \\
d\vphi &= d\tilde{\vphi} -(a/\D)dr. \label{vphih}
\end{align}
(In Schwarzschild spacetime $a=0$, $v$ is the ingoing Eddington-Finklekstein coordinate.) The Kerr metric becomes
\begin{align}
ds^2 & = - \left(1-\frac{2Mr}{\Sigma}\right)dv^2 + 2 dv dr \nonumber - 2 a \sin^2 \! \theta dr d\tilde{\vphi} \\ & \qquad - \frac{4Mar \sin^2\! \theta}{\Sigma} dv d\tilde{\vphi} + \frac{A}{\Sigma} \sin^2\! \theta d\tilde{\vphi}^2 + \Sigma d\theta^2 \label{kerrin1} \\
& = -dv^2 + 2 dr (dv-a \sin^2\! \theta d\tilde{\vphi}) + (r^2 + a^2) \sin^2\!\theta d\tilde{\vphi}^2 \nonumber \\ & \qquad + \Sigma d \theta^2 + \frac{2Mr}{\Sigma}(dv-a\sin^2\!\theta d\tilde{\vphi})^2. \label{kerrin2}
\end{align}
It follows immediately by inspection of \eqref{kerrin2} that the 1-form $dv-a\sin^2\!\theta d\tilde{\vphi}$ is equal to $(\partial_r)_a$, is null, and is orthogonal to $dv$, $d\tilde{\vphi}$, and $d\theta$.  We note that $(\partial_r)^a$ is proportional to the ingoing principal congruence.  The inverse metric components are
\begin{align}
g^{vv} = (a^2 \sin^2\!\theta)/\Sigma, \ g^{vr} = (r^2+a^2)/\Sigma, \ g^{v\tilde{\vphi}}=a/\Sigma, \nonumber \\
g^{rr} = \Delta/\Sigma, \ g^{r\tilde{\vphi}} = a/\Sigma, \ g^{\theta \theta}=1/\Sigma, g^{\tilde{\vphi}\tilde{\vphi}}=1/(\Sigma \sin^2\!\theta).
\end{align}

Alternatively one may use outgoing Kerr coordinates $u$ and $\bar{\vphi}$, defined by
\begin{align}
dt &= du +[(r^2 + a^2)/\D]dr, \label{u} \\
d\vphi &= d\bar{\vphi} + (a/\D)dr. \label{vphit}
\end{align}
(In Schwarzschild spacetime $a=0$, $u$ is the outgoing Eddington-Finklekstein coordinate.)
These coordinates are regular on the past horizon (but not the future horizon). They are useful for describing outgoing radiation processes, such as the Poynting flux solution discussed in the text. Analogous formulae for the metric may be obtained by exploiting the time reversal symmetry $t \rightarrow -t$ and $\vphi \rightarrow -\vphi$ of the Kerr metric. That is, one sends $v \rightarrow - u$ and $\tilde{\vphi} \rightarrow - \bar{\vphi}$ in \eqref{kerrin1} and \eqref{kerrin2}, which yields
\begin{align}
ds^2 & = - \left(1-\frac{2Mr}{\Sigma}\right)du^2 - 2 du dr \nonumber + 2 a \sin^2 \!\theta dr d\bar{\vphi} \\ & \qquad - \frac{4Mar \sin^2 \!\theta}{\Sigma} d u d\bar{\vphi} + \frac{A}{\Sigma} \sin^2\!\theta d\bar{\vphi}^2 + \Sigma d\theta^2 \label{kerrout1} \\
& = -du^2 - 2 dr (du-a \sin^2 \!\theta d\bar{\vphi}) + (r^2 + a^2) \sin^2\!\theta d\bar{\vphi}^2 \nonumber \\ & \qquad + \Sigma d \theta^2 + \frac{2Mr}{\Sigma}(du-a\sin^2\!\theta d\bar{\vphi})^2. \label{kerrout2}
\end{align}
As in the ingoing case, we see that the 1-form $du - a \sin^2 \!\theta d\bar{\vphi}$ is equal to $-(\partial_r)_a$, is null, and is orthogonal to $du$, $d\bar{\vphi}$, and $d\theta$.  The vector $(\partial_r)^a$ is proportional to the outgoing principal congruence.  The inverse metric components are
\begin{align}
g^{uu} = (a^2 \sin^2 \!\theta)/\Sigma, \ g^{ur} = - (r^2+a^2)/\Sigma, \ g^{u\bar{\vphi}}=a/\Sigma, \nonumber \\
g^{rr} = \Delta/\Sigma, \ g^{r\bar{\vphi}} = -a/\Sigma, \ g^{\theta \theta}=1/\Sigma, g^{\bar{\vphi}\bar{\vphi}}=1/(\Sigma \sin^2 \!\theta).
\end{align}

\section{Euler potentials with symmetry}
\label{sec:appEPsymm}

When a degenerate electromagnetic field has a symmetry, the corresponding 
Euler potentials do not in general have the same symmetry, but the form 
of their dependence on the ignorable coordinates 
is very constrained. \cite{uchida1997symmetry} solved the problem of finding their 
form in the presence of one symmetry or two commuting symmetries. In this 
appendix we follow his treatment, making use of 
differential forms to streamline the analysis. In the first and second
subsections we consider the case of one and two symmetries respectively, 
and in the final subsection we apply the results to
the case of stationary axisymmetry.

\subsection{One symmetry}\label{sec:onesymmetry}
Suppose the vector field $X$ generates a symmetry
of the field, $\cL_X F = 0$. Since the force-free equations
involve the metric, $X$ should presumably also generate a symmetry of the
metric, i.e.\ it should be a Killing vector. However the 
arguments in this section rely only on the symmetry properties
of $F$ and the metric-independent subset of Maxwell's equations, 
Faraday law, $dF=0$. 

Cartan's ``magic formula" [$\cL_X = (X{\cdot})d + d(X{\cdot})$]
and $dF=0$ imply that this symmetry condition is equivalent to $d (X{\cdot}   F) = 0$,
which implies (modulo homological obstructions) that $X{\cdot}   F$ is exact, i.e.\
\beq\label{invX}
X{\cdot}   F = df,
\eeq
for some function $f$. 
A degenerate field can be expressed in 
terms of Euler potentials as $F=d\phi_1\w d\phi_2$, in terms of which
(\ref{invX}) becomes
\beq\label{invXphi}
(X{\cdot}     d\phi_1)\, d\phi_2 - (X{\cdot}     d\phi_2)\, d\phi_1 = df.
\eeq
Since the differential of $f$ can be written as a linear combination of 
$d\phi_1$ and $d\phi_2$, evidently $f = f(\phi_1,\phi_2)$. 
If $X{\cdot}  F =0$, then both potentials are invariant under the symmetry,
but in general that is not the case.

Now recall that the Euler potentials are not unique: we may choose
any other pair ($\tilde{\phi}_1$,  $\tilde{\phi}_2$) such that 
$d\tilde{\phi}_1\w d\tilde{\phi}_2 = d\phi_1\w d\phi_2$, which amounts to
the requirement that the map $(\phi_1,\phi_2)\rightarrow
(\tilde{\phi}_1, \tilde{\phi}_2)$ have unit Jacobian determinant.
We may exploit this freedom to choose potentials that are 
adapted to the symmetry. 
In fact, if $df\ne0$ we may choose $\tilde\phi_1=-f$ to be one of a new pair of potentials. 
The unit Jacobian requirement then imposes a single, first order
partial differential equation on the other potential $\tilde\phi_2(\phi_1,\phi_2)$, 
which can be satisfied by integration with respect to $\phi_2$ provided
$\partial f/\partial\phi_1\ne0$, or by integration with respect to $\phi_1$ provided
$\partial f/\partial\phi_2\ne0$. 
The new pair will satisfy the analog of (\ref{invXphi}), with the
same function $f$ since that was defined in (\ref{invX}) without
reference to the potentials. In terms of this new pair, (\ref{invXphi}) becomes
\beq\label{invXphit}
(X{\cdot}     d\tilde\phi_1)\, d\tilde\phi_2 - (X{\cdot}     d\tilde\phi_2)\, d\tilde\phi_1 = -d\tilde\phi_1,
\eeq
from which we read off the symmetry properties
\beq\label{symEuler1}
X{\cdot}     d\tilde\phi_1 = 0, \quad X{\cdot}     d\tilde\phi_2=1.
\eeq
One of the potentials can thus be taken to be invariant, 
while the other has a constant, unit derivative along the symmetry flow.

\subsection{Two commuting symmetries}\label{sec:twosymmetries}

Suppose now that there are two commuting symmetry vectors, $X$ and $Y$, such that
$[X,Y]=\cL_XY=-\cL_Y X=0$, and $\cL_XF=\cL_YF=0$.  
Then, as in (\ref{invX}), we also have
\beq\label{invY}
Y{\cdot}  F = dg.
\eeq
for some $g$. To assess the relation between $df$ and $dg$ we compute their wedge product:
\beq\label{dfdg}
df\w dg = (X{\cdot}  F)\w (Y{\cdot}  F) = (Y{\cdot} X{\cdot}   F)F.
\eeq
The scalar $Y{\cdot}   X{\cdot}    F$ (=$F_{ab}X^aY^b$) must be constant:
\beq\label{lambdaconstant}
d(Y{\cdot}   X{\cdot}    F) =\cL_Y (X{\cdot}    F) - Y{\cdot}   d (X{\cdot}    F)=0.
\eeq
(The first term can be expanded using the Leibniz rule for the Lie derivative,
and both the resulting terms vanish, while the second term vanishes since
$d(X{\cdot}  F)=\cL_XF$.) 
Hence there are two cases to consider:
$Y{\cdot}   X{\cdot}    F=0$, which 
Uchida called Case I,  and $Y{\cdot}   X{\cdot}    F\ne0$, which he called Case II.
Case II will not be relevant when the two Killing fields are time translation
and rotations around an axis since, as explained below, no such field configuration
is regular on the axis, even if it is not force-free everywhere.

If both $X{\cdot}  F$ and $Y{\cdot} F$ vanish, 
then both potentials are simply invariant under 
both symmetries. Suppose now that $X{\cdot}  F\ne0$.  
Considering first only the vector field $X$, we may then conclude, as
in the one-symmetry case, that one of the Euler potentials may be chosen
to be $\phi_1= -f$.
In Case I we have $df\w dg=0$, from which it follows
that $g=g(\phi_1)$. The symmetry condition (\ref{invY}) for $Y$ then reads
\beq\label{invYphi}
Y{\cdot}  F = (Y{\cdot}    d\phi_1)\, d\phi_2 - (Y{\cdot}    d\phi_2)\, d\phi_1 = g'(\phi_1) d\phi_1.
\eeq
It follows from (\ref{invYphi}) and (\ref{symEuler1})
(without the tildes)
that the potentials have the symmetry properties
\begin{align}
X{\cdot}     d\phi_1 = 0,\quad & X{\cdot}    d\phi_2=1,\label{symEulerIX}\\
 Y{\cdot}    d\phi_1 = 0 ,\quad &Y{\cdot}    d\phi_2=\kappa(\phi_1),\label{symEulerIY}
\end{align}
where $\kappa(\phi_1)=-g'(\phi_1)$.

The function $\kappa(\phi_1)$ has an interesting geometric interpretation.
The potentials are both invariant with respect to the flow of the vector field
\beq
Z=Y-\kappa(\phi_1)X,
\eeq
hence $Z$ generates a symmetry of the field
and is tangent to the field surfaces. 
If $X$ and $Y$ 
are spacetime Killing vectors then $Z$ is also 
only if $\kappa(\phi_1)$ is constant, but it 
is always a Killing vector of the induced metric
on the field sheets because $\phi_1$ is constant on each field sheet. 
That is, $Z$ is a \textit{field sheet Killing vector}.
 
In case II both $X{\cdot}  F$ and $Y{\cdot} F$ must be nonvanishing, 
and according to (\ref{dfdg}) and (\ref{lambdaconstant}), we may choose the 
Euler potentials to be $\phi_1=- f$ and $\phi_2 =  g/\lambda$,
where $\lambda = Y{\cdot} X{\cdot}   F$. 
Then (\ref{invXphi}) (without the tildes) and (\ref{invYphi}) imply 
the symmetry conditions
\begin{align}\label{symEulerII}
X{\cdot}     d\phi_1 = 0, \quad &X{\cdot}     d\phi_2=1,\\
Y{\cdot}    d\phi_1 = \lambda, \quad &Y{\cdot}    d\phi_2=0.
\end{align}
In this case, no linear combination of $X$ and $Y$ is 
tangent to the field surfaces.

\subsection{Stationary axisymmetry}

In stationary axisymmetry,
there are two commuting Killing fields, $\partial_t$
and $\partial_\vphi$, where $t$ and $\vphi$ are Killing coordinates in some coordinate system. (For example, they could be the usual Boyer-Lindquist coordinates for the Kerr metric, but they could also be, say, the ingoing Kerr coordinate $v$ and azimuthal angle $\tilde\vphi$ respectively. These differ by the addition of functions of the coordinate $r$, so yield different choices for the potentials below.)
Case II does not occur for such fields,
since $\partial_\vphi$ vanishes
on the symmetry axis, so the constant 
$\partial_t\cdot\partial_\vphi\cdot F$
always vanishes. Thus we consider only Case I.

If $\partial_t{\cdot}F$ and $\partial_\vphi{\cdot}F$ both vanish, 
then both the potentials are independent of $t$ and $\vphi$, 
hence the field tensor is purely poloidal ($F\sim dr\w d\theta$).

Now let $X=\partial_\vphi$ and 
$Y=\partial_t$, in the notation of the previous
subsection. 
If $\partial_\vphi{\cdot}F\ne0$
we have from (\ref{symEulerIX}) and 
(\ref{symEulerIY}) that for
stationary, axisymmetric fields, the Euler potentials may always
be taken to have the form 
\beq\label{EPsymmAppendix}
\phi_1 = \psi(r,\theta), \quad \phi_2 = \psi_2(r,\theta) + \varphi - \Omega_F(\psi) t,
\eeq
for some function $\O_F(\psi)$. We have replaced the notation $\kappa$ by $-\O_F$
since, as explained in the text, $\O_F$ corresponds to the ``angular velocity of the
field lines".

An exceptional case not mentioned by Uchida occurs if 
instead $\partial_\vphi{\cdot}F=0$, i.e.\ if 
there is no poloidal magnetic field.
Then we must reverse the 
roles of $X$ and $Y$ before invoking the results of the previous subsection,
and the Euler potentials may always
be taken to have the form 
\beq\label{EPsymmAppendixEx}
\phi_1 = \psi(r,\theta), \quad \phi_2 = \psi_2(r,\theta) + t.
\eeq
This can be viewed as a singular limit of \eqref{EPsymmAppendix}
in which $\O_F$ and $\psi_2$ go to infinity while $\psi$ goes to zero,
with the products held finite.

\section{Conserved Noether current associated with a symmetry}
\label{Noether}

Let $L$ be the Lagrangian 4-form of some field theory, i.e.\ it depends on various dynamical
fields and perhaps on some background fields. 
Suppose the vector field $\xi^a$ generates a symmetry of the dynamics, in the sense that
when the dynamical fields are varied by their Lie derivative with respect to $\xi^a$, the net 
induced variation of $L$ is simply the Lie derivative of $L$ itself as 
a 4-form, $\cL_\xi L=d(\xi\cdot L)$. Since this is a total derivative, 
the variation of the action $\int L$ will be at most a boundary term. 
Typically this will be the case if the background fields in $L$ have zero Lie derivative
with respect to $\xi$. In this case, there is a Noether current 3-form $\cJ_\xi  $ that is closed (conserved) when
the dynamical equations of motion are satisfied. To see how this comes about, and how 
$J_{\xi}$ is defined, we can just make the two variations in question. 

Let $\Phi$ stand for
all the dynamical fields, and let $E$ be their equations of motion form. Then the two variations 
are:
\begin{align}
\d_{\rm dynamical} L &= E \, \cL_\xi \Phi + d\theta(\cL_\xi \Phi)\\
\d_{\rm total} L &= \cL_\xi L = d(\xi\cdot L).
\end{align}
The 3-form $\theta$ depends linearly on the variation $\cL_\xi \Phi$, and also
on the fields. It is called the symplectic potential, and its integral over a 
spacelike (Cauchy) surface is the field-theoretic analog
of $p_i\,dq^i$ in mechanics. Setting these two variations equal, we have 
\beq\label{dJ}
d\bigl(\theta(\cL_\xi \Phi) -\xi\cdot L\bigr) = - E \, \cL_\xi \Phi. 
\eeq
When the equations of motion are satisfied, $E=0$, the Noether current
$\cJ_\xi  $ is closed, where 
\beq\label{Jxi}
\cJ_\xi   = \theta(\cL_\xi \Phi) -\xi\cdot L
\eeq
For instance when $\xi$ is the time-translation vector $\partial_t$, 
the current
$\cJ_\xi  $ is the field theory analog of $p\dot q - L$, the canonical 
Hamiltonian.

If the only background field is the spacetime metric, then we get a 
conserved current for every Killing vector, and perhaps more conserved
currents, if $L$ does not depend on all aspects of the metric. For example,
in vacuum or force-free electromagnetism, $L$ depends only on the 
conformal structure, so also conformal Killing vectors produce conserved currents.
Note that for a field configuration such that the total variation $ \cL_\xi L = d(\xi\cdot L)$
vanishes, the second term in the Noether current (\ref{Jxi}) is 
automatically conserved by itself, without appeal to field equations.
In this case, the first term also is conserved by itself, when the equations
of motion hold.

\subsection{Noether currents for electromagnetic field}

In the usual Lagrangian formulation of electrodynamics, 
the dynamical field is the vector potential $A$, the field
strength is $F=dA$, and the Lagrangian 4-form is $-\half F\w *F$,
plus any interaction terms.  The Lagrangian is invariant (possibly only
up to addition of an exact form, i.e.\ a total derivative) under 
gauge transformations of the potential, $A\rightarrow A + d\l$, 
where $\l$ is any scalar function $\l$. 

The electromagnetic Noether current 
associated with a vector field $\xi$ can be constructed as above,
yielding
\beq\label{JA1}
\cJ_\xi   = -\cL_\xi A\w *F +\half\xi\cdot(F\w*F).
\eeq
If $A$ is treated as an ordinary 1-form, substituting 
\beq\label{LieA}
\cL_\xi A = \xi\cdot F + d(\xi\cdot A)
\eeq
into (\ref{JA1}), the result is not gauge-invariant,
although it is still a correct contribution to a conserved current. 
The terms that violate gauge-invariance consist of one that vanishes by the 
equations of motion, and an exact form that is automatically 
conserved by itself.
One could drop those terms to arrive at a gauge-invariant Noether current.
A more insightful way to arrive at the same current
is to note that the response of $A$ to the diffeomorphism generated by $\xi$ is
defined only up to a gauge transformation.
We can define a gauge-invariant response by omitting the gauge transformation term
$d(\xi\cdot A)$ from \eqref{LieA}, yielding a ``gauge-invariant Lie derivative",
\beq
\cL_\xi' A = \xi\cdot F.
\eeq
The reasoning leading to the 
conserved Noether current \eqref{Jxi} can be applied using this variation, which leads directly to
the Noether current
\beq\label{JA2}
\cJ_\xi   = -(\xi\cdot F)\w *F +\half\xi\cdot(F\w*F). 
\eeq
Note that for configurations that share the Killing symmetry, i.e. 
such that $\cL_\xi F=0$, the second term in
\eqref{JA2} is conserved independently of field equations,
so the first term is conserved by itself when the field equations hold.

\bibliography{spacetimeFF}

\begin{thebibliography}{82}
\expandafter\ifx\csname natexlab\endcsname\relax\def\natexlab#1{#1}\fi

\bibitem[{{Alfv{\'e}n}(1942)}]{alfven1942}
{Alfv{\'e}n} H., 1942, \nat, 150, 405

\bibitem[{{Alic} {et~al}\mbox{.}(2012){Alic}, {Moesta}, {Rezzolla}, {Zanotti}, \& {Jaramillo}}]{alic-etal2012}
{Alic} D., {Moesta} P., {Rezzolla} L., {Zanotti} O., {Jaramillo} J.~L., 2012, \apj, 754, 36

\bibitem[{{Bardeen}, {Carter} \& {Hawking}(1973){Bardeen}, {Carter}, \& {Hawking}}]{bardeen-carter-hawking1973}
{Bardeen} J.~M., {Carter} B., {Hawking} S.~W., 1973, Communications in Mathematical Physics, 31, 161

\bibitem[{{Bekenstein}(1973)}]{bekenstein1973}
{Bekenstein} J.~D., 1973, \prd, 7, 2333

\bibitem[{{Beskin}(2010)}]{beskin-book2010}
{Beskin} V.~S., 2010, {MHD Flows in Compact Astrophysical Objects: Accretion, Winds and Jets}. Springer-Verlag, Berlin

\bibitem[{{Blandford}(1976)}]{blandford1976}
{Blandford} R.~D., 1976, \mnras, 176, 465

\bibitem[{{Blandford}(2002)}]{blandford-lighthouse}
{Blandford} R.~D., 2002, in Lighthouses of the Universe: The Most Luminous Celestial Objects and Their Use for Cosmology, {Gilfanov} M., {Sunyeav} R., {Churazov} E., eds., p. 381

\bibitem[{{Blandford} \& {Znajek}(1977)}]{blandford-znajek1977}
{Blandford} R.~D., {Znajek} R.~L., 1977, \mnras, 179, 433

\bibitem[{{Bogovalov}(1999)}]{bogovalov1999}
{Bogovalov} S.~V., 1999, \aap, 349, 1017

\bibitem[{{Brennan} \& {Gralla}(2014)}]{brennan-gralla2014}
{Brennan} T.~D., {Gralla} S.~E., 2014, \prd, 89, 103013

\bibitem[{{Brennan}, {Gralla} \& {Jacobson}(2013){Brennan}, {Gralla}, \& {Jacobson}}]{brennan-gralla-jacobson2013}
{Brennan} T.~D., {Gralla} S.~E., {Jacobson} T., 2013, Classical and Quantum Gravity, 30, 195012

\bibitem[{{Buniy} \& {Kephart}(2014)}]{buniy-kephart2014}
{Buniy} R.~V., {Kephart} T.~W., 2014, Annals of Physics, 344, 179

\bibitem[{{Carter}(1970)}]{carter1970}
{Carter} B., 1970, Communications in Mathematical Physics, 17, 233

\bibitem[{{Carter}(1979)}]{carter1979}
{Carter} B., 1979, in {General Relativity: An Einstein Centenary Survey, edited by S. W. Hawking and W. Israel}, Vol. 179, Cambridge University Press, Cambridge, England, pp. 457--472

\bibitem[{{Choquet-Bruhat} \& {Dewitt-Morette}(1982)}]{choquetbruhat-dewitt1982}
{Choquet-Bruhat} Y., {Dewitt-Morette} C., 1982, Analysis, manifolds and physics, Part I: basics, Revised edition. North Holland

\bibitem[{{Contopoulos}, {Kalapotharakos} \& {Kazanas}(2014){Contopoulos}, {Kalapotharakos}, \& {Kazanas}}]{contopoulos-kalapotharakos-kazanas2014}
{Contopoulos} I., {Kalapotharakos} C., {Kazanas} D., 2014, \apj, 781, 46

\bibitem[{{Contopoulos}, {Kazanas} \& {Fendt}(1999){Contopoulos}, {Kazanas}, \& {Fendt}}]{contopoulos-kazanas-fendt1999}
{Contopoulos} I., {Kazanas} D., {Fendt} C., 1999, \apj, 511, 351

\bibitem[{{Contopoulos}, {Kazanas} \& {Papadopoulos}(2013){Contopoulos}, {Kazanas}, \& {Papadopoulos}}]{contopoulos-kazanas-papadopoulos2013}
{Contopoulos} I., {Kazanas} D., {Papadopoulos} D.~B., 2013, \apj, 765, 113

\bibitem[{{Contopoulos}(1995)}]{contopoulos1995}
{Contopoulos} J., 1995, \apj, 450, 616

\bibitem[{{Galili} \& {Goihbarg}(2005)}]{galili-goihbarg2005}
{Galili} I., {Goihbarg} E., 2005, American Journal of Physics, 73, 141

\bibitem[{{Gold}(1968)}]{gold1968}
{Gold} T., 1968, \nat, 218, 731

\bibitem[{{Goldreich} \& {Julian}(1969)}]{goldreich-julian1969}
{Goldreich} P., {Julian} W.~H., 1969, \apj, 157, 869

\bibitem[{{Gourgoulhon} \& {Bonazzola}(1993)}]{gourgoulhon-bonazzola1993}
{Gourgoulhon} E., {Bonazzola} S., 1993, \prd, 48, 2635

\bibitem[{{Gourgoulhon} {et~al}\mbox{.}(2011){Gourgoulhon}, {Markakis}, {Ury{\= u}}, \& {Eriguchi}}]{gourgoulhon-etal2011}
{Gourgoulhon} E., {Markakis} C., {Ury{\= u}} K., {Eriguchi} Y., 2011, \prd, 83, 104007

\bibitem[{{Gruzinov}(2006)}]{gruzinov2006}
{Gruzinov} A., 2006, astro-ph/0604364

\bibitem[{{Gruzinov}(2011)}]{gruzinov2011}
{Gruzinov} A., 2011, arXiv:1101.3100

\bibitem[{{Harris}(1962)}]{harris1962}
{Harris} E.~G., 1962, Il Nuovo Cimento Series 10, 23, 115

\bibitem[{{Hartle} \& {Thorne}(1968)}]{hartle-thorne1968}
{Hartle} J.~B., {Thorne} K.~S., 1968, \apj, 153, 807

\bibitem[{{Hawking}(1972)}]{hawking1972}
{Hawking} S.~W., 1972, Communications in Mathematical Physics, 25, 152

\bibitem[{{Hawking} \& {Ellis}(1973)}]{hawking-ellis1973}
{Hawking} S.~W., {Ellis} G.~F.~R., 1973, {The large-scale structure of space-time.} Cambridge (UK): Cambridge University Press, 11 + 391 p.

\bibitem[{{Hewish} {et~al}\mbox{.}(1968){Hewish}, {Bell}, {Pilkington}, {Scott}, \& {Collins}}]{hewish-etal1968}
{Hewish} A., {Bell} S.~J., {Pilkington} J.~D.~H., {Scott} P.~F., {Collins} R.~A., 1968, \nat, 217, 709

\bibitem[{{Hirose} {et~al}\mbox{.}(2004){Hirose}, {Krolik}, {De Villiers}, \& {Hawley}}]{hirose-etal2004}
{Hirose} S., {Krolik} J.~H., {De Villiers} J.-P., {Hawley} J.~F., 2004, \apj, 606, 1083

\bibitem[{{Kalapotharakos}, {Contopoulos} \& {Kazanas}(2012){Kalapotharakos}, {Contopoulos}, \& {Kazanas}}]{kalapotharakos-contopoulos-kazanas2012}
{Kalapotharakos} C., {Contopoulos} I., {Kazanas} D., 2012, \mnras, 420, 2793

\bibitem[{{Komissarov}(2001)}]{komissarov2001}
{Komissarov} S.~S., 2001, \mnras, 326, L41

\bibitem[{{Komissarov}(2002)}]{komissarov2002}
{Komissarov} S.~S., 2002, \mnras, 336, 759

\bibitem[{{Komissarov}(2004)}]{komissarov2004}
{Komissarov} S.~S., 2004, \mnras, 350, 427

\bibitem[{{Komissarov}(2009)}]{komissarov2009}
{Komissarov} S.~S., 2009, Journal of Korean Physical Society, 54, 2503

\bibitem[{{Komissarov} \& {McKinney}(2007)}]{komissarov-mckinney2007}
{Komissarov} S.~S., {McKinney} J.~C., 2007, \mnras, 377, L49

\bibitem[{{Lasota} {et~al}\mbox{.}(2014){Lasota}, {Gourgoulhon}, {Abramowicz}, {Tchekhovskoy}, \& {Narayan}}]{lasota-etal2014}
{Lasota} J.-P., {Gourgoulhon} E., {Abramowicz} M., {Tchekhovskoy} A., {Narayan} R., 2014, \prd, 89, 024041

\bibitem[{{Lyutikov}(2011)}]{lyutikov2011}
{Lyutikov} M., 2011, \prd, 83, 124035

\bibitem[{{Lyutikov} \& {McKinney}(2011)}]{lyutikov-mckinney2011}
{Lyutikov} M., {McKinney} J.~C., 2011, \prd, 84, 084019

\bibitem[{{MacDonald} \& {Thorne}(1982)}]{macdonald-thorne1982}
{MacDonald} D., {Thorne} K.~S., 1982, \mnras, 198, 345

\bibitem[{{McKinney}(2005)}]{mckinney2005}
{McKinney} J.~C., 2005, \apjl, 630, L5

\bibitem[{{McKinney}(2006)}]{mckinney2006}
{McKinney} J.~C., 2006, \mnras, 368, L30

\bibitem[{{McKinney} \& {Gammie}(2004)}]{mckinney-gammie2004}
{McKinney} J.~C., {Gammie} C.~F., 2004, \apj, 611, 977

\bibitem[{{McKinney}, {Tchekhovskoy} \& {Blandford}(2012){McKinney}, {Tchekhovskoy}, \& {Blandford}}]{mckinney-tchekhovskoy-blandford2012}
{McKinney} J.~C., {Tchekhovskoy} A., {Blandford} R.~D., 2012, \mnras, 423, 3083

\bibitem[{{Menon} \& {Dermer}(2007)}]{menon-dermer2007}
{Menon} G., {Dermer} C.~D., 2007, General Relativity and Gravitation, 39, 785

\bibitem[{{Menon} \& {Dermer}(2011)}]{menon-dermer2011}
{Menon} G., {Dermer} C.~D., 2011, \mnras, 417, 1098

\bibitem[{{Michel}(1973)}]{michel1973}
{Michel} F.~C., 1973, \apj, 180, 207

\bibitem[{{Nathanail} \& {Contopoulos}(2014)}]{nathanail-contopoulos2014}
{Nathanail} A., {Contopoulos} I., 2014, \apj, 788, 186

\bibitem[{{Northrop} \& {Teller}(1960)}]{northrop-teller1960}
{Northrop} T.~G., {Teller} E., 1960, Physical Review, 117, 215

\bibitem[{{Okamoto}(1974)}]{okamoto1974}
{Okamoto} I., 1974, \mnras, 167, 457

\bibitem[{{Pacini}(1968)}]{pacini1968}
{Pacini} F., 1968, \nat, 219, 145

\bibitem[{{Palenzuela} {et~al}\mbox{.}(2011){Palenzuela}, {Bona}, {Lehner}, \& {Reula}}]{palenzuela-etal2011}
{Palenzuela} C., {Bona} C., {Lehner} L., {Reula} O., 2011, Classical and Quantum Gravity, 28, 134007

\bibitem[{{Palenzuela}, {Lehner} \& {Liebling}(2010){Palenzuela}, {Lehner}, \& {Liebling}}]{palenzuela-lehner-liebling2010}
{Palenzuela} C., {Lehner} L., {Liebling} S.~L., 2010, Science, 329, 927

\bibitem[{{Palenzuela} {et~al}\mbox{.}(2013){Palenzuela}, {Lehner}, {Ponce}, {Liebling}, {Anderson}, {Neilsen}, \& {Motl}}]{palenzuela-etal2013}
{Palenzuela} C., {Lehner} L., {Ponce} M., {Liebling} S.~L., {Anderson} M., {Neilsen} D., {Motl} P., 2013, Physical Review Letters, 111, 061105

\bibitem[{{Paschalidis}, {Etienne} \& {Shapiro}(2013){Paschalidis}, {Etienne}, \& {Shapiro}}]{paschalidis-etienne-shapiro2013}
{Paschalidis} V., {Etienne} Z.~B., {Shapiro} S.~L., 2013, \prd, 88, 021504

\bibitem[{{Penrose}(1969)}]{penrose1969}
{Penrose} R., 1969, Nuovo Cimento Rivista Serie, 1, 252

\bibitem[{{Penrose}(2002)}]{penrose19692002}
{Penrose} R., 2002, General Relativity and Gravitation, 34, 1141

\bibitem[{{Penrose} \& {Floyd}(1971)}]{penrose-floyd1971}
{Penrose} R., {Floyd} R.~M., 1971, Nature Physical Science, 229, 177

\bibitem[{{Penrose} \& {Rindler}(1984)}]{penrose-rindler-book1}
{Penrose} R., {Rindler} W., 1984, {Spinors and space-time. Vol. 1: Two-spinor calculus and relativistic fields.} Cambridge University Press, Cambridge (1984).

\bibitem[{{Pfeiffer} \& {MacFadyen}(2013)}]{pfeiffer-macfadyen2013}
{Pfeiffer} H.~P., {MacFadyen} A.~I., 2013, arXiv:1307.7782

\bibitem[{{Poisson}(2004)}]{poisson-book}
{Poisson} E., 2004, {A relativist's toolkit: the mathematics of black-hole mechanics}. Cambridge, UK: Cambridge University Press, 2004

\bibitem[{{Punsly}(2008)}]{punsly2008}
{Punsly} B., 2008, {Black hole gravitohydromagnetics}. New York: Springer, c2008.~Astronomy and astrophysics library, 0941-7834

\bibitem[{{Robinson}(1961)}]{robinson1961}
{Robinson} I., 1961, Journal of Mathematical Physics, 2, 290

\bibitem[{{Ruderman} \& {Sutherland}(1975)}]{ruderman-sutherland1975}
{Ruderman} M.~A., {Sutherland} P.~G., 1975, \apj, 196, 51

\bibitem[{{Scharlemann} \& {Wagoner}(1973)}]{scharlemann-wagoner1973}
{Scharlemann} E.~T., {Wagoner} R.~V., 1973, \apj, 182, 951

\bibitem[{{Schmidt}(1963)}]{schmidt1963}
{Schmidt} M., 1963, \nat, 197, 1040

\bibitem[{{Spitkovsky}(2006)}]{spitkovsky2006}
{Spitkovsky} A., 2006, \apjl, 648, L51

\bibitem[{{Takamori} {et~al}\mbox{.}(2011){Takamori}, {Nakao}, {Ishihara}, {Kimura}, \& {Yoo}}]{takamori-etal2011}
{Takamori} Y., {Nakao} K.-I., {Ishihara} H., {Kimura} M., {Yoo} C.-M., 2011, \mnras, 412, 2417

\bibitem[{{Thompson} \& {Blaes}(1998)}]{thompson-blaes1998}
{Thompson} C., {Blaes} O., 1998, \prd, 57, 3219

\bibitem[{{Thorne}, {Price} \& {MacDonald}(1986){Thorne}, {Price}, \& {MacDonald}}]{membrane-book}
{Thorne} K.~S., {Price} R.~H., {MacDonald} D.~A., 1986, {Black holes: The membrane paradigm}

\bibitem[{{Timokhin}(2006)}]{timokhin2006}
{Timokhin} A.~N., 2006, \mnras, 368, 1055

\bibitem[{{Uchida}(1997{\natexlab{a}})}]{uchida1997general}
{Uchida} T., 1997{\natexlab{a}}, \pre, 56, 2181

\bibitem[{{Uchida}(1997{\natexlab{b}})}]{uchida1997symmetry}
{Uchida} T., 1997{\natexlab{b}}, \pre, 56, 2198

\bibitem[{{Uchida}(1997{\natexlab{c}})}]{uchida1997linear1}
{Uchida} T., 1997{\natexlab{c}}, \mnras, 286, 931

\bibitem[{{Uchida}(1997{\natexlab{d}})}]{uchida1997linear2}
{Uchida} T., 1997{\natexlab{d}}, \mnras, 291, 125

\bibitem[{{Uchida}(1998)}]{uchida1998oblique}
{Uchida} T., 1998, \mnras, 297, 315

\bibitem[{{Uzdensky}(2005)}]{uzdensky2005}
{Uzdensky} D.~A., 2005, \apj, 620, 889

\bibitem[{{Wald}(1974)}]{wald1974}
{Wald} R.~M., 1974, \prd, 10, 1680

\bibitem[{{Wald}(1984)}]{wald-book1984}
{Wald} R.~M., 1984, {General relativity}. Chicago, University of Chicago Press, 1984, 504 p.

\bibitem[{{Znajek}(1977)}]{znajek1977}
{Znajek} R.~L., 1977, \mnras, 179, 457

\end{thebibliography}
\bibliographystyle{mn2e}

\tableofcontents

\end{document}